\documentclass{ar-1col} 

\usepackage{url, natbib, verbatim}
\usepackage{amsmath,amssymb}
\usepackage[flushleft]{threeparttable}

\newcommand{\mnras}{MNRAS}
\newcommand{\apj}{ApJ}
\newcommand{\apjs}{ApJS}
\newcommand{\apjl}{ApJL}

\newcommand{\nat}{Nature}
\newcommand{\aap}{A\&A}
\newcommand{\aj}{AJ}
\newcommand{\araa}{ARA\&A}

\newcommand{\pasa}{PASA}

\newcommand{\Msun}{${\rm M}_{\odot}$}

\newcommand\ion[2]{#1$\;${\scshape{#2}}}%
\newcommand{\lya}{Ly$\alpha$}

\newcommand{\hi}{\ion{H}{i}}

\newcommand{\cii}{\ion{C}{ii}}
\newcommand{\ciii}{\ion{C}{iii}}
\newcommand{\civ}{\ion{C}{iv}}
\newcommand{\feii}{\ion{Fe}{ii}}
\newcommand{\nii}{\ion{N}{ii}}
\newcommand{\niii}{\ion{N}{iii}}
\newcommand{\nv}{\ion{N}{v}}

\newcommand{\siii}{\ion{Si}{ii}}
\newcommand{\siiii}{\ion{Si}{iii}}

\newcommand{\mgii}{\ion{Mg}{ii}}
\newcommand{\caii}{\ion{Ca}{ii}}

\newcommand{\ovi}{\ion{O}{vi}}
\newcommand{\ovii}{\ion{O}{vii}}
\newcommand{\neviii}{\ion{Ne}{viii}}
\newcommand{\msfr}{\dot{M}_{\rm{sfr}}}
\newcommand{\mhalo}{M_{\rm halo}}
\newcommand{\mstar}{M_{\star}}

\newcommand{\rvir}{R_{\rm vir}}

\jname{Annual Review of Astronomy and Astrophysics}
\jvol{AA}
\jyear{2017}
\doi{/10.1146/annurev-astro-091916-055240}

\begin{document}

\markboth{Tumlinson, Peeples, \&\ Werk}{The Circumgalactic Medium}

\title{The Circumgalactic Medium}

\author{Jason Tumlinson$^1$, \\
Molly S. Peeples$^1$, \\
\& Jessica K. Werk$^2$
\affil{$^1$Space Telescope Science Institute and Johns Hopkins University, Baltimore, Maryland; email: tumlinson@stsci.edu, molly@stsci.edu}
\affil{$^2$University of Washington, Seattle, Washington, email: jwerk@uw.edu}} 

\begin{abstract}
The gas surrounding galaxies outside their disks or ISM and inside their virial radii is known as the ``circumgalactic medium'' (CGM). In recent years this component of galaxies has assumed an important role in our understanding of galaxy evolution owing to rapid advances in observational access to this diffuse, nearly invisible material. Observations and simulations of this component of galaxies suggest that it is a  multiphase medium characterized by rich dynamics and complex ionization states. The CGM is a source for a galaxy's star-forming fuel, the venue for galactic feedback and recycling, and perhaps the key regulator of the galactic gas supply. We review our evolving knowledge of the CGM with emphasis on its mass, dynamical state, and co-evolution with galaxies. Observations from all redshifts and from across the electromagnetic spectrum indicate that CGM gas has a key role in galaxy evolution. We summarize the state of this field and pose unanswered questions for future research. 
\end{abstract}


\maketitle

\tableofcontents

\section{A Very Brief History}

\begin{figure}[t]
\includegraphics[width=6.3in]{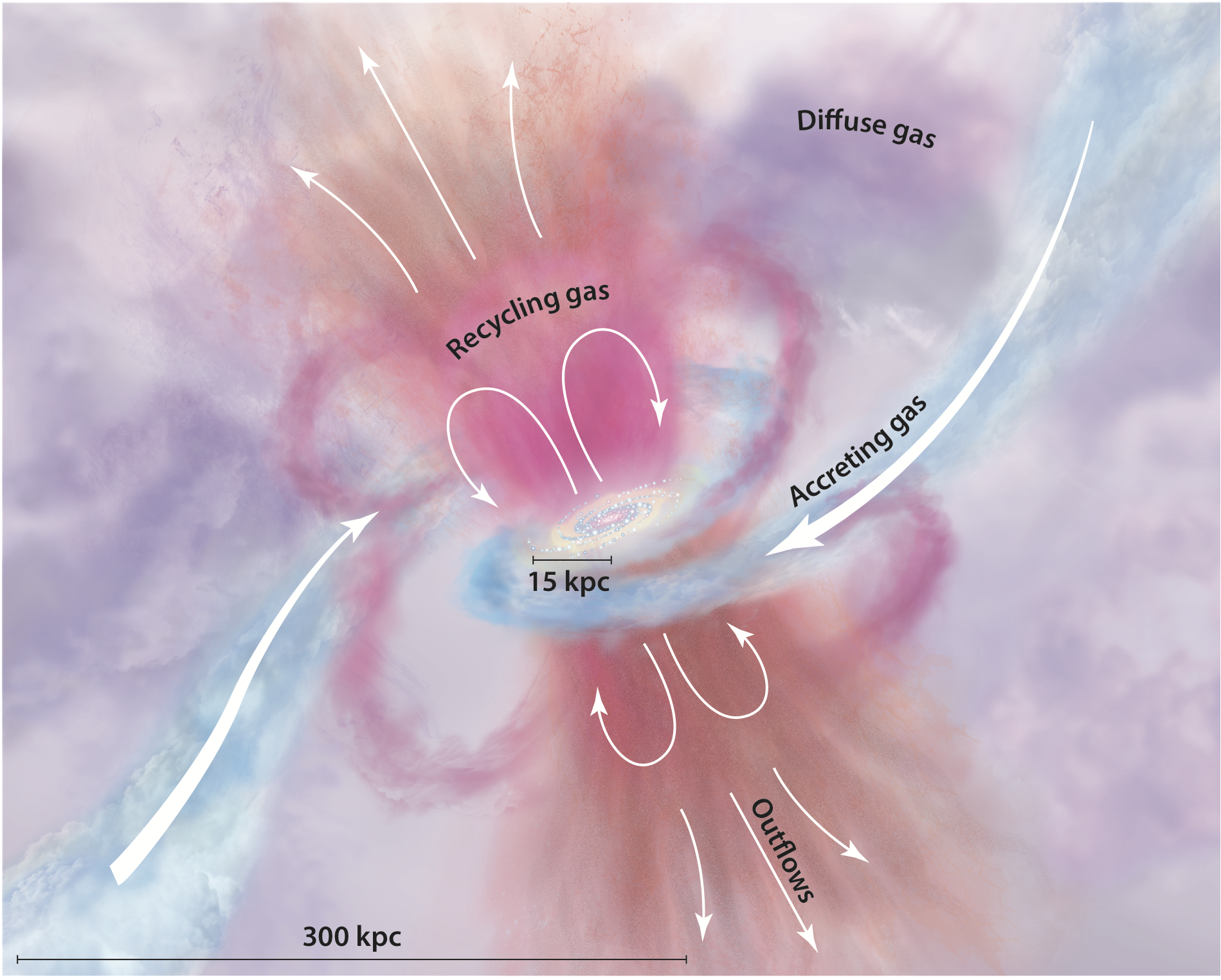}
\caption{A cartoon view of the CGM. The galaxy's red central bulge and blue gaseous disk are fed by filamentary accretion from the IGM (blue). Outflows emerge from the disk in pink and orange, while gas that was previously ejected is recycling. The ``diffuse gas" halo in varying tones of purple includes gas that is likely contributed by all these sources and mixed together over time. }
\label{fig_cartoon}
\end{figure}

In the mid-1950s, Guido M{\"u}nch observed neutral sodium (\ion{Na}{i}) and singly-ionized calcium absorption (\ion{Ca}{ii}) in the spectra of hot stars at high Galactic latitudes. Before these data were published as \citet{Munch:1961}, M{\"u}nch showed them to Lyman Spitzer, who interpreted the lines as evidence for diffuse, extraplanar hot gas ($T \sim 10^6$ K), which keeps the colder clouds traced by \ion{Na}{i} and \ion{Ca}{ii} in pressure confinement \citep{Spitzer:1956}. And so was born the idea of the ``Galactic corona'' and its exploration by absorption lines in the spectra of background objects. Following Schmidt's 1963 discovery of quasars, studies of ``extragalactic'' gas rapidly progressed with spectroscopy of the intervening absorption lines by J.~Bahcall, M.~Burbidge, J.~Greenstein, W.~Sargent, and others. \cite{Bahcall:1969} then proposed that ``most of the absorption lines observed in quasi-stellar sources with multiple absorption redshifts are caused by gas in extended halos of normal galaxies''. In the 1980s, subsets of the QSO absorption lines were associated with galaxies \citep{Bergeron:1986a, Bergeron:1991} while the "Lyman alpha forest" emerged as their IGM counterpart \citep{Sargent:1980}. Spurred by these developments, {\it Hubble} and Keck made great leaps in the 1990s towards a broader characterization of the number density and column density distribution of the IGM and CGM back to $z \sim 3$. Pioneering studies from Hubble's Key Project on QSO absorption lines demonstrated that galaxy halos give rise to strong \lya, \civ, and other metal lines \cite[e.g.][]{Lanzetta:1995, Chen:1998} in a gaseous medium that is richly structured in density, temperature, and ionization (Figure~\ref{fig_cartoon}). 

\begin{marginnote}[]
\entry{CGM}{Circumgalactic Medium}
\entry{IGM}{Intergalactic Medium}
\entry{ISM}{Interstellar Medium}
\entry{SDSS}{Sloan Digital Sky Survey}
\entry{CMD}{Color-Magnitude Diagram}
\end{marginnote}

In the 2000s, large galaxy surveys such as SDSS uncovered the galactic baryon deficit, the mass metallicity relation, and quenching problems (\S\,2). Meanwhile theorists implemented new physical prescriptions for gas accretion and feedback with new numerical methods and faster computers. It soon became impossible even to address these big mysteries of galaxies without appealing to gas flows between the ISM, the IGM, and by implication, the CGM. Yet most such models of gas flows were, and are still, tested against observations of starlight---the same observations that first posed the problems. By the mid-2000s, models and observations of gas flows in and out of galaxies had reached the point that the former were in urgent need of direct observations of the gas flows themselves. CGM studies leaped forward in the late 2000s with the installation of Hubble's {\it Cosmic Origins Spectrograph}, which was designed for reaching diffuse gas with $30 \times$ the sensitivity of its predecessors, and with new techniques for stacking and combining X-ray and optical spectra. This, then, is the context in which our review occurs. We aim to survey recent progress in observing and modeling the gas flows that drive galaxy evolution and thus to tell the story of galaxy evolution writ large, from the perspective of the CGM. 

For additional perspective on the issues raised here from a more Galactic point of view, we recommend the recent Annual Review on halo gas by \cite{Putman:2012}. For an up-to-date survey of accretion, see the forthcoming volume ``Gas Accretion onto Galaxies'' \citep{Fox:2017}.


\section{Galaxies in Gaseous Halos}
\label{section_problems}

\subsection{The Major Problems of Galaxy Evolution}

\begin{figure}[t]
\includegraphics[width=4in]{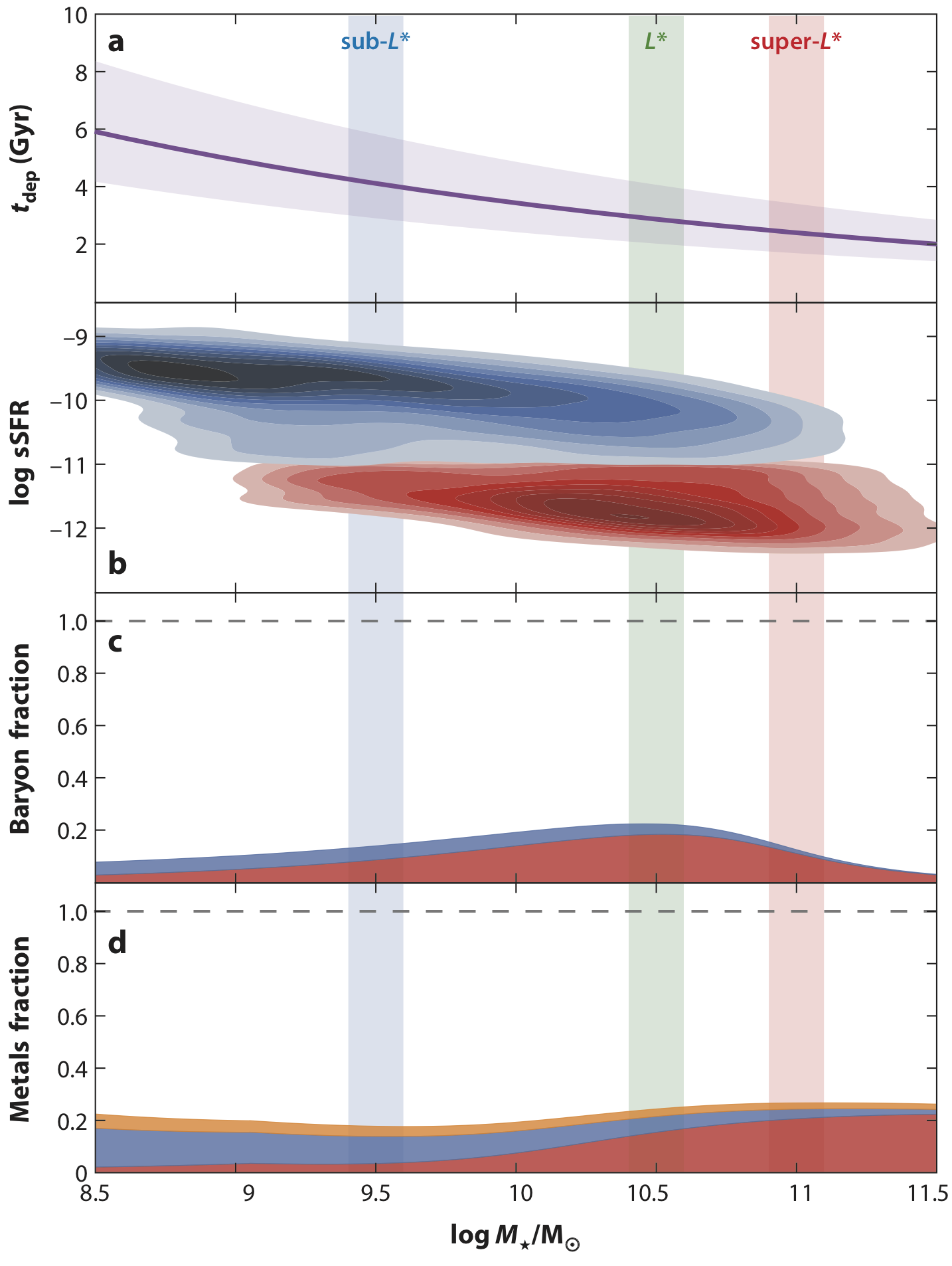}
\caption{Four important problems in galaxy evolution viewed with respect to $\mstar$. (a) the gas depletion timescale $\tau_{\rm dep} \sim M_{\rm gas}/\msfr$ for star-forming galaxies at $z\sim 0$, with $M_{\rm gas}$ from \citet{Peeples:2014} and $\msfr$ from \citet{whitaker12}; the shading denotes $\pm 0.15$\,dex scatter in $\msfr$. (b) the galaxy bimodality in terms of $\mstar$ and specific star formation rate \citep{Schiminovich:2010}. (c) the galactic baryon fraction, $\mstar/((\Omega_{b}/\Omega_{m})\mhalo)$ from \citet{Behroozi:2010}, with stars in red and interstellar gas in blue (from \citet{Peeples:2014}. (d) the ``retained metals fraction'', metals for several galactic components relative to all the metals a galaxy has produced \citet{Peeples:2014}, with stars in red, interstellar gas in blue, and interstellar dust in orange. Vertical bars mark the properties of sub-$L^*$, $L^*$, and super-$L^*$ galaxies at $\log M_{\star}/{\rm M}_{\odot} = 9.5$ (blue), 10.5 (green), and 11.0 (red), respectively.  
\label{fig_problems}
}
\end{figure}

We will motivate and organize our review with four major galaxy evolution problems in which the CGM is implicated (Figure~\ref{fig_problems}).
Why do dark matter halos of different masses give rise to galaxies with drastically different star formation and chemical histories (\S\,2.1.1, 2.1.2)? Why do such a small fraction of cosmic baryons and metals reside in the galaxies (\S\,2.1.3, 2.1.4)?
The prevailing answers to these questions all feature the regulation of gas flows into and out of galaxies---which necessarily pass into and through the CGM. We initially pose these problems at low redshift, but they all have high-$z$ counterparts, and their solutions require understanding the CGM and the flows that feed it at all cosmic epochs.

\subsubsection{How do galaxies sustain their star formation?}

Star-forming galaxies pose a conundrum: their ISM gas can last for only a small fraction of the time they have been forming stars (Figure~\ref{fig_problems}a), implying an external supply of gas that keeps the ISM in a quasi-equilibrium state. The depletion time, $\tau_{\rm dep} \sim M_{\rm gas}/\msfr$ changes by only $\sim 2\times$ over the factor of 30 between sub-$L^*$ and super-$L^*$ galaxies. More generally, sub-$L^*$ galaxies generally have extended bursty star formation histories, as opposed to the more continuous star formation found in more massive galaxies, suggesting differences in how and when these galaxies acquire their star forming fuel. As this fuel is from the CGM, we must explain how sub-$L^*$ and $L^*$ galaxies fuel star formation for longer than their $\tau_{\rm dep}$.

\subsubsection{What quenches galaxies and what keeps them that way?}

How galaxies become and remain passive is one of the largest unsolved problems in galaxy evolution (Figure~\ref{fig_problems}b). Proposed solutions to this problem involve controlling the gas supply, either by shutting off IGM accretion or keeping the CGM hot enough that it cannot cool and enter the ISM. Low-mass galaxies tend to continue forming stars unless they are a satellite of or near a more massive galaxy \citep{Geha:2012}. This finding suggests that the central galaxy's gaseous halo strips the satellite with ram pressure or  ``starves'' the satellite of fresh fuel. These ideas have specific testable implications for the physical state of the CGM.

\subsubsection{Why do galaxies lack their fair share of baryons?}

The $\Lambda$CDM model predicts that baryons follow gravitationally-dominant dark matter into halos, where the gas dissipates energy as radiation and cools into the center of the halo. Observed galaxies, however, harbor only small share of the halo's expected baryons in their stars and ISM, with $M_{\rm b}\ll(\Omega_{\rm b}/\Omega_{\rm m}) M_{\rm h}$ (Figure~\ref{fig_problems}c). Even at their most ``efficient'', $L^*$ galaxies have converted only $\sim 20$\% of their halos' baryons into stars (Figure~\ref{fig_problems}c), with values of only about 5-10\% in sub-$L^*$ and super-$L^*$ galaxies \citep{Behroozi:2010, McGaugh:2010}. There are three basic possibilities: the baryons are in the halo but not yet detected, such as hot or diffuse gas; the baryons have been accreted and then ejected from the halo altogether; or the baryons have been prevented from accreting into the halo in the first place. While reality probably combines aspects of all three, in any combination they strongly suggest that the CGM is an excellent place to look for missing halo baryons in cold or hot gas, or for direct evidence of past ejection.

\begin{marginnote}[]
\entry{$\Lambda$CDM}{Cold-Dark-Matter Cosmology with a Cosmological Constant}
\end{marginnote}

\begin{figure}[t]
\includegraphics[width=4.8in]{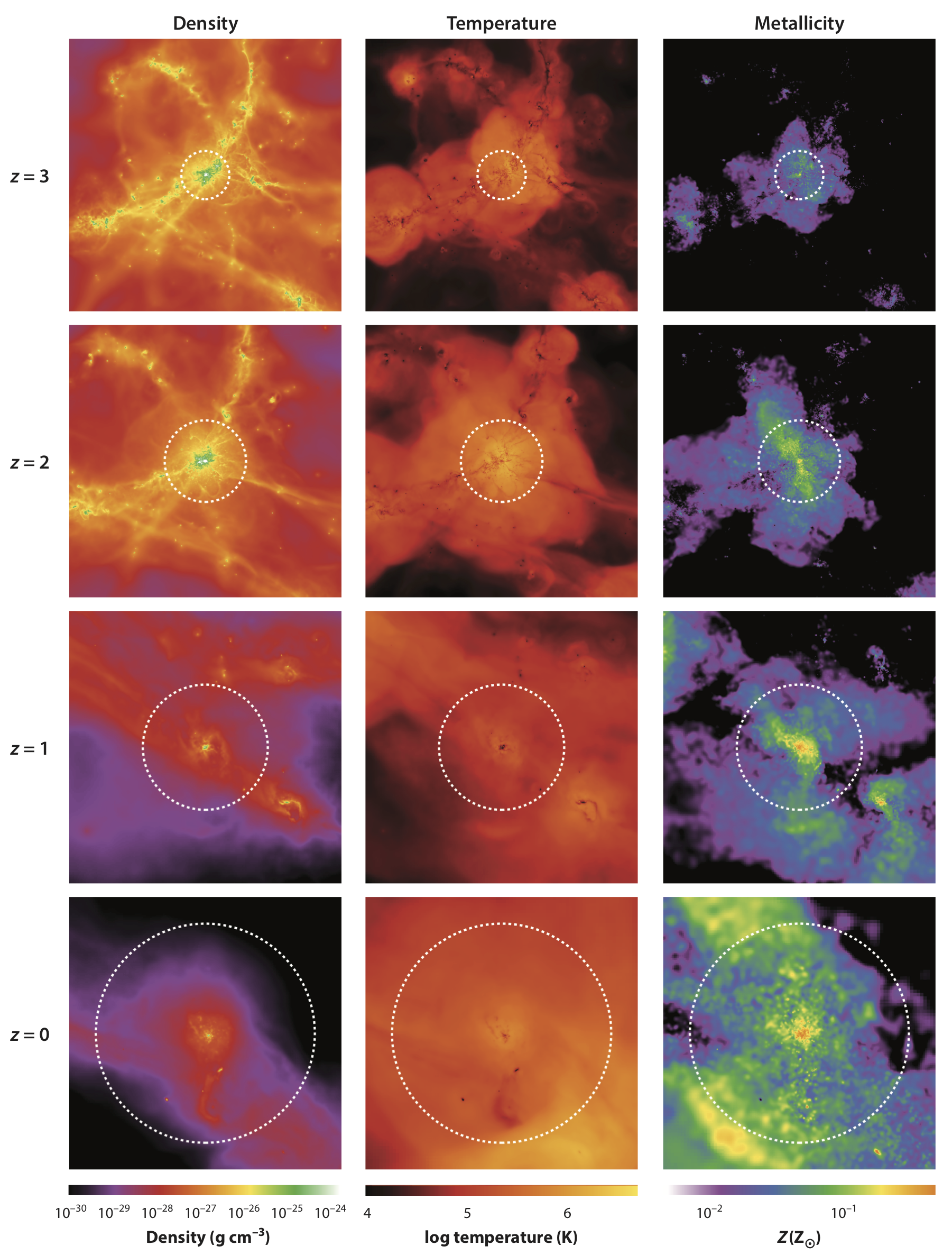}
\caption{These simulated views (from EAGLE, \citealp{Schaye:2015,Oppenheimer:2016}) of the CGM are more sophisticated but possibly just as uncertain as Figure \ref{fig_cartoon}.  The four columns render a single galaxy with $\mstar=2.5\times 10^{10}$\Msun\ at $z=0$ in density (left), temperature (middle) and metallicity (right). The galaxy is shown at redshifts $z =$ 3, 2, 1, and 0 from top to bottom. The dotted white circle encloses the virial radius at each epoch. }
\label{fig_sims_showcase}
\end{figure}

\subsubsection{Where are the metals?}

While baryons come from outside the halo, metals are sourced locally by stars and the deaths of stars. Star-forming galaxies over $\sim 3$ decades in stellar mass retain a surprisingly flat $\sim 20$--$25$\% of the metals they have ever produced \citep{Peeples:2014} in their stars, ISM gas, and dust. Metals have clearly been lost to outflows \citep{Tremonti:2004}, but how these outflows scale with galaxy mass is unclear. Models that already struggle to reproduce the observed steep mass-metallicity relation \citep{Somerville:2015b} fail to retain the low, flat fraction of metals produced (e.g., \citealp{Muratov:2015, Zahid:2012b, Oppenheimer:2016}). While ``missing baryons'' concern accretion and feedback through the outer boundary of the CGM, metals address the disk/halo interface:
do they leave the halo altogether, or recycle back into the galaxy's ISM on long timescales as a ``halo fountain'' \citep{Oppenheimer:2008} On what timescales are ejected metals recycled? How metal-enriched is outflowing material relative to the ambient ISM, i.e., what are the entrainment fractions and metal-loading factors? How does dust survive the journey out of galaxies, and what chemical clues does it hide? As we will show, following the metals as ``Nature's tracer particles'' is a fruitful and revealing route to understanding of the CGM.

\subsection{Our Point of View}

How galaxies acquire, eject, and recycle their gas are core issues in galaxy evolution, on par with how they evolve in their shapes and how star formation works. To a large extent these gas flows {\em are} galaxy evolution. The CGM is a main venue for these flows: it is potentially the galactic fuel tank, waste dump, and recycling center all at the same time. This review will approach the growing body of empirical results and theoretical insights from the direction of these four major questions. Rather than asking, for example, ``what are the \mgii\ absorbers?'', we will ask ``what do the \mgii\ absorbers tell us about the mass and kinematics of galactic outflows?''. We will thus favor physical insights and synthesis of discoveries over detailed discussions of methods, compilations of data, or exhaustive cataloging of the literature. We hope that this approach will improve understanding between those who study gas and galaxies (still disparate groups) and more effectively highlight open issues to be pursued in the future. 

For the purposes of our discussion, we define the CGM to be bounded at the outside by the virial radius $R_{vir}$ of a galaxy's dark matter halo, and on the inside by the disk or ISM. Neither boundary is well-defined, and precisely defining when gas passes through one of these boundaries can be either a valuable research contribution or a fruitless semantic exercise depending on circumstances. We focus on the physics of gas that fills out halos without too much attention to these exact definitions. 

\begin{marginnote}[]
\entry{Physics}{underlying physical properties and processes}
\entry{Phenomenology}{emergent properties and scaling relations}
\end{marginnote}


\section{How We Study the CGM}
\label{major_section_techniques}

\subsection{Transverse Absorption-Line Studies}
\label{subsection_absorption}

Viewing the CGM in absorption against a bright background source like a quasar offers three major advantages over other methods: (1) sensitivity to extremely low column density, $N \simeq 10^{12}$ cm$^{-2}$, (2) access to a wide range of densities, unlike emission-line measures that scale as density squared, and (3) invariance of detection limits to redshift and the luminosity of the host galaxy. These advantages come at a cost, however: absorption provides only projected, pencil-beam measures of gas surface density, usually limited to one sightline per galaxy by the rarity of background quasars. Within the local Universe (a few Mpc) it is possible to use multiple sightlines \citep[e.g.,][]{Lehner:2015, Bowen:2016}, and at higher redshift, multiply-lensed images from background quasars \citep[e.g.,][]{Rauch:2011, Rubin:2015} to constrain the sizes of absorbers.
In general, however, CGM maps made from absorption-line measurements are a statistical sampling of gas aggregated from many galaxies. With massive optical spectroscopic surveys, samples have grown to hundreds or thousands in low ions like \mgii\ and \caii\ \citep[e.g.,][]{Zhu:2013b}. Quasar/galaxy pairings have now been extended out to $z \sim 4$ and beyond \citep{Turner:2014, Matejek:2012}.

There are three basic ways of building absorber samples. First, ``blind'' surveys select background quasars on brightness and/or redshift and so are optimal for samples that are unbiased with respect to foreground structure. Ground-based redshift surveys around previously observed quasar sightlines are now a time-honored method for constructing samples of quasar/galaxy pairs
(e.g.,\ \citealp{Chen:1998,Stocke:2006,Rudie:2012}).
The second, ``targeted'', approach chooses background sources {\em because} they probe particular foreground structures, such as $L^*$ galaxies \citep{Tumlinson:2013}, sub-$L^*$ galaxies \citep{Bordoloi:2014a}, galaxies with known ISM content \citep{Borthakur:2015}, or groups and filaments \citep{Wakker:2015,Tejos:2016}, 
by cross-matching the observable quasar with catalogs of these structures. Finally, maps of absorption in the Milky Way's CGM use essentially any quasar (or UV-bright halo stars), sometimes chosen to pass through known halo gas structures and sometimes not. Though most absorption-line work has been in the UV and optical,
{\em Chandra} and {\em XMM} have been used to search for X-ray gas in individual absorbers, constraining the extent of CGM and IGM hot gas 
\citep{nicastro05}.

\begin{marginnote}[]
\entry{LLS}{Lyman Limit Systems, $N_{\rm HI} > 10^{16.2}$\,cm$^{-2}$, the ``dense'' CGM}
\entry{DLA}{Damped Lyman-$\alpha$ Systems, $N_{\rm HI}>2\times10^{20}$\,cm$^{-2}$, generally ISM}
\end{marginnote}

It is useful to distinguish between \ion{H}{i} column density regimes that must be, or can be, treated differently in analysis. Lines up to $\log N \simeq 15$ can usually be analyzed with equivalent widths or Voigt profile fitting. The value $\log N \simeq 15$ is high for the \lya\ forest but low for the CGM (there are of course a few exceptions, \cite{Tumlinson:2013, Johnson:2014}, where \ion{H}{i} is not seen at $<$ 100 kpc even to low limits). At $\log N \simeq 16$, saturation becomes a major factor and robust column densities (as opposed to lower limits) must come from profile fitting or from the higher Lyman series lines, if the system is redshifted enough. Systems with $\log N \simeq 16$, are partial or complete LLSs. If the Lyman limit is covered ($z > 0.24$ for Hubble), the flux decrement at $\lambda = 912 (1+z)$ \AA\ allows a precise measurement of $\log N_{HI}$ and improved ionization and metallicity diagnostics. Above $\log N_{HI} \simeq 18$ (where $N_{HI}$ is the HI column density in cm$^{-2}$), the Lyman limit is totally opaque, the highest Lyman series lines are saturated, and genuine column densities must come from fitting the \lya\ profile for LLS and DLAs.

\subsection{Stacking Analyses}

Massive spectroscopic surveys have enabled another novel method for examining halo gas. ``Stacking'' of hundreds or thousands of spectra is a powerful way to extract faint signals from absorption-line datasets. This technique requires catalogs of redshifts, for either foreground galaxies or absorbers, so that the spectra of background objects can be shifted to their rest frames and continuum-normalized and then co-added together. The co-addition beats down statistical noise, enabling measurements of weak absorption at the cost of averaging over individual absorber profiles.  When the catalogs of foreground galaxies include  properties such as mass, radius, star formation rate, color, environment, or orientation, the stacks can be performed with subsets of the data to examine the variation of mean profiles with these properties \citep{York:2006, Zhu:2013c, Bordoloi:2011}. Stacking experiments that correlate the reddening of quasars due to foreground galaxy halos in the SDSS survey have revealed large quantities of dust in the CGM of galaxies \citep{Menard:2010, Peek:2015}. Stacking techniques can also exploit more numerous, but fainter, sources; for example, \citet{Steidel:2010} characterized the CGM of $z\sim 3$ galaxies by stacking the spectra of background galaxies. Stacking can detect weak signals in the mean properties of gas absorbers, but at the cost of averaging out kinematic and ionization structure that may contain significant physical meaning. 

\begin{figure}[t]
\includegraphics[width=6in]{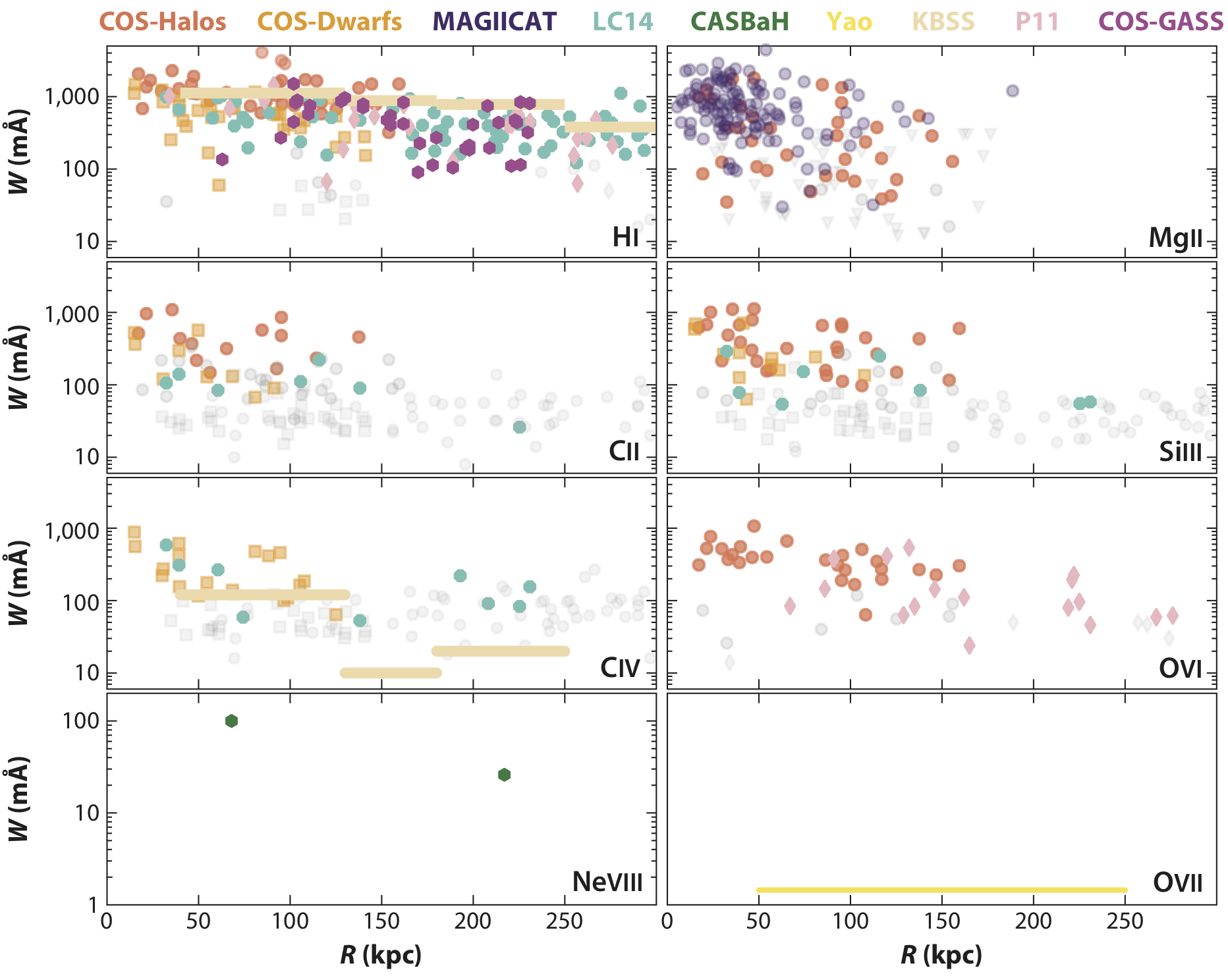}
\caption{A range of ion equivalent width (rest-frame) measurements for a compilation of published surveys. We progress from \hi\ though seven metallic ions of increasing ionization potential. The surveys are COS-Halos \cite{Tumlinson:2013,Werk:2013}, COS-Dwarfs \citep{Bordoloi:2014a}, COS-GASS \citep{Borthakur:2015}, MAGIICAT \cite{Nielsen:2013a}, \cite{Liang:2014}, the Keck Baryonic Structure Survey \citep{Rudie:2012,Turner:2015}, CASBaH \citep{Tripp:2011}, \cite{Prochaska:2011c}, and the X-ray study of \cite{Yao:2012} that imposes as stacked upper limit on \ion{O}{vii}.
}
\label{fig_multiphase_ions}
\end{figure}

\subsection{Down the Barrel}
``Down-the-barrel'' spectroscopy uses a galaxy's own starlight as a background source for detecting absorption. This method has been a fruitful one for studying galactic inflows and outflows from spectroscopy of star-forming galaxies. This method is commonly used in optical and near-UV lines such as \caii, \ion{Na}{i}, \mgii, and \feii\ \citep{Martin:2005,Kornei:2012a,Bordoloi:2011,Rubin:2014} to study outflows from galaxies out to $z \sim 1$, in UV lines for low-redshift star-forming galaxies \citep{Henry:2015,Heckman:2015}, or even to examine accretion \citep{Rubin:2012}. Down-the-barrel measurements are critical pieces of the CGM puzzle because they directly trace current outflows at galactocentric radii that are inefficiently covered by background sources (because of the $R^2$ scaling of foreground cross-section). While down-the-barrel spectra are key for tracing the accretion and outflows that dominate CGM kinematics, they have the key limitation that the galactocentric radius of any detected absorption is unconstrained---it could be anywhere along the line of sight---complicating mass and covering fraction estimates inferred from these spectra. 

\subsection{Emission-line maps}

Emission-line observations search for photons emitted directly from CGM gas. As the emission measure scales as $n^2$, and the CGM has $n_H \sim 10^{-2}$ or less, this is a stiff challenge. The MW halo has been extensively mapped for HVCs and other halo structure using radio emission at 21 cm. This technique has been applied to external galaxies \citep{Putman:2012} but detections are limited to within $\sim 10-20$ kpc of the targeted galaxies. The soft X-ray band is optimal for gas at $\gtrsim 1$ million K. The extremely low surface brightness of the gas makes these observations challenging and expensive, but a few individual halos have been detected and their hot gas budgets measured by Chandra and/or ROSAT \citep[e.g.,][]{Humphrey:2011,Anderson:2016}. Stacking of individual galaxies techniques has also yielded mass density profiles for hot gas around nearby galaxies \cite{Anderson:2013}. When combined with halo size, density, and metallicity constraints from soft X-ray absorption line techniques, these maps have aided in the assessment of the total mass and baryon fraction of the hot CGM.

Emission line maps are also possible at UV/optical wavelengths, though no less challenging than in the X-ray. Recent reports claim a detection of an extended \ovi\ halo ($R \sim 20$ kpc) around a low-redshift starburst galaxy \citep{Hayes:2016}. Extended Ly$\alpha$ emission has been seen out to $\sim 100$ kpc away from $z \sim 2.5$ galaxies and QSOs \citep{Cantalupo:2014, Prescott:2015}. In another case, an extended filamentary structure connected to a galactic disk was detected using diffuse emission in the optical \citep{Martin:2015}. Emission maps can constrain the density profile, morphology, and physical extent of the gas more directly than aggregated pencil-beam sightlines \citep{Corlies:2016}. For X-ray emission from fully ionized gas, masses can be inferred more directly, avoiding the uncertain ionization corrections that plague absorption-line measurements (\S~4); indeed, the CGM's more massive cousin, galaxy clusters' intracluster medium, has been studied in detail via X-ray emission for decades \citep{Vikhlinin:2005}. On the downside, emission line maps are still challenging technically; the surface brightnesses are extremely small compared to sky and detector backgrounds, and surface brightness dimming has a steep increase with redshift. In a recent study using stacks of fiber spectra from SDSS, \cite{Zhang:2016} achieved detections of H$\alpha$ at $50-100$ kpc around low-redshift galaxies, demonstrating that very sensitive limits can be reached on galaxies in the aggregate. These observations remain challenging, but as ``taking a picture'' of an astrophysical object remains the ideal, efforts to improve instrument technology and enable emission line mapping to reach samples of hundreds of galaxies across cosmic time is an important goal.

\subsection{Hydrodynamic Simulations}
Physical models and simulations are essential tools for understanding the CGM. In contrast to observations, they provide for controlled environments where physical properties, histories, and futures of gas are all known and can be manipulated to tease insights out of the otherwise unmanageable complexity of a multiphase gaseous medium. As reviewed by \citet{Somerville:2015a}, there are many schemes for simulating the development of the cosmic web and galaxies under the influence of dark matter, gravity, and hydrodynamics. The major methods at present are smoothed particle hydrodynamics (SPH, such as Gadget,  \citealp{Ford:2013,Oppenheimer:2016}, Gasoline, \citealp{Christensen:2016,Gutcke:2017}, and GIZMO \citealp{Muratov:2016}), adaptive mesh refinement (AMR, such as Enzo, \citealp{Hummels:2013,Corlies:2016}), and moving mesh (Arepo and the Illustris simulation, \citealp{Suresh:2015a}). Large-scale cosmological simulations in Mpc-scale boxes can simulate hundreds of galaxies in their proper $\Lambda$CDM context (e.g., \citealp{Oppenheimer:2006, Vogelsberger:2014, Ford:2014}). At the opposite end of the scale, very high resolution simulations focused on the interaction between dense clouds and diffuse halos (e.g., \citealp{heitsch09, armillotta16}) that can reach scales at $\ll$ parsec. Spanning these two regimes are the so-called ``zoom'' simulations, which resolve enough of the large scale structure to accurately trace a single galaxy or a subset of galaxies selected out of larger boxes (Figure 2, \citealp{Schaye:2015}). Even zooms must make assumptions about physics that they do not resolve, using ``sub-grid'' prescriptions to stand in for such complex phenomena as star formation, metal mixing and transport, supernova and AGN feedback, and others. Sub-grid models are parameterized and tuned to yield specific metrics---like the stellar mass function at $z = 0$---and then the properties that emerge---such as SFRs, morphology, quenching, and the CGM---are analyzed and compared to data to constrain the physical prescriptions that went in. We will use simulations from a broad range of techniques and groups to look for insights into how the CGM participates in galaxy evolution, and to help interpret data.


\section{The Physical State of the CGM}

\label{major_section_physical_state}

We now turn to the density profile, phase structure, and kinematics of the CGM. We first present the data that show the various ionization states and velocity distributions of the CGM absorption (\S~4.1). Next, we describe how the absorption line measurements may be translated into physical parameters such as density, temperature, and size in (\S~4.2). We then draw lessons from kinematics (\S~4.3) before considering the physical complexities and challenges inherent in the interpretation of these data (\S~4.4 and \S~4.5).

\subsection{The Complex, Multiphase CGM}


As a matter of empirical inference, the CGM is ``multiphase'' in its ionization structure and complex in its dynamics. The ionization structure is seen in Figure \ref{fig_multiphase_ions}, which compiles measurements for six diagnostic ions as a function of impact parameter (a proxy for radius). These data indicate a wide range of density and ionization conditions up to a few $10^5$ K with very little interpretation required. Observationally, ``multiphase'' means many of these metal ions spanning an order of magnitude in ionization potential energy are commonly found within the same ``absorber system'' occupying a galaxy's halo. An open question in the physics of circumgalactic gas is what this observed mulitphase ionization structure reveals about the small-scale multiphase density, temperature, and metallicity structure of the CGM.

Over the last 20 years, the practice of using such empirical inputs in analytic arguments to infer the physical state and structure of the diffuse plasma has matured greatly \citep{Mo:1996, Maller:2004}. To produce an extended, multiphase CGM, authors have proposed several scenarios which we categorize as follows: (1) massive inward cooling flows driven by local thermal instabilities \citep[e.g.][]{McCourt:2012}; (2) boundary layers between moving cool clouds in a hot atmosphere \citep[e.g.][]{begelman90}; and (3) the continual shocking and mixing of diffuse halo gas by galactic outflows \citep[e.g.][]{fielding16, Thompson:2016}.   We discuss the applicability of some of these analytic models in \S~4.4 and \S~4.5. 

Direct evidence for a hot component ($\log T \gtrsim 6$) in the multiphase CGM comes from diffuse soft X-ray emission \citep{Anderson:2010, Anderson:2013}, and in absorption along QSO sightines \citep{williams05, Gupta:2012a} for the Milky Way and external galaxies. Indirect evidence for a hot phase comes from highly ionized metals that correlate with the low-ionization HVCs \citep{sembach03, fox06, lehner09, wakker12}, suggesting boundary layers between a hot medium and the colder HVCs. Milky Way HVCs also show head-tail morphologies indicative of cool clouds moving through a hot medium \citep[e.g.,][]{bruns00}. Finally, the multiphase CGM is clearly manifested in hydrodynamic simulations, which exhibit a mixture of cool (10$^4$ K) and warm-hot (10$^{5.5}$--10$^6$ K) gas within a galaxy virial radius with a density profile that drops with increased distance from the central host galaxy (e.g., \citealp{Shen:2013, Stinson:2012, Ford:2013, suresh17}, Figure 3). For practical purposes we can regard the outer boundary of the CGM to correspond to $R_{vir}$, but there is no empirical reason to believe that any special behavior occurs at that radius; current observations favor trends in column densities that scale with $R_{vir}$ but do not change in form at that arbitrary boundary. 

\begin{marginnote}[]
\entry{Low Ions}{\\ IP $< 40$ eV, \\ $T = 10^{4-4.5}$ K}
\entry{Intermediate Ions}{\\ 40 $\gtrsim$ IP (eV) $\lesssim$ 100, \\ $T = 10^{4.5 - 5.5}$ K}
\entry{High Ions}{\\ IP $\gtrsim$ 100 eV,\\ $T >$ 10$^{5.5}$ K}
\end{marginnote}

\begin{figure}[t]
\includegraphics[width=6.0in]{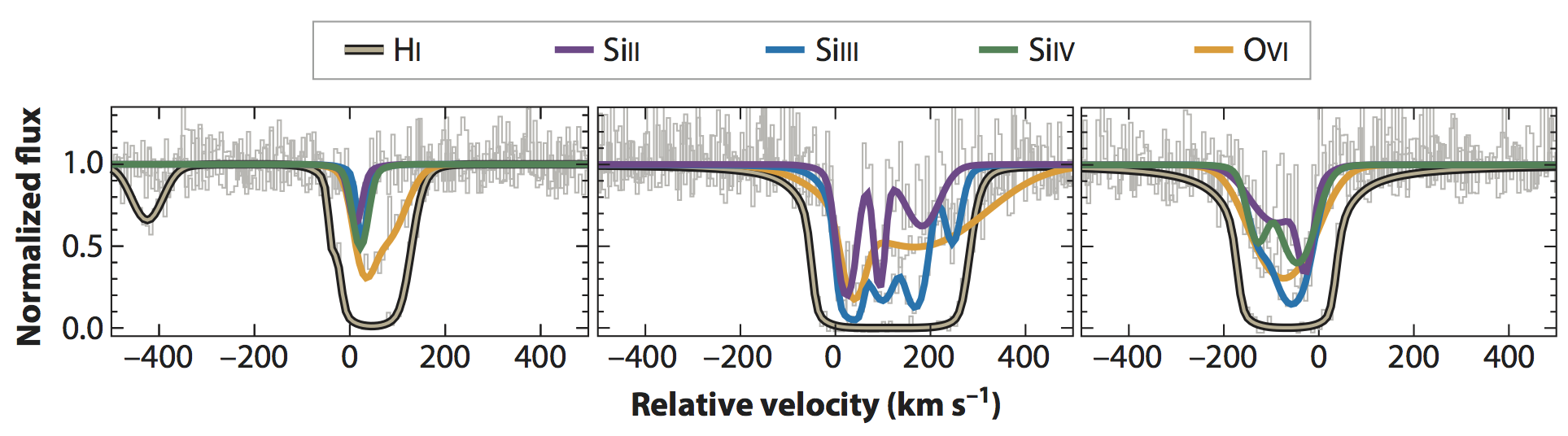}
\caption{A selection of absorption-line data and Voigt profile fits from the COS-Halos survey \citep{werk16}, showing a range of metal ions and HI on a common velocity scale with the galaxy at $v=0$ km/s on the x-axis. The black outlined beige curve traces \ion{H}{i}, the purple \ion{Si}{ii}, the blue  \ion{Si}{iii}, the green  \ion{Si}{iv}, and the orange shows \ion{O}{vi}. }
\label{fig_examplelines}
\end{figure}

Evidence for kinematic complexity is revealed as the detected
ion species breaking into different ``components'' with distinct velocities and linewidths. Shown in Figure \ref{fig_examplelines}, the various metal ions show significant but varied correspondence in their component structure. The combination of both aligned and misaligned components between ionization states may reflect clouds or streams with density structure or a population of clouds with different ionization states projected together along the line of sight to the same range of observed velocities. Cloud sizes are difficult to constrain in a model independent way, but  multiply-lensed images from background quasars \citep{rauch01,Rauch:2011} prefer 1--10 kiloparsec scales.  Fitting Voigt profiles to multi-component absorption yields column density $N$, Doppler $b$ parameter, and velocity offset $v$ for each component from the galaxy systemic redshift, as well as the total kinematic spread of gas in a halo (but this fitting is subject to issues caused by finite instrumetal resolution). Generally, the kinematic breadth of an absorber system is thought to reflect the influence of the galaxy's gravitational potential, bulk flows, and turbulence in the CGM.

\subsection{From Basic Observables to Physical Properties}
\label{sec_obs_to_physical}
We must characterize the ionization states, chemical composition, and density to properly describe the  symbiotic relationship with the gas and stars in the central galaxy disk and the CGM. If it were feasible to obtain precise measurements for every ion of every abundant element, in all velocity components, then the gas flows, metallicity, and baryon budget of the multiphase CGM would be  well-constrained. However, atomic physics dictates that only a subset of the ionization states of each element lie at accessible wavelengths. Taking oxygen as an example, \ion{O}{i} and \ovi\ place strong lines in the far-UV, while \ion{O}{ii}--\ion{O}{v} lines appear in the extreme-UV (400--800\AA). \ion{O}{vii} and \ion{O}{viii}, arising in hot gas, have strong transitions  in the soft X-ray ($\sim 20$ \AA). While it is therefore possible in principle to detect (or limit) every stage of oxygen,  this potential has yet to be realized. 

\begin{marginnote}[]
\entry{NUV}{Near UltraViolet, $2000\lesssim\lambda\lesssim 3400$\AA}
\entry{FUV}{Far UltraViolet, $900\lesssim\lambda\lesssim 2000$\AA}
\entry{EUV}{Extreme UltraViolet, $400\lesssim\lambda\lesssim 900$\AA}
\entry{X-ray}{$\lambda\lesssim 30$\AA}
\end{marginnote}

\begin{figure}[t]
\includegraphics[width=6.0in]{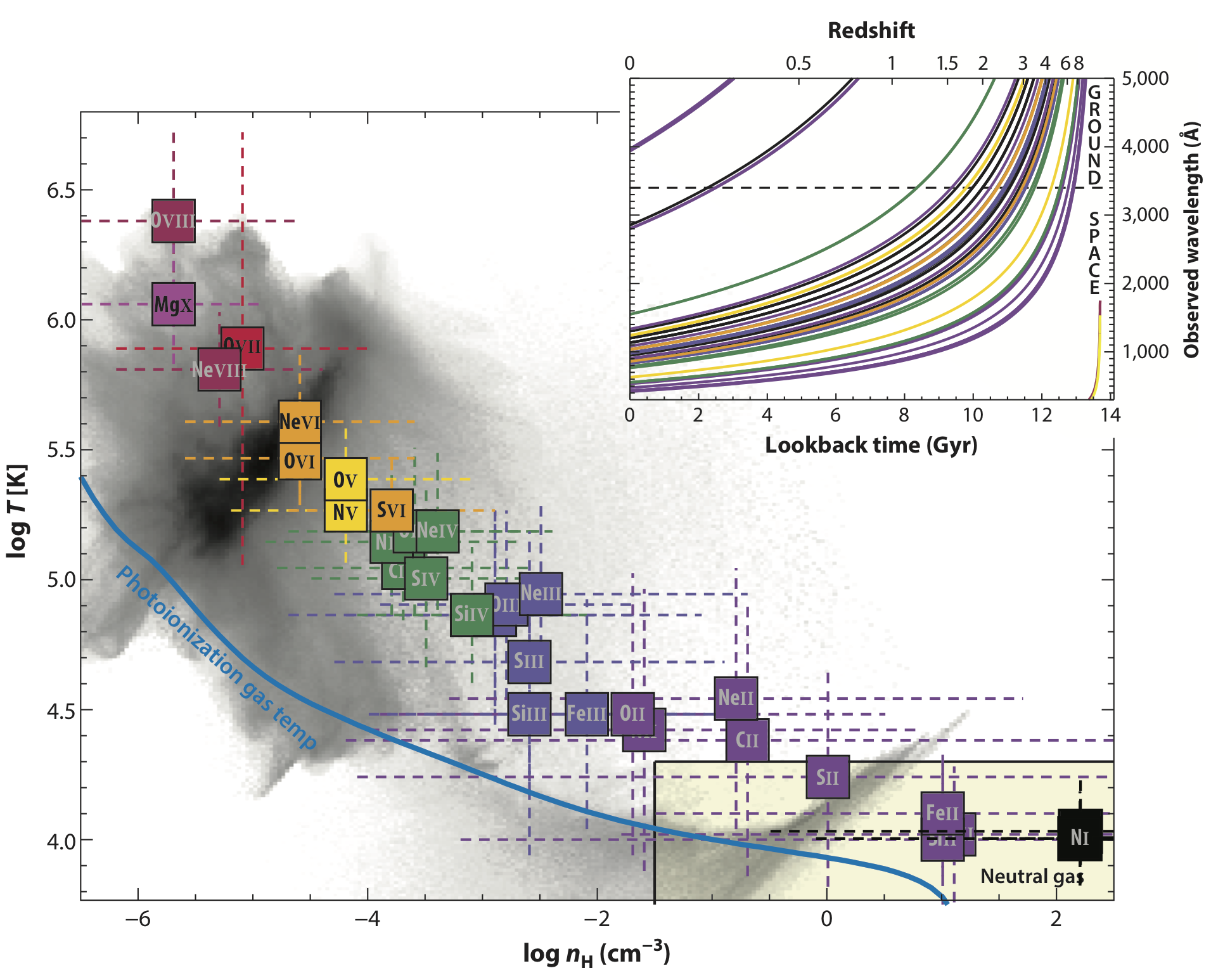}
\caption{Metal absorption lines (ions) of the CGM from \ion{Mg}{i} to \ion{O}{viii} having 19 $< \lambda_{\rm rest} < 6000$ \AA\ shown on a phase ($T$-$n_H$) diagram within $\rvir$ of the $z=0$ EAGLE simulation shown in Figure~2. The points are colored according to ionization state, ranging from neutral (I; black) to highly ionized (X; magenta). The  position of each point is set on each axis where
its ionization fraction peaks in CIE (temperature axis) and a standard PIE model (density axis) \citep{Gnat:2007, Oppenheimer:2013}; the range bars show the $T$ and $n$ range over which each species has an ionization fraction over half its maximum value (i.e., the FWHM).  Complete line lists are available in \cite{Morton:2003}. \label{fig_usual_ions}} \end{figure}

Figure \ref{fig_usual_ions} shows the basic schema for constraining CGM gas properties with these ``multiphase'' ions. The grey-scale phase diagram renders the properties of all $<\rvir$ gas from a Milky Way mass EAGLE zoom simulation \citep{Oppenheimer:2016}. Accessible ions at each temperature and density are marked with colored squares and dashed lines. This plot is intended to be a useful guide for finding the most likely tracers of a given CGM gas phase. It cannot be used to extract precise temperatures and densities for any given ion since the metal ion positions on this phase diagram are model-dependent. The inset shows the most common strong lines from these species plotted as {\it observed} wavelength versus redshift; the rest frame wavelength is where each intercepts $z=0$. Practically, FUV lines are available at $z < 1$ with Hubble and $z > 2$ from the ground, the EUV lines can be reached at $z \gtrsim 0.5-1$ with Hubble ($\lambda_{\rm obs} \gtrsim 1100$ \AA), and the X-ray lines can currently only be detected toward the small number of bright QSOs and blazars with reach of the sensitivity of {\em Chandra} and {\em XMM}. As a result,  most CGM measurements rely on  heterogeneous ion sets--- several low ions from C, N,  Si, and Mg, a few intermediate ions from C and Si, and a high ion or two from Ne and O. Therefore, the gas density and temperature can only be understood in the context of a model for its ionization state (and abundance patterns). 

\begin{marginnote}[]
\entry{CIE}{Collisional Ionization Equilibrium}
\entry{PIE}{PhotoIonization Equilibrium}
\entry{EUVB}{Extragalactic UltraViolet Background}
\end{marginnote}

Many assumptions are necessary to make progress toward physical models of the CGM. The two most generic classes of models are PIE and CIE. Generally, low and intermediate ions can be accommodated within PIE models, while high ions require CIE models. Species at intermediate ionization potentials, such as \civ\ and \ovi, will sometimes show a preference for one or the other or have contributions from both. These two classes of model are not mutually exclusive: a gas that is collisionally ionized may have the ion ratios further affected by incident radiation, and there are numerous possible departures from equilibrium that further complicate modeling \citep[e.g.][]{Gnat:2007}. Generally, having access to more metal ion tracers means one is able to place more refined constraints on the models, while results from models with fewer ions are more model-dependent. 

Radiative transfer models like Cloudy \citep{ferland13} are used to build PIE models \citep[e.g.][]{bergeron:1986b,prochaska04, Lehnert:2013,  Werk:2014, Turner:2015}, which are parametrized by density $n_H$, or equivalently the ionization parameter $\log U \equiv \Phi/n_{\rm H} c$, the observed neutral gas column density $N_{\rm HI}$, and a gas-phase metallicity, log [Z/H]. Here, $\Phi$ is the number of photons at the Lyman edge (i.e., the number ionizing photons), set by the assumed incident radiation field with a given flux of ionizing photons.  Besides ionization and thermal equilibrium, another major underlying assumption of photoionization modeling is that the included metal ions arise from a single gas phase with the same origin (i.e., are co-spatial). The single cloud, single density approximation for PIE modeling of low-ions leads to uncertain ``cloud'' sizes, determined by $N_{\rm H}/n_{\rm H}$ ranging from 0.1--100 kpc \citep[][]{Stocke:2013, Werk:2014}. In response, some models have begun to explore internal cloud density structure \citep{Stern:2016} or local sources of radiation (e.g. star-formation in the galaxy, the hot ISM, \citealp{fox05, werk16}).  PIE models generally fail for highly-ionized metal species like \ovi, sometimes \civ, and certainly for X-ray ions. For those we turn to CIE, where temperature controls the ionization fractions and a metallicity must be assumed or constrained to derive total hydrogen column $N_H$. 

Beyond PIE and CIE, there are non-equilibrium ionization mechanisms that may reproduce the intermediate- and high-ion states that generally fail for PIE (e.g. \civ, \nv, \ovi). These models include: (1) radiative cooling flows that introduce gas dynamics and self-photoionization to CI models \citep{edgar86,  benjamin94, wakker12}, (2) turbulent mixing layers, in which cool clouds develop skins of warm gas in Kelvin-Helmholtz instabilities \citep{begelman90, slavin93, kwak10}, (3) conductive interfaces, in which cool clouds evaporate and hot gas condenses in the surface layer where electron collisions transport heat across the boundary \citep{gnat10, armillotta16}, and (4) ionized gas behind radiative shocks, perhaps produced by strong galactic winds \citep{dopita96, Heckman:2002, allen08, gnat09}. These models all modify the column density ratios given by pure CIE, but do not change the basic conclusion that gas bearing these ionic species {\it must} be highly ionized, i.e. with a neutral fraction $\ll$  1\%. These large and unavoidable ionization corrections, when applied to \hi\ column densities of $\log N_{HI} \sim 15$--18, entail surface densities and total masses that are significant for the galactic budgets (\S~5). It is likely that combinations of PIE and CIE into these more complex models are more accurate descriptions of Nature than either basic process considered in isolation.

\subsection{Line Profiles and Gas Kinematics}

Linewidths, given by the Doppler $b$ parameter, illuminate the CGM temperature structure and gas dynamics. The gas temperature, $T$, and any internal non-thermal motions are captured in the following parameterization:  $b^{2} = (2kT/m_i) + b^{2}_{\rm nt}$, for a species with atomic mass $m_i$. When the low and high ions are assessed via Voigt profile fitting, the low ions are usually consistent with gas temperatures $<$ 10$^{5}$ K, with a contribution from non-thermal broadening ($<20$ km s$^{-1}$, \citealp{Tumlinson:2013, Churchill:2015, werk16}). ``Broad Lyman alpha'' (BLA; $b \gtrsim$ 100 km s$^{-1}$) and \neviii\ systems have been detected in QSO spectra at high S/N that directly probe gas at $\log T \sim 5.7$ \citep{Narayanan:2011, Savage:2011a, Tripp:2011, Meiring:2013}. These UV absorption surveys indicate that the CGM contains a mixture of photoionized and/or collisionally ionized gas in a low-density medium at 10$^{4}$ - 10$^{5.5}$ K \citep[e.g.,][]{Adelberger:2003, richter04, fox05, Narayanan:2010, Matejek:2012, Stocke:2013, Werk:2013, Savage:2014, Lehner:2014a, Turner:2015}. 

The velocity dispersion and number of components reveals the kinematic substructure of the CGM. Most significantly, gas near low-$z$ galaxies across the full range of $\log \mstar = 8.5$--$11.5$ show the projected line-of-sight velocity spreads that are less than the inferred halo escape velocity, even accounting for velocity projection. Thus most of the detected CGM absorption is consistent with being bound to the host galaxy, with implications for outflows and recycling (\S~7). This is true for all the observed species from \hi\ \citep{Tumlinson:2013} to \mgii\ \citep{Bergeron:1991,Nielsen:2015a, Johnson:2015b} to \ovi\ \citep{Tumlinson:2011, Mathes:2014a}. The strongest absorption seen in \ion{H}{i} and low ions are heavily concentrated within $\pm 100$\,km\,s$^{-1}$. For low ionization gas, internal turbulent / non-thermal motions are $b_{\rm nt} \sim 20$\,km\,s$^{-1}$, while for high ionization gas the non-thermal/turbulent contributions to the line widths are 50--75 km s$^{-1}$ \citep{werk16, faerman17}. Similar total linewidths are seen in the $z > 2$ KODIAQ sample, possibly indicating similar physical origins at different epochs \cite{Lehner:2014a}. 

Mis-alignments of the high and low-ion absorption profiles in velocity space may indicate that the gas phases bearing high and low-ions are not co-spatial and thus that the gas is multiphase \citep[e.g.][]{fox13}. Some systems, however, show close alignment between low and high ionization gas \citep{Tripp:2011} in a fashion that suggests each detected cloud is itself multiphase, perhaps in a low-ion cloud / high ion skin configuration. \cite{Heckman:2002} and others \citep[e.g.][]{grimes09, bordoloi16} have argued that the relationship between \ovi\ column density and absorption-line width for a wide range of physically diverse environments indicates a generic origin of \ovi\ in collisionally-ionized gas. However, the  relationship exhibits considerable scatter, is impacted significantly by blending of multiple unresolved components (at least at the moderate $R \sim 20,000$ resolution of COS), and may arise from other physical scenarios such as turbulent mixing \citep[e.g.][]{Tripp:2008, Lehner:2014a}. Generally, high-ions like  \ovi\ in the CGM exhibit systematically broader line widths than low and intermediate ions \citep[e.g.][]{werk16}. Though complex and varied, absorber kinematics may provide important observational constraints on both ionization and hydrodynamic modeling, but new methods of analysis and new statistical tools will be required to realize their full potential.

\subsection{Challenges in Characterizing the Multiphase CGM} 
\label{subsection_challenges}

Ionization modeling is limited by what might be considered ``sub-grid'' processes that investigators must cope with to get from line measurements to useful constraints on models. The most basic of these arise in the data themselves. CGM absorption observations are generally not photon-noise limited, but line saturation is a major issue particularly for the most commonly detected species. Only lower limits can be derived from the equivalent widths of saturated lines; line profile fitting helps where the saturation is not too severe. Reliable columns of the crucial \hi\ ion are often challenging except where the Lyman limit is available. Moreover, the blending of narrow components with small velocity offsets in data with finite spatial resolution make all line measurements somewhat ambiguous. It is often necessary to model an entire line profile as a single nominal cloud, though sometimes the ionization state can be constrained on a component-by-component basis.

There is often ambiguity about whether to adopt PIE, CIE, or combination non-equilibrium models. These issues are compounded by uncertainties in the additional model inputs. These include the relative elemental abundances, which need not be solar but are usually assumed to be.  The EUVB is a particular problem as it may be uncertain especially at low redshift \citep{kollmeier14}, introducing up to an order of magnitude systematic error into some ionic abundances \citep{oppenheimer13}. 

Though \ovi\ is among the strongest and most frequently detected  CGM metal absorption line, it amply demonstrates the problems encountered in precisely constraining the exact physical origins of ionized gas.  
For example, absorption-line studies in high-resolution and high-S/N QSO spectra and complementary studies of HVCs around the MW show that the ionization mechanisms of \ovi\ are both varied and complex over a wide range of environments \citep[e.g.][]{sembach04, Tripp:2008, Savage:2014}.
Ionic column density ratios and line profiles sometimes support a common photoionized origin for \ovi, \nv, and low-ion gas \citep[e.g.][]{muzahid15}, while other systems require \ovi\ to be collisionally ionized in a $\sim$10$^{5.5}$ K plasma \citep[e.g.][]{tumlinson05, fox09, Tripp:2011, wakker12,  Narayanan:2011, Meiring:2013, Turner:2016}.  Often, the multiple components for a single absorber show both narrow and broad absorption lines consistent with both scenarios. 

All these thorny issues with ionization modeling highlight the difficultly of getting at the detailed ``sub-grid'' physics of a complex, dynamic, ionized medium. We should maintain a cautious posture toward conclusions that depend sensitively on exact ionization states. Much of the detailed physics is still at scales that we cannot yet resolve. Nevertheless, in Section 5 we will see what we can learn by simplifying the situation to the most basic classes of models and proceeding from there.

\subsection{Gastrophysical Models}
\label{sec_gastrophysics}
The ``Galactic Corona'' began with Spitzer's insight that cold clouds could be confined by a hot surrounding medium. This model has matured over the years into a strong line of theoretical research focused on the detailed physics of how the thermal, hydrodynamic, and ionization state of CGM gas evolves in dark matter halos. 
Placing multiphase gas into the context of the dark matter halo, \cite{Maller:2004} suggested cold clouds cooled out of thermal instabilities in a hot medium, while maintaining rough pressure equilibrium (though see \cite{Binney:2009} for a counterpoint). Accretion may also be seeded by gas ejected from the disk, as in the ``galactic fountain'' or ``precipitation'' model \citep[e.g.,]{fraternali08, Voit:2015b}. These scenarios start with very simple assumptions--such as hydrostatic hot halos, diffuse clouds in photoionization equilibrium, or particular radial entropy profiles. These simplifying assumptions are necessary because we do not know the large-scale pjhysical state of the CGM as a whole. Photoionization modeling of the low-ionization CGM using only the EUVB \citep{hm01} strongly disfavors hydrostatic equilibrium with hot gas at $T_{\rm vir}$ \citep{Werk:2014}; the cool and hot phases appear to have similar densities, rather than similar pressures. Furthermore, if \ovi-traced gas follows a  hydrostatic profile at the temperature where its ionization fraction peaks, T $\sim$ $10^{5.5}$\,K, then its column density profile would be significantly steeper than observed \citep{Tumlinson:2011}. There may be other means of supporting this gas, such as turbulence \citep{fielding16}, cosmic rays \citep{Salem:2016}, or magnetic fields. 

Adding to the uncertain physical conditions in the CGM is the fact that \ovi\ likely represents a massive reservoir of warm gas (\S\,5.2.3). Such a massive reservoir is apparently at odds with the short cooling times for \ovi\ given by typical CI models; these timescales are often much shorter than the dynamical time, on the order of $\lesssim 10^8$\,yr. Yet, the short cooling times for \ovi\ are in fact characteristic of many models for the multiphase CGM. In many formulations, the cooler low-ion traced gas precipitates out of the warmer \ovi-traced phase, owing to thermal instabilities (\citealp{shapiro76, McCourt:2012, Voit:2015b,Thompson:2016}; see also \citealp{wang95}), while the \ovi-traced gas may be continually replenished by a hot galactic outflow. In a similar vein, the \ovi-traced warm gas could be cooling isochorically out of a hotter halo \citep[e.g.][]{edgar86, faerman17} but overcome its short expected lifetime by extra energy injection from star formation or AGN. 

Fully understanding the broader context and origin of the multiphase CGM will require more than microphysical and phenomenological models alone can offer. Cosmological hydrodynamic simulations with self-consistent cosmic accretion and multiphase outflows are key to deciphering the panoply of observed absorption lines (\S\,7). Moreover, much of the microphysics proposed as a natural source or maintainer of multiphase gas (e.g., thermal instabilities and turbulence) requires resolutions much higher than can be achieved by simulations that must simultaneously model a the enormous dynamic range required for galactic assembly. Yet essentially all cosmological hydrodynamic simulations {\em do} produce a multiphase CGM (see, e.g., Figure 3). In general, the combination of the simulated density and temperature profiles of the CGM results in different ions preferentially residing at different galactocentric radii, with low-ions preferring the denser, cooler inner CGM and higher ions filling the lower density, hotter outer CGM (\citealp{Hummels:2013,Ford:2014, Suresh:2015a}; see also \citet{Stern:2016}). Yet inhomogeneous mixing of the different gas phases complicates predictions for gas cooling rates and the small-scale metal mixing which depend crucially on the unknown diffusion coefficient \citep{schaye07}. 

Hydrodynamic simulations may be compared directly to observations via synthetic spectra, potentially helping to disentangle the degeneracy between physical space and observed velocity space. Constructing these synthetic spectra, however, faces many of the same challenges as modeling the ionization states of the observed gas: while the density, temperature, and metallicity of the simulated gas may be known, the EUVB and ionization mechanism must still be assumed in order to calculate ionization states (see, e.g., \citealt{hummels16}). Most simulations rely on the same radiative transfer codes (e.g., \texttt{Cloudy}, \citealp{ferland13}) that observational analyses do, though non-equilibrium chemistry and cooling are being included as computation power increases \citep{Oppenheimer:2013, Silvia:2013}. If these assumptions are incorrect, comparisons of derived results (such as masses) rather than observables (such as column densities) may lead to simulations getting the ``right answer'' for the wrong reasons.


\section{The Baryonic Mass Distribution of the CGM}

\label{major_section_mass} 

\subsection{The Missing Baryons Budget}

Empirically constraining the total CGM mass as a function of stellar and/or halo mass is essential to quantifying models of galactic fueling and feedback. Under the condition $\Omega_{b}$/$\Omega_{m} = 0.16$ \citep{Planck:2013}, the total baryonic budget of sub-$L^*$ to super-$L^*$ galaxies spans two orders of magnitude, ranging from 10$^{10.3}$ $-$  10$^{12.3}$ M$_{\odot}$.  Although the stars and ISM for super-$L^*$ galaxies are similar {\it fractions} of the total ($\sim$ 5\%), the absolute amount of mass that must be found is around $100\times$ larger for sub-$L^*$ galaxies and $10\times$ larger for $L^*$ galaxies. How much of this 80--90\% missing mass is in the CGM?  We organize this subsection by temperature, and review the observations, assumptions, and uncertainties in each calculation, using Figures~\ref{fig_surface_density} and \ref{fig_baryons} to synthesize current results. We note that a recent review by \cite{JBH:2016} performed a similar radially-varying mass-budget compilation for the Milky Way and its halo and incorporates some of these same results. 

The baryon census as presented here relies on the assumption that galaxies fall along well-defined scaling relations of ISM and CGM gas mass as a function of stellar mass, and that the scatter in these scaling relations is uncorrelated. We caution that there is tentative evidence that this is not necessarily the case: COS-GASS has shown galaxies with more cold gas in their ISM have more cold gas in their CGM \citep{Borthakur:2015}. While the correlation between CGM and ISM exhibits a high degree of scatter, likely from patchiness in the CGM, it exists at $>99.5$\%\ confidence, and stacked \lya\ profiles for low and high ISM masses clearly show the effect. The large-scale environment and gaseous interstellar content are difficult to explicitly account for in overall baryon budgets, and may account for some of the scatter in the various estimates. For example, \cite{Burchett:2015b} find that the detection of \civ\ around galaxies with $\mstar > 10^{9.5}$\Msun~drops significantly for galaxies in high-density regions \citep[see also, ][]{Johnson:2015a}. Future work should control for these properties. 

\subsection{CGM Masses by Phase}

\subsubsection{Cold Gas, T $<$ 10$^{4}$ K}  

Cold-gas tracers consist of neutral and low ions like \ion{H}{i}, \ion{Na}{i}, \ion{Ca}{ii}, and dust. This is material that may have cooled from hotter phases that experienced thermal instability, or may arise in clouds entrained in multiphase outflows. \cite{Putman:2012} estimated the total cold gas mass traced by HVCs in the Milky Way halo to be $M = 2.6 \times 10^7$ \Msun\ (including only HVCs detected via 21 cm emission, and excluding the Magellanic Stream system). The Magellanic Stream  provides an additional contribution of $\sim 3 \times 10^8$ \Msun, but it cannot  be assumed to be a generic feature of galaxies. Thus, the total contribution from cold gas is $M \lesssim 10^9$ \Msun\ even if the ISM of the Clouds are included, making up less than 1\% of the missing baryons for a Milky-Way like halo. We further note that while dust masses have been estimated from stacks of reddened background QSOs \citep{Menard:2010} and  galaxies as ``standard crayons'' \citep{Peek:2015} indicating values comparable to the dust in the ISM of these galaxies (see \S\,\ref{subsection_dust}), both ISM and and CGM dust are at most only $\sim$1\% of the missing halo baryons. Finally, using stacked optical spectra from SDSS, \cite{Zhu:2013a} derived a column density profile for gas bearing \ion{Ca}{ii} H and K around $\sim L^*$ galaxies. For the purposes of Figure 5 we have converted this to a mass density profile, conservatively assuming that the calcium is entirely in \ion{Ca}{ii} and $Z = Z_{\odot}$. The total mass for \ion{Ca}{ii} itself is 5000 \Msun, and when we scale to $Z = Z_{\odot}$ we derive $M = 2 \times 10^8$ \Msun for the cold component, again $\sim 1$\% or less of the baryons budgets.

\subsubsection{UV Absorption Lines and the Cool 10$^{4-5}$ K CGM}

The mass of the cool CGM ($\sim 10^{4-5}$ K) is perhaps the best constrained of all the phases at low redshift, owing to the rich set of UV lines in this temperature range. Prior to COS, estimate for this phase were based on single ions with very simple ionization and metallicity corrections to arrive at rough estimates. \cite{Prochaska:2011a} estimated $M_{\rm cool} \approx$ 3 $\times$ 10$^{10}$ M$_{\odot}$ for {\it{all}} galaxies from 0.01 $L^{*}$ to $L^{*}$, assuming a constant $N_{\rm H} = 10^{19}$ cm$^{-2}$ out to 300 kpc. Using a ``blind'' sample of \ion{Mg}{ii} absorbers, \cite{Chen:2010c} estimated $M_{\rm cool} \approx$ 6 $\times$ 10$^{9}$ M$_{\odot}$ for the \ion{Mg}{ii}-bearing clouds alone. The former estimate simply took a characteristic ionization correction, while the latter counted velocity components as clouds and converted from a metal column density to $N_H$ using a metallicity, because neither study had the multiphase diagnostic line sets that could be used to self-consistently constrain gas density and metallicity. Both  $L^*$ and super-$L^*$ galaxies have provided the most reliable constraints, mainly due to their relative ease of detection in photometric and spectroscopic surveys at $z < 0.5$ \citep{chen09,Prochaska:2011a, Werk:2012, Stocke:2013}. 

\begin{figure}[t]
\includegraphics[width=\textwidth]{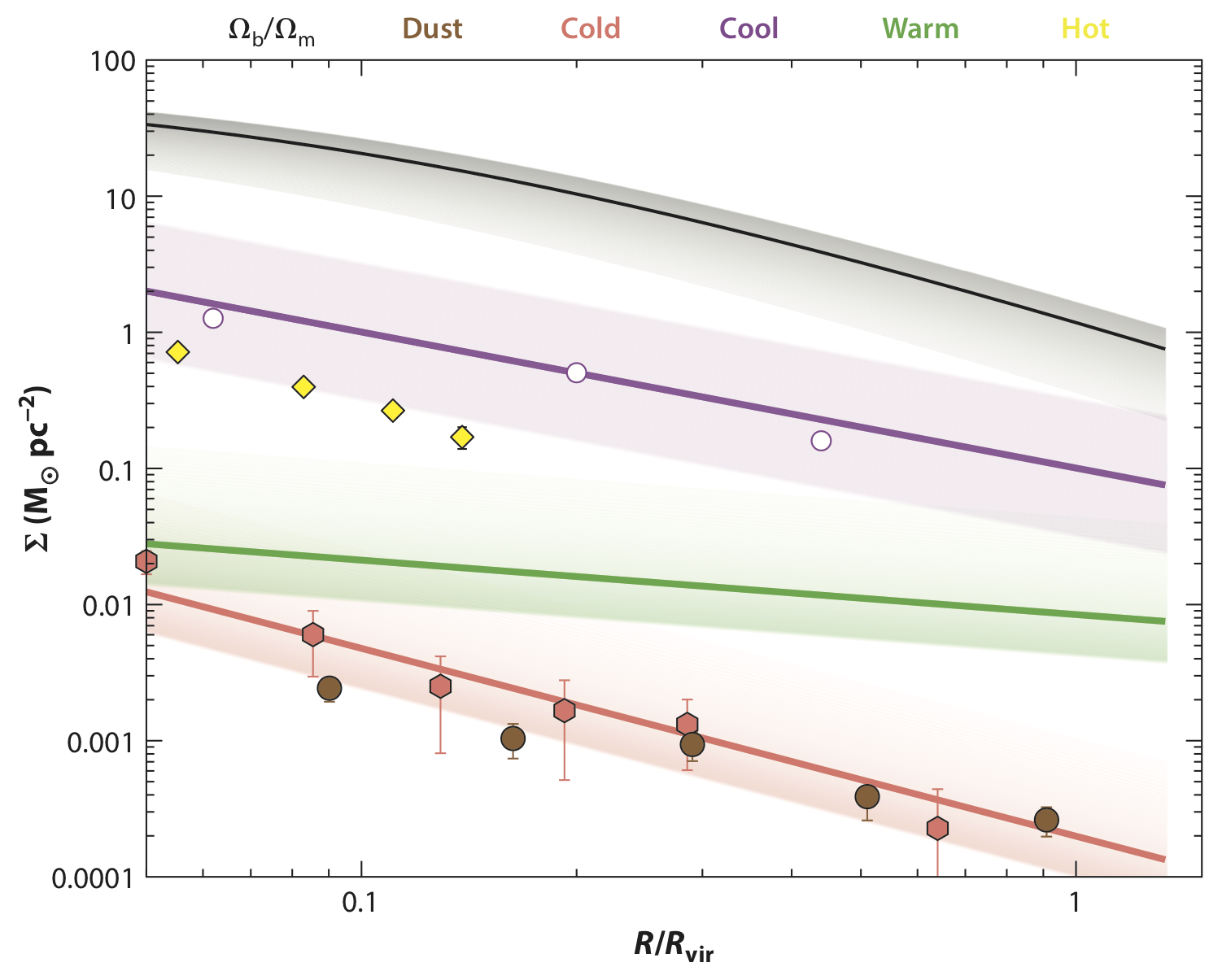}
\caption{A synthesis of CGM mass density results for  
``cold gas'' (pink, \citealp{Zhu:2013c}),  ``cool gas'' (purple, \citealp{Werk:2014}), ``warm gas'' traced by \ovi\ (green, \citealp{Tumlinson:2011, Peeples:2014}), X-ray emitting gas (yellow, NGC1961, \citealp{Anderson:2016}), and dust (brown, \citealp{Menard:2010}). An NFW profile for $M_{\rm DM} = 2 \times 10^{12}$ \Msun\ is at the top in black. \label{fig_surface_density}}
\end{figure}

With COS, it became practical to build statistically significant samples of absorbers that cover a broader range of ions. These estimates still rely on photoionization modeling, carried out under the standard assumption that the low-ions and \hi\ trace cool ($T< 10^{5}$\,K) gas and the primary source of ionizing radiation is the extragalactic UV background (UVB). Using the COS-Halos survey, \cite{Werk:2014} addressed the mass density profile and total mass for $L\approx L^{*}$ galaxies with PIE models that derive self-consistent $n_H$ and $Z$ using a range of adjacent ionization states of low-ion absorption lines (primarily \cii, \ciii, \siii, \siiii, \nii, and \niii).  The resulting surface density profile appears in Figures~\ref{fig_surface_density}, and yields $M_{\rm cool} = 6.5 \times 10^{10} M_{\odot}$ for $L^{*}$ galaxies out to $\rvir$.  Using the same COS-Halos sample with new COS spectra covering the Lyman limit, and taking a non-parametric approach with a robust treatment of uncertainties,  \cite{Prochaska:2017} recently refined the cool CGM mass estimate to be 9.2$\pm$4.3 $\times$ $10^{10} M_{\odot}$ out to 160 kpc. \cite{Stocke:2013} used the complementary approach of estimating of individual cloud sizes and masses, along with their average volume filling factor, for galaxies in three luminosity bins ($< 0.1 L^{*}$, $0.1-1 L^{*}$, and $L > L^{*}$). They find volume filling factors that range from 3-5\% for their modeled clouds, with length scales (N$_{\rm H}$ / n$_{\rm H}$) ranging from 0.1--30 kpc, totaling $\log M_{\rm cool} = 7.8 - 8.3$, $9.5 - 9.9$, and $10 - 10.4$, respectively. Finally, \cite{Stern:2016} determine the total mass in the cool (and possibly warm CGM) of $1.3 \pm 0.4 \times 10^{10}$\Msun\ for $L^{*}$ galaxies given their ``universal'' cloud density profile. In this phenomenological model each ion occupies a shell of a given $n$ and $T$ such that the fraction of gas in that particular ionization state is maximized.  Thus, this calculation represents a conservative minimum of baryons that must be present.  These ranges are shown in Figure~\ref{fig_baryons}. 

For super-$L^{*}$ galaxies, \cite{Zhu:2014} use stacking techniques to estimate the correlation function between luminous red galaxies with a mean stellar mass of 10$^{11.5}$ M$_{\odot}$ and cool gas traced by \ion{Mg}{ii} absorption in SDSS data for $\sim$ 850,000 galaxies with  $0.4 <z< 0.75$. The cool CGM around massive galaxies calculated in this way appears to completely close the CGM baryon budget for super-$L^*$ galaxies, at 17\% of the total halo mass. The assumptions for metallicity and ionization corrections, however, make it uncertain.

\subsubsection{UV Absorption Lines and the Warm 10$^{5-6}$ K CGM}

In Figure \ref{fig_usual_ions}, it appears as though ions like \ion{C}{iv}, \ion{N}{v}, \ovi, and \ion{Ne}{vii} trace the warm CGM at $T\approx 10^{5-6}$\,K. However, this temperature range in particular is burdened by significant uncertainty in the precise ionization mechanism responsible for its purported ionic tracers (see \S~4.4).  If high-ions are partially photoionized,  \ovi\, for example, may trace a non-negligible fraction of $T < 10^{5}$\,K gas that has already been counted toward the total baryon census in the previous section. For gas traced by \ovi, \cite{werk16} point out that typical photoionization models like those used for the low-ions have difficulty accounting for the total column of \ovi\ and column density ratios of \ion{N}{v} / \ovi\ without the need for path lengths in excess of 100 kpc.  However, significant additional ionizing radiation at $\sim$ 100 eV may reduce this requirement. 

\begin{figure}[t]
\includegraphics[width=6in]{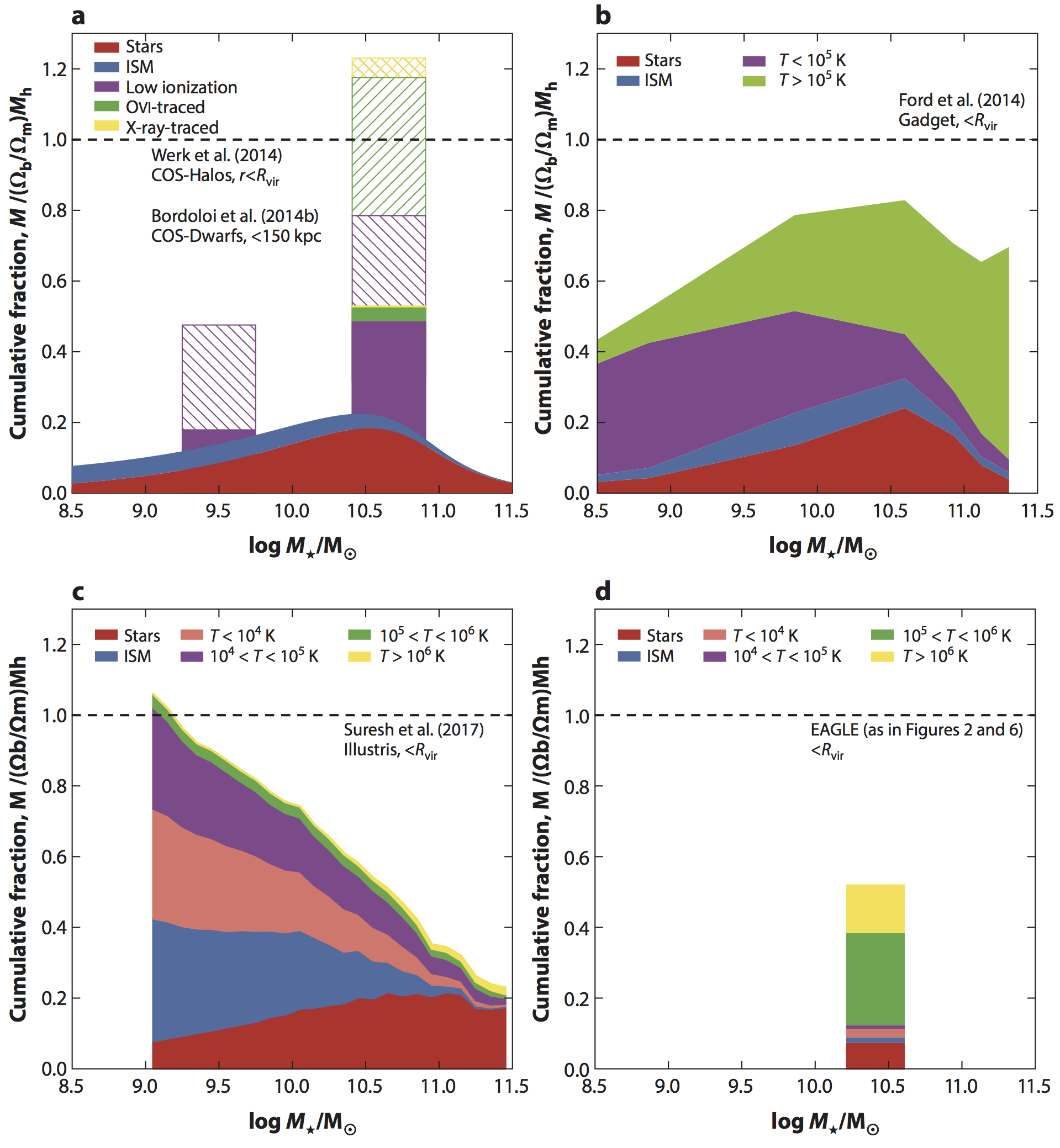}
\caption{Upper left: An accounting of CGM baryon budgets for all physical phases. The solid bars show the minimum values, while the hatched regions show the maximal values. The other three panels show simulated baryon budgets from  \citet{Ford:2014} in the upper-right, Illustris \citep{suresh17} in the bottom left, and in the bottom right, the EAGLE halo shown in Figures 2 and 6 \citep{Schaye:2015,Oppenheimer:2016}.  
}
\label{fig_baryons}
\end{figure}

In general, CIE models require a very narrow range of temperature to reproduce the \ovi\ observations, $T = 10^{5.3-5.6}$ K \citep{Tumlinson:2011, werk16}.
Furthermore, the kinematics of \ovi\ relative to the low-ions, in particular large $b$ values, seem to naturally support the idea that the \ovi\ is in a hotter phase (\citealp{Tripp:2011, muzahid12}; see also \citealp{tripp01, Stern:2016}). \cite{Tumlinson:2011} found that \ovi\ traces a warm CGM component that contributes  $>2 \times 10^{9}$ M$_{\odot}$ of gas to the $L^{*}$ baryon budget.  This mass estimate is strictly a lower limit due to the conservative assumptions adopted: (1) solar metallicity; (2) the maximum fraction of oxygen in \ovi\ allowed by CIE models, 0.2, and (3)  the CGM sharply ends at 150 kpc.
We adopt $\log M_{\rm warm} = 10.0$ in Figure~\ref{fig_baryons} for the COS-Halos galaxies (see also \citealt{faerman17}). 

For sub-$L^*$ galaxies, \cite{Bordoloi:2014a} estimate $M_{\rm warm}$ using \civ. As these galaxies are at $z<0.1$, the COS spectra do not cover the full range of Lyman series lines and ions available at $z > 0.1$, hindering detailed ionization modeling. 
COS only covers \ovi\ at $z > 0.2$, where it is difficult to assemble statistically significant samples of confirmed sub-$L^{*}$ galaxies, so an \ovi-based mass estimate for low-mass galaxies is not currently possible. 
With these caveats in mind, assuming a limiting ionization fraction for \civ, \citeauthor{Bordoloi:2014a} derive $\log M_{\rm warm} = 9.5$, if the gas typically has solar metallicity. For gas with lower metallicity, e.g., 0.1 solar, the value is 10 times higher and rather closer to baryonic closure for sub-$L^*$ galaxies (Figure~\ref{fig_baryons}). We caution that for \civ,
detailed photoionization often
places \civ\  with low-ionization state gas rather than with high-ionization state gas (e.g., \citealp{Narayanan:2011}). Thus, the \civ-derived mass for sub-$L^{*}$ galaxies is highly uncertain without detections of additional ionization states. 

One of the most surprising results to emerge from \cite{Tumlinson:2011} is that \ovi\ appears to be absent around the non--star-forming, more massive galaxies in the COS-Halos sample. Thus, there is tentative evidence that $\sim 10^{5.5}$ K gas is not a major component of the CGM of super-$L^{*}$ galaxies, which may be a result of massive galaxies having generally hotter halos or non-equilibrium cooling \citep{Oppenheimer:2016}. Thus, we do not have a good observational constraint for the warm CGM baryonic content for super-$L^{*}$ galaxies. The extreme-UV ion \ion{Ne}{viii} redshifts into the COS band at $z > 0.5$, where a few detections \citep{Tripp:2011, Meiring:2013} hint that it may be present in halos out to 100--200 kpc. However, the number of absorbers associated with particular galaxies is not yet sufficient to include it in mass estimates for the warm phase. 

\subsubsection{The Hot T $> 10^{6}$ K Phase}

Hot gas at the virial temperature ($T_{\rm vir}$= G$\mhalo$m$_{\rm p}$/$k R_{\rm vir}$) is a long-standing prediction. For $M_{\rm halo} \gtrsim 10^{12}$\Msun, the temperature should be $T \gtrsim 10^6$ K, and observable at X-ray wavelengths, although there are extreme-UV tracers such as \ion{Mg}{x} and \ion{Si}{xii} that have yet to yield positive detections (Figure~\ref{fig_multiphase_ions}). Only a few very luminous spirals and ellipticals have had their halos detected \citep{Anderson:2011, Dai:2012, Bogdan:2013, Walker:2015, Anderson:2016}, and independent constraints the temperature, density, and metallicity profiles from soft X-ray spectroscopy is rarer still. Thus the fraction of baryons residing in the hot phase, and its dependence on stellar and or halo mass, are not yet determined. 

Three sets of constraints are relevant: the Milky Way, individual external galaxies, and stacked samples of external galaxies.  
\cite{Anderson:2010} addressed directly the problem of whether hot gas could close the baryon budget for the Milky Way. From indirect constraints such as pulsar dispersion measures toward the LMC, cold gas cloud morphology, and the diffuse X-ray background, they limited the hot gas mass to $M \lesssim 0.5-1.5 \times 10^{10}$ \Msun, or only 2-5\% of the missing mass. The choice of an NFW profile for the hot gas is a key assumption: if the density profile is assumed to be flatter ($\beta \sim 0.5$), the mass can be 3-5 times higher, but still only 6-13\% of the missing baryons. The \citet{Gupta:2012a} claims that the baryon budget is closed for the Milky Way, based on the assumption of an isothermal, uniform density medium, have been questioned by evidence that the gas is neither isothermal nor of uniform density \citep{Wang:2012}. 

The well-studied case of NGC 1961 \citep{Anderson:2016} constrains the hot gas surface density out to $R \simeq 40$ kpc, inside which $M_{\rm hot} = 7 \times 10^9$ \Msun\ compared with the stellar mass of $3 \times 10^{11}$ \Msun\ and far from baryonic closure. Extrapolating to 400 kpc yields $M_{\rm hot} = 4 \times 10^{11}$ \Msun, but given the declining temperature profile it is likely that it declines to more intermediate temperatures, $T \lesssim 10^6$ K, where EUV and FUV indicators provide the best diagnostics. Stacked emission maps of nearby galaxies provide the strongest evidence for extended hot halos. In a stack of 2165 isolated, K-selected galaxies from ROSAT, \cite{Anderson:2013} found strong evidence for X-ray emission around early type galaxies and extremely luminous galaxies of both early and late type. The X-ray luminosity depends more on galaxy luminosity than on morphological type. Luminous galaxies show $M = 4 \times 10^9$ \Msun\ within 50 kpc, and $M = 1.5-3.3 \times 10^{10}$ \Msun\ if extrapolated out to 200 kpc, comparable to the stellar masses. Yet high amounts of hot gas this far out would appear to be excluded by \cite{Yao:2010a}, who stacked Chandra spectra at the redshifts of foreground galaxies and placed strict ($\lesssim 1$ m\AA) limits on \ovii\ and \ion{O}{viii}. The limits are also consistent with the limits on nearby galaxy emissivity earlier derived by \cite{Anderson:2010}. The key uncertainty is how far out the hot gas extends with the flat, $\beta \sim 0.5$ density profile seen at $R \lesssim 50$ kpc, but the \cite{Yao:2010a} limits imply that hot gas halos around nearby galaxies appear to host at most $\simeq 10^{10}$ \Msun. In their summary of the X-ray results, \cite{Werk:2014} adopted $M_{\rm hot} = 1$--$14 \times 10^9$ \Msun\ from \cite{Anderson:2013}. 

\begin{marginnote}[]
\entry{CMB}{Cosmic Microwave Background}
\entry{SZ}{Sunyaev-Zeldovich}
\end{marginnote}

The thermal SZ effect---scattering of CMB photons by free electrons in a plasma---may constrain the hot gas content of galaxy clusters and halos down to the galactic scale. \cite{Planck:2013} and \cite{Greco:2015} claim detections down to $\mstar = 2 \times 10^{11}$ \Msun\ and a possible signal down to $\mstar =  6 \times 10^{10}$ \Msun. These results create tension with the X-ray measurements, since the SZ detections imply a ``self-similar'' relation between $\mhalo$ and $M_{\rm hot}$ down from the cluster scale ($\mhalo \sim 10^{14}$ \Msun), where we know hot baryons close the budgets, into the galactic range where this is much less clear. It may be that the hot gas extends well beyond the X-ray surface brightness limits at 50 kpc up to the Mpc scales where the SZ effect is measured. On the other hand, if every $\geq L^*$ halo was filled with $T_{\rm vir}$ gas, it would violate constraints from the soft X-ray background \citep{Wu:2001}. If halos with $\mhalo\lesssim 10^{11}$ \Msun\ depart from self-similarity, the cause could be the cooling and feedback that cause prevent halos from reaching their cosmic share of baryons. The kinematic SZ effect---in which photons receive a Doppler shift when scattering of a plasma with bulk motion---may be able to reach even lower masses for halo gas measurements \citep{Hill:2016}. This work is in its early stages and we look forward to more progress that complements the UV and X-ray.

\subsubsection{Theoretical Considerations}
From the discussion above and the synthesis in the top left panel of Figure~\ref{fig_baryons}, we see that CGM measurements have added significantly to the baryon budgets for galaxies, and may complete those budgets under some assumptions. There has been theoretical progress as well: hydrodynamical simulations generally agree that the CGM contains a budget of baryons at the same order of magnitude as the stellar masses. In the other three panels of Figure~\ref{fig_baryons}, we show there is less quantitative agreement for the temperature partitioning of the CGM as a function of stellar mass, despite these models having approximately the same predictions for the baryonic content of galaxies.

A promising aspect of this quantitative disagreement is that different physical treatments of energetic and/or kinetic feedback do indeed lead to different total baryon fractions, and in particular to different trends in the fraction by phase.  Thus, observations of how CGM gas masses are distributed by phase can favor or disfavor particular physical prescriptions, and thus already offer phenomenological tests of models. However, these comparisons additionally show how challenging it will be to perform stringent tests. Even where simulations with radically different physical prescriptions yield opposite trends, at any particular mass they only different by factors of $\lesssim 2$ in the fraction of any phase. At present, this range is comparable to the systematic errors remaining in the observational characterization of the phases. Thus any claims that the data favors or disfavors any particular model should be made and interpreted carefully. As discussed in \S\,\ref{sec_gastrophysics}, comparing the models to observations by using synthetic data and directly comparing {\em observables} such as column densities and line kinematics have the benefit of shifting the myriad assumptions discussed in \S\,\ref{sec_obs_to_physical} onto the simulations.


\section{Metals: Nature's Tracer Particles}
\label{major_section_metals}

\subsection{The Metals Census}

\begin{figure}[t]
\hspace{-0.5in}
\includegraphics[width=6in]{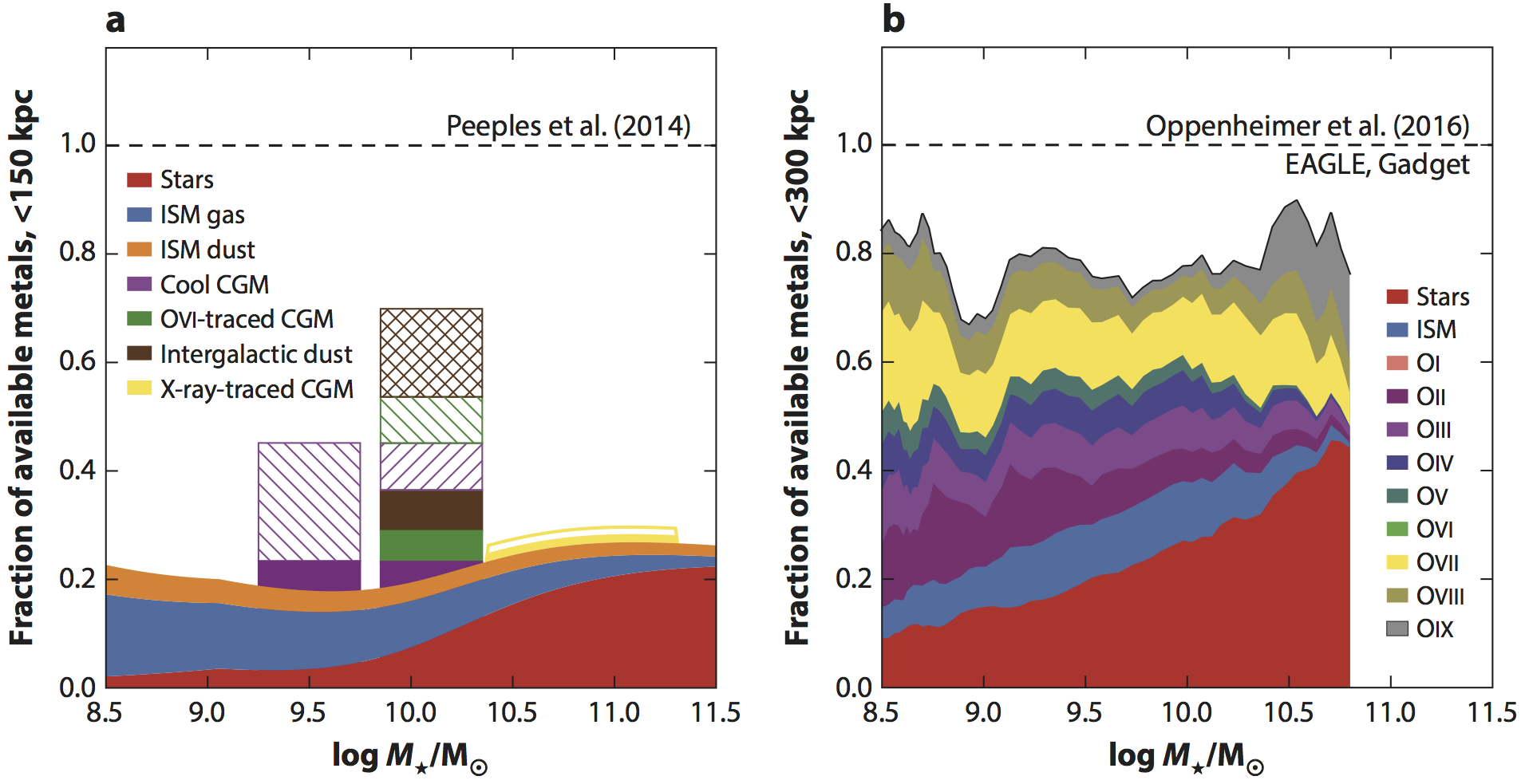}
\caption{Left: A metals census of the CGM around star-forming $z\sim 0$ galaxies following \cite{Peeples:2014}, including a sub-$L^*$ budget from \citet{Bordoloi:2014a}. As in Figure 7, stars are red, ISM gas is blue, ISM dust is orange the cool CGM is purple, the \ovi-traced CGM is green, the X-ray traced CGM is yellow, and intergalactic dust is in brown.  Right: A simulated budget from 55 relatively isolated $\log\mstar \geq 8.5$ star-forming EAGLE halos, with a moving average smoothing \citep{Oppenheimer:2016}. In both panels, the denominator is the total mass of metals ever produced by the central galaxy; the CGM may have contributions from, e.g., satellites.
}
\label{fig_metals_census}
\end{figure}

Total mass budgets by themselves do not fully reveal the {\it flows} that govern galaxy evolution. However, there is a ready means of distinguishing inflows from outflows: stars produce heavy elements sending passively-advecting ``tracer particles'' out into the ISM, CGM, and IGM from stellar winds and supernovae. The metal content of galactic flows can help identify their origins and determine their fate, and break degeneracies between models matched to the four galaxy problems. The galactic metals census (\S~2) requires that we compute the total budget of ``available metals'' produced by the galaxy by $z = 0$. This census was performed by \cite{Peeples:2014} by compiling measurements on stars, ISM and CGM gas, and dust. As shown in Figure~\ref{fig_metals_census}, the contributions bound in stars (red), interstellar gas (blue), and interstellar dust (orange)---the metals inside galaxies---add up to
only consistently 20--30\% over a factor of $\sim 1000$ in stellar mass.\footnote{While the overall level of the fraction of metals retained in galaxies is uncertain, primarily owing to uncertainties in nucleosynthetic yields, the {\em flatness} of this relation is fairly robust; see \citet{Peeples:2014} for a thorough discussion of the uncertainties in this calculation.} Ideally, this census would be done for each element individually, with the CGM divided into each ionization state of that element, e.g., oxygen \citep{Oppenheimer:2016}, but as that is observationally not yet generally feasible, the ionization corrections discussed in earlier sections must instead be done to account for unobservable ionization states. Qualitatively similar results are seen in simulations that have addressed this problem in particular \citep{Muratov:2016}. This striking invariance must offer some important clues to the operation of galactic outflows and inflows, with potentially large implications for the processes of galaxy fueling, feedback, and recycling.

\subsection{Metals Observed as Gas}
\label{subsection_metals_gas}

Even Lyman Spitzer might have recognized that the heavy elements observed in the CGM are in some sense the cause of, and solution to, all our problems. Apart from the (problematic) series of Lyman lines in the rest-frame FUV, virtually all our knowledge of the physical state, mass, kinematics, and evolution of the CGM gas come from lines of C, N, O, Si, Fe, Mg, Ca, and so on, whether they appear in the UV or X-ray. Yet, as described in Section \ref{major_section_physical_state}, these critical diagnostics also present many problems of analysis and interpretation. To work through this, it helps to distinguish between measurements of {\it metal content} or {\it metal mass} on the one hand and {\it metallicity} on the other. This distinction hinges on whether or not the hydrogen content can be measured, which is notoriously difficult. Measurements of hydrogen suffer severe \hi\ saturation effects, and juggling both metals and hydrogen compounds the difficulties of ionization corrections. When considering metal mass, we can often tolerate simpler ionization corrections or even direct sums of metal ion surface densities, sidestepping the large ionization corrections for \hi\ (\S~\ref{subsection_challenges}). 

The COS-Halos survey \citep{Tumlinson:2011} used the \ovi\ line observed with COS in a way that typifies measurements of metal content rather than metallicity. Their basic empirical finding is that \ovi\ appears at column densities of $\log N_{\rm OVI} \simeq 14$--$14.5$ out to the 150 kpc limits of the survey. Since \ovi\ does not reach more than 20\% of the total oxygen in most ionization conditions, they were able to place a robust lower limit of $> 10^7$\Msun\ of total oxygen for star forming galaxies. As it comes from direct integration of surface densities for a heavy element, does not refer to H, and uses a limiting ionization correction, this estimate avoids some of the trickiest aspects of metallicity measurements, and yet has significant implications for the budgets of galactic metals \citep{Peeples:2014}. The \ovi\ traces a high ionization component of the CGM gas; adding lower ionization gas to the budget requires the more complex ionization corrections and assumed relative abundances of oxygen and, e.g. Mg and Si, though it does not require the \hi-dependent metallicity corrections that plague the baryon census. Altogether, 20--30\% of available metals have been located in the $R < 150$ kpc CGM around $\sim L^*$ galaxies. 

By contrast with the measurements of total metal mass, {\it bona fide} metallicities require robust measurements of the hydrogen surface density, which entails accurate measurements of $N_{\rm HI}$ and reliable ionization corrections. For most strong CGM absorbers at $z \lesssim 0.2$, the Lyman series lines are saturated and do not yield reliable \ion{H}{i} column densities. However, beyond this redshift, and at $\log N_{\rm HI} > 16.2$, Lyman limit systems enable adequately precise ($\pm 0.2-0.3$) measurements of $N_{\rm HI}$ and the ionization corrections are manageable.

\begin{figure}[t]
\includegraphics[width=6in]{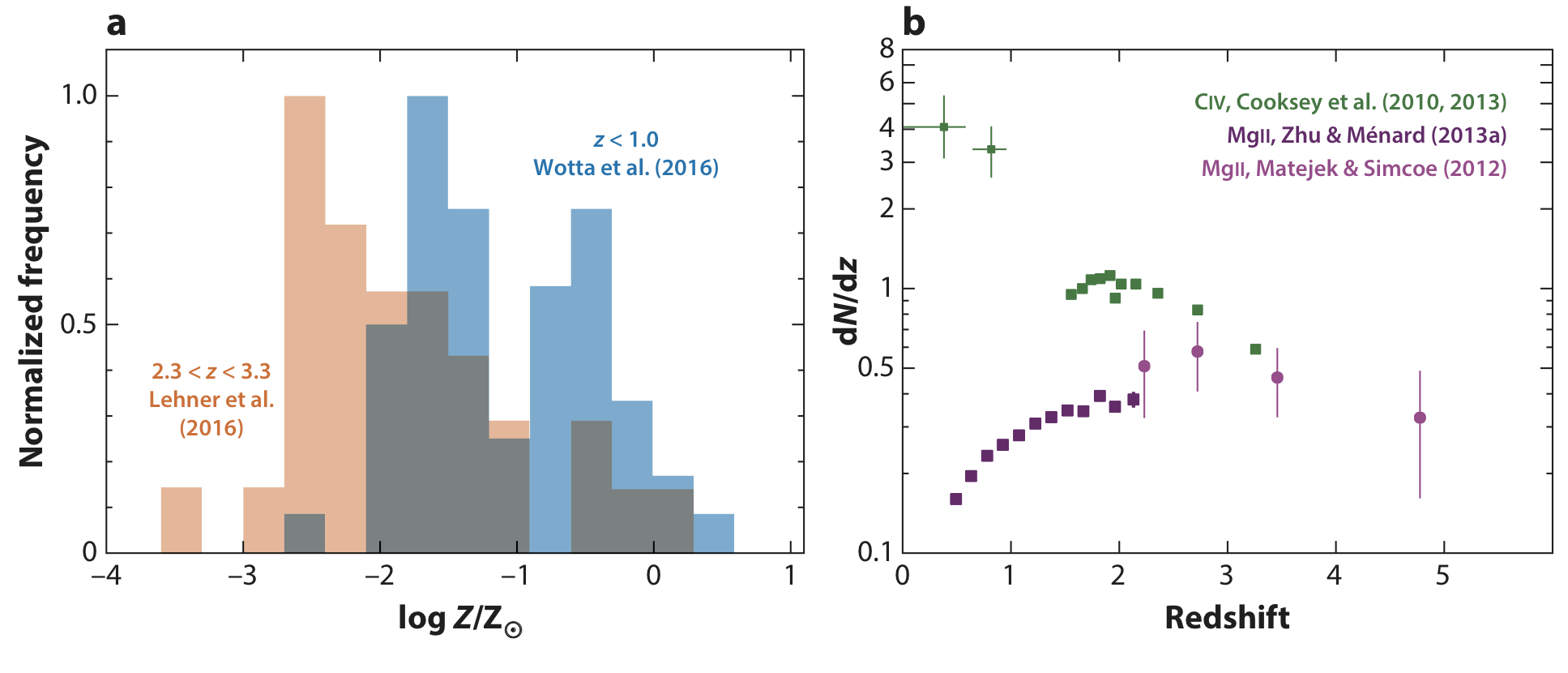}
\caption{Two views of CGM metallicity: (a) Two LLS distributions from \citet{Lehner:2013} and \citet{Wotta:2016}. This comparison clearly shows evolution in the LLS metallicities over time. (b) Trends in \mgii\ and \civ\ line density per unit redshift: the low-ion \mgii\ traces the cosmic star formation history, while \civ\ continually becomes more abundant.}
\label{fig_metallicities}
\end{figure}

By building a sample of LLSs from high-quality COS sightlines, \cite{Lehner:2013} and \cite{Wotta:2016} found that the distribution of metals in LLS clearly exhibits two peaks near 4\% solar and 50\% solar metallicity (Figure~\ref{fig_metallicities}a). The metallicities are constrained by detections of low-intermediate ions such as C II-IV, Si II-IV, OII-IV, and \mgii. This bimodal distribution qualitatively matches with expectations that accretion from the IGM into halos will have low metallicity, while accretion of gas previously ejected will have higher metallicity. The relative absence of intermediate values challenges our intuition that gas should naturally mix over time into a continuous distribution, and has posed a challenge to simulations \citep[][etc.]{Hafen:2016}. But most of these systems have not yet been identified with galaxies. In contrast to the Lehner bimodality, \citet{Prochaska:2017} find a unimodal distribution of metallicities within 160 kpc of L* galaxies with a median of $\sim$ 30\% solar. These metallicities derive from tight constraints on $N_{\rm HI}$ around $L^*$ COS-Halos galaxies with well-defined masses and distance to the absorber.  The contrast between the absorber-selected Lehner et al.\ sample and the galaxy-selected COS-Halos sample may indicate that they arise in other selection effects, but it may also indicate variation in CGM metallicity in different subsets of the galaxy population. 

By mining the Keck database of high-$z$ QSO absorbers, the KODIAQ survey studied a sample of LLSs at $z > 2$ \citep{Lehner:2016}. This sample is shown in the left panel of Figure~\ref{fig_metallicities} compared to the expanded low-z sample of \cite{Wotta:2016}. The $z > 2$ distribution is unimodal and centered at [X/H] $\sim -2$. A similar result was obtained for two samples of LLSs at still higher redshift, $z = 3.5-4$, with unimodal distributions centered at [X/H] $\sim -2.5$ \citep{Glidden:2016, Cooper:2015}. This is near the bottom edge of the low-metallicity peak at $z < 1$, indicating evolution in the average metallicity of high-column CGM over the few Gyr interval. Somehow, the bimodality emerges long after the initial buildup of metals, and is noticeable only in the $z < 1$ sample. Note that neither of these samples has specific galaxies attached---both are selected based on HI alone and the galaxies will have to be identified later. It is also possible that the column density range used for selection traces different galaxy masses, radii, and total column densities at the different redshifts, and so the apparent evolution does not occur in the same type of physical system (owing to a higher mean cosmic density). Nevertheless, it is now possible to compare the distribution of CGM metallicities over $\sim$6--10 Gyr of cosmic time.

In particular, there are ever-increasing samples of $z>2$ absorbers that {\em do} have associated galaxy information, allowing for a more direct comparison to the low-$z$ COS studies (Figure~\ref{fig_multiphase_ions}). The Keck Baryonic Structure Survey \citep[KBSS;][]{Rudie:2012} has engaged in a long campaign to characterize the CGM of star-forming galaxies at $z \sim 2.2$, going back to pioneering studies of absorption associated with Lyman-break galaxies \citep{Adelberger:2003}. These data show ion sets that overlap strongly with the low-$z$ studies. Both \ion{H}{i} and metals (\ion{O}{iv}, \nv, \ciii, \civ, and \ovi) show strong statistical correlations with galaxies out to 100-300 kpc. Using stacking, \cite{Steidel:2010} and \cite{Turner:2015} examined the relative kinematics of metals and galaxies, finding essentially all outflow kinematics and little sign of inflow; there must be gas flowing in to mainain the observed star formation rates, but it may be occuring in thin filaments with low covering fraction. 

These results across redshift can be viewed a different way, by examining the redshift evolution of strong lines that are likely to trace CGM gas. Figure~\ref{fig_metallicities}b shows the comoving sightline density of \mgii\ ($W_{\rm rest} \geq 1$ \AA) and \civ\ ($W_{\rm rest} \geq100$ m\AA), which follow different trends at $z < 2$. The number density of strong \mgii\ absorbers rises and then declines again toward $z = 0$. Absorbers above this limit occur within $\sim 100$ kpc of galaxies (see Figure~\ref{fig_multiphase_ions}), so the resemblance of this curve to the cosmic SFR density \citep{Hopkins:2006} suggests that the strong \ion{Mg}{ii} absorbers are linked to the fueling or feedback of star formation. Indeed, other evidence suggests that we are seeing the rise and decline of galactic superwinds (See 7.3). In contrast to the \mgii, strong \civ\ absorbers continue their march upwards at low redshift. This trend in moderate-to-high ionization gas may indicate that ionized gas in occupying the bulk of the CGM volume becomes more common even as strong winds creating \mgii\ absorbers decline with the cosmic SFR density. 

\subsection{Metals Observed as Dust}
\label{subsection_dust}

The interstellar medium is a mixture of gas and dust; this is no less true of the CGM. In a pioneering study, \cite{York:2006} stacked a sample of 800 strong Mg II absorbers to find evidence of SMC-like dust reddening. \cite{Menard:2010} added the SDSS photometric galaxy catalogs to this style of analysis and found that the reddening extends over angular scales consistent with distances hundreds of kpc away from the luminous galaxies (\ref{fig_surface_density}). To tie dust to specific galaxies and precise physical scales, \cite{Peek:2015} used passively evolving galaxies from SDSS as ``standard crayons'' to examine the reddening imposed by foreground SDSS spectroscopic galaxies. They found a strong reddening effect out to 150 kpc in the bluest bands and a steeper drop past that radius than in the angular correlations of \cite{Menard:2010}. The correlations with physical radius allow \cite{Peek:2015} to further estimate the typical total mass of dust for galaxies between $0.1$--$1L^*$ of $M_{\rm dust} \simeq 6 \pm 2 \times 10^7$ \Msun. They found only a weak trends with stellar mass, $M_{\rm dust} \propto \mstar^{0.2}$ and no discernible trend with the galaxies' specific star formation rates. Thus the presence of dust in the CGM out to 100 kpc scales provides unambiguous evidence that the CGM is fed by galactic outflows, accounting for approximately 10\% of the metals budget near $L^*$ (Figure \ref{fig_metals_census}). This degree of reddening can be explained by outflows from normal star forming galaxies in simulations, provided the dust-to-gas ratio is similar to the Galactic value and the dust survives the trip \citep{Zu:2011}. It is not yet clear why the dust properties show so little dependence on galaxy stellar mass, resembling the CGM \hi\ and low ions more than the CGM high-ionization gas. It might be that the increasing reddening at low redshift indicate a steady buildup of metals in the CGM and a relative lack of recycling into future star formation. Dust observations could also be used to test the physical models of galactic outflows that employ radiation pressure on dust to drag gas out of galaxies \citep{Murray:2005, Murray:2011}. Further explorations of CGM dust promise to constrain galactic outflows and recycling in ways that complement studies of gas.

\section{Inflows, Outflows, and Recycling}
\label{major_section_flows}

\subsection{The Problems: Galaxy Fueling and ``Missing'' Metals}
Recent findings show that the CGM possesses a significant budget of baryons, but how are they feeding galaxies across the spectrum of galaxy masses (Figure~\ref{fig_problems})? Accreting gas passes through the CGM on its journey from the IGM to galaxies, where it presumably leaves some observable signatures that we can use to characterize the inflows. The rates of accretion onto galaxies and of outflow out of galaxies are crucial parameters in most models of galaxy evolution \citep{tinsley80a}. However, there is not agreement about where and how a galaxy's fuel source is regulated. It is often assumed gas inflow from the IGM is balanced by the sum of star formation, gas ejection as outflows, and any net buildup of gas in the ISM \citep{Lilly:2013,Dekel:2014,Somerville:2015b}. This formulation completely omits the role of the CGM, even at the phenomenological level, but this ``bathtub'' model appears to nonetheless describe the many broad trends in galaxy scaling relations with redshift \citep{Dekel:2014}.Col
These models, though they do not explicitly address the CGM's composition or physical state, nonetheless have specific implications for its content and evolution (e.g., \citealp{Shattow:2015}). Conversely, models that use physical principles to describe the regulation of flows between the CGM and ISM \citep{Voit:2015b} can reproduce the same phenomenological galaxy scaling relations without detailed treatments of star formation inside galaxies. With observations of the CGM and its dynamics, we can potentially assess whether its role in regulating star formation is trivial, as the former models assume, or essential, as the latter models assume. Ideally, CGM observations would not only answer this question, but also reveal how it fuels star formation and manages outflows as a function of galaxy mass.

The observations we have discussed up to this point reveal the CGM (at low $z$) as a massive gaseous medium with a rich internal kinematic structure that is, in bulk, consistent with being bound to the host galaxies. Yet the degeneracy between kinematics and the physical location of absorbing gas can easily get lost in transverse sightline observations.

In simulations the CGM can appear to have obvious and well-ordered large-scale structure, with accreting and outflowing gas occupying physically distinct regions such as filaments and biconical outflows (\citealp{Shen:2012a,Corlies:2016}, see also Figure~\ref{fig_sims_showcase}), but at low redshift, circumgalactic gas tends to be more well mixed, with instantaneous velocities having little bearing on the origin or fate of a particular pocket of gas \citep{Ford:2014,Muratov:2015,Christensen:2016}, though this is also seen at $z=3$ \citep{vandeVoort:2012b}. In light of the observational projection effects, and theoretical cautions, we will now consider what can be learned from observing inflow and outflow directly in down-the-barrel observations, in which we interpret gas blueshifted relative to the galaxy as outflowing and redshifted gas as inflowing. These observations are better at probing gas in or near the disk-halo interface rather than the ``proper'' CGM out in the halo. Considering them in conjunction with CGM finding from transverse sightlines promises insights into the dynamics of the CGM that are not otherwise available.


\subsection{Empirical Signs of Fueling and Inflows}
\label{subsection:inflow}

Gas accretion is perhaps the most fundamental process in their formation \citep{Fox:2017}, as they must acquire gas, but feedback is optional. In the prevailing theoretical paradigm, gas flowing into galaxies at $\lesssim 10^{12}$ \Msun\ should be dynamically and thermally cold, while more massive halos receive most of their baryons as hotter ($T>10^5$) gas (\citealp{Dekel:2003,Dekel:2006,Keres:2005,Keres:2009a,Stewart:2011a}, though see \citealp{Nelson:2013}). Thus cold, dense, metal-poor CGM gas is often interpreted as direct evidence of accretion. First, cool, dense CGM gas is abundant in the form of LLSs. A large fraction of these are metal-poor at all redshifts \cite{Lehner:2013, Glidden:2016, Cooper:2015}. Metal-poor LLSs are evident as tracers of accretion in high resolution simulations \citep{Fumagalli:2011b, Hafen:2016}. The cool, bound \hi\ seen in the CGM of $z\sim 0.25$ galaxies \citep{Tumlinson:2013} should have a short cooling time. Finally, the finding from COS-GASS that there is a correlation between interstellar and circumgalactic \hi\ \citep{Borthakur:2015} implies a connection between circumgalactic fuel and star forming fuel. Though sub-$L^*$ and dwarf galaxies have not yet had their ``cool'' CGM masses measured directly, the widespread presence of \lya\ at similar strength suggests they too possess significant budgets of cold halo gas. 

All this evidence taken together strongly indicates that galaxies possess large reservoirs of CGM gas eligible for accretion. Yet evidence for {\em fuel} does not automatically constitute evidence for {\em fueling}: bound, cold gas has turned up in halos where its presence is surprising, such as the CGM of passive galaxies \citep{Thom:2012}. The actual fate of this material is unclear: how can we claim the bound cold gas is fueling star forming galaxies but not the passive galaxies? We therefore seek direct signatures of gas accretion onto galaxies. Yet these signatures are notoriously difficult to observe as incoming material may be metal poor, ionized, and obscured by outflowing material. Once gas is near the disk, proving empirically that it is accreting can be extremely difficult when it is seen in projection and its kinematics are easily confused with disk material. 

The Milky Way itself provides direct and unambiguous evidence for inflow in the form of its blueshifted high-velocity clouds (HVCs) and the striking Magellanic Stream. The HVCs arise in many complexes of clouds lying within $\sim 10$ kpc of the disk and have 100-300 km s$^{-1}$ blueshifted radial velocities that indicate they will reach the disk within $10^{7-8}$ yr. Their mass inflow rate falls between $0.1$--$0.5$ \Msun\ yr$^{-1}$, compared to the 1--2\Msun yr$^{-1}$ of star formation \citep{Putman:2012}. These clouds are all detectable in 21 cm emission, meaning that they occupy the tip of the column density distribution of CGM gas seen around other galaxies. The inflow rate inferred for ionized gas is much larger than for the classical HVCs, $\dot{M} \simeq 0.8 - 1.4$ \Msun\ yr$^{-1}$ \citep{Lehner:2011}, more comparable to the Milky Way's star formation rate. The Magellanic Stream is estimated to contain around $2 \times 10^9$ \Msun\ of gas in neutral and ionized form \cite{JBH:2007, Tepper-Garcia:2015}, and could provide $\sim 5$ \Msun\ yr$^{-1}$ of gas to the Milky Way disk as it accretes \citep{Fox:2014a}. Unfortunately, HVCs both above and below the radio-detection threshold are difficult to detect in external galaxies, despite intensive searches \citep{Putman:2012}, and satellites like the Magellanic Clouds and their Stream are not very common in L* galaxies. So we cannot generalize this result to mainstream galaxy populations. 

Down-the-barrel spectroscopy provides complementary information on inflows. Using this technique on $z\sim 0.5$ galaxies with Keck spectroscopy and {\em HST} imaging, \cite{Rubin:2012} detected clear signs of inflow at $80-200$ km s$^{-1}$ in star forming galaxies of $\log \mstar/M_{\odot} = 9.5 - 10.5$, inferring mass inflow rates of $\dot{M} \gtrsim 0.2$--3\Msun\ yr$^{-1}$. It seems likely that these estimates significantly undercount inflow, since inflowing (redshifted) gas is often obscured by outflows (blueshifted) or by emission from the galaxy's ISM (this problem is esepcially noticeable at higher redshift, \citealt{Steidel:2010}). Even if outflow is not present, the profiles are not sensitive to accretion from the lower half of the bimodal LLS metallicity distribution \citep{Lehner:2013}, which could make up a large fraction of the available cold CGM gas. Recently, \cite{zheng17} reported the detection of enriched, accreting gas at the disk-halo interface of M33 via COS observations of SiIV absorption along several sightlines to bright O stars in the disk. Their kinematic modeling of the observed absorption features implies an accretion rate of 2.9 \Msun\ yr$^{-1}$. While these results  provide evidence for accretion of cold, metal-enriched gas directly into galaxy disks, evidence for more metal-poor ``cold-mode'' accretion, and for gas entering further out in the disk (``on-ramp'', Figure 1), is still lacking (though see \citealp{Bouche:2013}), as is empirical characterization of how accretion rates vary with galaxy mass.


\subsection{The Preeminence of Outflows}

By consensus, outflows are an accomplice if not the perpetrator in each of the problems outlined in \S\,2. The existence of outflows is not in question: the large share of metals outside galaxies provides incontrovertible evidence for them (\S\,\ref{major_section_metals}). COS-Halos found widespread \ion{O}{vi} around star-forming galaxies---extended to $\sim 300$ kpc by \cite{Johnson:2015b}---but could not show that this ion becomes more prevalent with SFR. Even so, simulations found that robust outflows were necessary to produce the observed reservoir of metals \cite[e.g.,][]{Hummels:2013, Ford:2013, Suresh:2015a}, such that the high metal ions provide a significant constraint on the time-integrated effects of outflows even if it does not show the effects of recent or ongoing outflows directly. After that, the important questions concern {\it how} they transport baryons, metals, momentum, energy, and angular momentum. There is empirical evidence and strong theoretical suggestions that the physical drivers and properties of galaxy winds---their velocity, mass loading, metal content, and likelihood of escape---depends on galaxy mass, circular velocity ($v_{\rm circ}$), star formation rate, and metallicity. Many investigators pursue CGM observations in the hope that they can help to constrain these outflows and how they scale with galaxy properties. 

Direct observational evidence for outflows is readily available at all redshifts \citep[see][for a review]{Veilleux:2005}. In the nearby universe, large-scale complex multiphase outflows are seen in starbursts (e.g., M82) and from the Milky Way's central regions \citep{Fox:2015}. Down-the-barrel spectroscopy of the \ion{Na}{i}\,D in local starbursts \citep{Martin:2005} found that outflow velocities depend linearly on $v_{\rm circ}$. \cite{Rubin:2012} and \cite{Bordoloi:2014b} characterized similar flows using \ion{Mg}{ii} at $z \sim 1$. At $z > 2$, where the FUV-band ions used at $z \sim 0$ appear at visible wavelengths, \citet{Steidel:2010} used down-the-barrel spectroscopy to detect nearly ``ubiquitous'' outflows in rapidly star-forming LBG galaxies, with no clear indications for redshifted inflow. While these results help constrain the mass loading and covering fraction of outflows, they do not show how far these winds propagate into the CGM. It may be that the bulk of the energy is transported out in the hot gas while the bulk of the mass leaves in the cold phase, but this is still an open question \citep{strickland09}. 

Absorbers on transverse sightlines can directly constrain the impact of winds on the CGM. Cross-correlations of \mgii\ absorbers with the orientation of galaxies on the sky at $z\lesssim 1$, from both samples of individual galaxies \citep{Kacprzak:2012, Mathes:2014a} and stacked spectroscopy \citep{Bordoloi:2011, Zhu:2013c} find that the strongest absorbers prefer the semi-minor axis of disk galaxies, as expected for biconical outflows emerging from the disk. The preference for the semi-minor axis disappears by $\sim 60-80$ kpc, indicating that winds propagate at least that far, or merge into the general medium near that radius (e.g., the $z=2$ example in Figure~3). Studies of outflow covering fractions at $z\sim 1$ reinforce a picture of outflows being roughly biconical, with little surface area ($\sim 5$\%) solely dedicated to inflow \citep{Martin:2012,Rubin:2014}. Another strong clue about outflows comes from examining the CGM of starburst and post-starburst galaxies. Using an SDSS-selected sample, \cite{Heckman:2016} found unusually strong \ion{H}{i} and multphase ions at $100-200$ kpc compared with the COS-Halos and COS-GASS samples of galaxies at lower SFRs. These studies collectively show that SFR is a factor in determining the content of the CGM, perhaps as far out as $\rvir$. 

Down-the-barrel measurements tell us that outflows are ubiquitous, and sightline measurements tell us that they reach 100 kpc scales. Together these findings suggest that a large part of the CGM is made of outflows, and to examine one is to illuminate the other. The open questions concern not only the basic scaling of velocity and mass loading with galaxy $v_{\rm circ}$---which has received much attention---but just as importantly the distribution of outflow temperatures, metallicities, and fate. These cannot (yet) be simulated  from first principles but can be constrained by the combination of CGM and down-the-barrel observations. The former constrain the radial extent and the velocity fields of multiphase gas far from the disk, while the latter constrain the initial velocities, mass loading, and (possibly) metallicities.  

A recent goal of models and simulations has been to discriminate between winds that are ``momentum-driven'' \citep{Murray:2005}, which appear to improve the match of simulations to the galaxy mass-metallicity relation \citep{Finlator:2008} and the metal content of the IGM \citep{Oppenheimer:2006,Oppenheimer:2008}, and those that are ``energy-driven'' \citep{Murray:2011}, which appear to better match the galaxy stellar mass function \citep{Dave:2012} and new COS data \citep{Ford:2016}. 
A momentum-driven outflow has a velocity $v_{\rm w}\propto v_{\rm circ}^{-1}$, while an energy-driven flow has much faster outflows for low-mass galaxies with $v_{\rm w}\propto v_{\rm circ}^{-2}$; with a fiducial wind speed of $\sim 100$km\,s$^{-1}$, an unimpeded flow reaches 100\,kpc in only 1\,Gyr, i.e., the scales on which metals are seen in the CGM (\S\,6). Thus understanding the history of CGM metals and the velocities and mass flow rates of galactic flows go hand-in-hand.
Real winds may depend less on the local potential well and more on the local star formation rate surface density \citep{Kornei:2012a, Heckman:2015}. New hydrodynamic simulations of galaxies that resolve the multiphase ISM and explicitly include radiation pressure and thermal pressure \citep{Hopkins:2012} support this picture. Like essentially every other simulation suite on the market, however, models with this feedback scheme have too little \ovi\ in the CGM while retaining too many metals in stars \citep{Muratov:2015}.

\subsection{Following the Metals: The Role of Recycling}

Inflow and outflow are necessary processes in galaxy and CGM evolution; can one become the other by the recycling of outflows into fresh accretion of ejected gas? We have already established that, at least at low-redshift, galaxies require a long-term source of fuel, and that their CGM gas and metals are massive and bound. Recycling is a natural consequence; this gas ``should'' reaccrete onto the galaxy if the cooling time is short. Indeed, the predominance of metal-enriched accretion is supported by essentially all cosmological simulations where the origins of gas joining the ISM has been tracked: significant fractions at gas accreting onto galaxies has previously been ISM gas---and often through multiple cycles \citep{Ford:2014,Christensen:2016,Muratov:2016}, with the majority of star formation at late times fueled by recycled gas \citep{Oppenheimer:2010}. \citet{Ford:2014} found 60\% of all star formation at $z=0$ is powered by gas that was in the CGM a billion years before. This idea has the intriguing implication that a substantial fraction of all heavy elements on Earth once cycled through the Milky Way's halo at 100 kpc scales. The timescales are unclear: \citet{Christensen:2016} find that half of outflow mass is recycled on timescale of 1 Gyr with a logarithmic tail, independent of halo mass, while \citet{Oppenheimer:2008} find that $t_{\rm rec} \propto \mhalo^{-1/2}\sim 10^{9\pm 0.5}$\,yr, a timescale so short for massive galaxies that it is like not having an outflow at all, and so long for dwarfs that it essentially escapes forever. 

Thus the idea of recycling is well-motivated, but the details are still murky. Is it a simple process in which gas launched at $v<v_{\rm esc}$ encounters hydrodynamic resistance and eventually succumbs to gravity to fall back into the galaxy as part of a large-scale halo fountain? Or is the CGM well-mixed but multi-phase, with metal-rich gas precipitating out of the hot halo and raining onto the galaxy \citep{Voit:2015d,Fraternali:2015,Thompson:2016}? Here too can metals help disentangle the ins and outs.
Intriguingly, dense CGM gas (\citealp{Lehner:2013, Wotta:2016}; \S\,\ref{subsection_metals_gas}) is roughly equally divided between gas at a few percent solar (metal-poor IGM accretion) and 40\% solar (recycling ejecta?). 

While gas ``accreting'' from the IGM generally has (or is assumed to have) very {\em low} metallicity \citep{Lehnert:2013, Cooper:2015, Glidden:2016}, cases with metallicity well below the IGM (Ly$\alpha$ forest) at the same redshift are rare \citep{Fumagalli:2011a, Crighton:2016}. That is, either pristine cosmic accretion entrains metal-enriched circumgalactic gas on its way into the galaxy \citep[e.g.,][]{Fraternali:2015}, or that even at the highest redshifts where accretion is potentially observable, it is at least partially comprised by material that has previously been in the ISM, i.e., that recycled mode accretion is critical to galaxy evolution even at early cosmic times. Yet most formulations of the ``bathtub model'' assume that the accreting gas is pristine (e.g., \citealp{Lu:2015a}, though see \citealp{dave12}). Entrainment is a commonly invoked phenomenon for galaxy outflows, where it refers to the wind fluid sweeping up ambient ISM and mixing it with the fresh supernova ejecta powering the outflow. (It is important to note that the metallicity of the outflowing material is {\em necessarily} higher than that of the ambient ISM, contrary to what is assumed in some popular simulation recipes, e.g., \citealt{Vogelsberger:2014}.) Does ``recycled accretion'' behave in a simular way but in the opposite direction, with pristine inflows sweeping up metal-polluted CGM material on its way from the IGM to the ISM? Or do galaxy winds preferentially re-accrete, sweeping up more pristine cosmic accretion? 

Taking all this evidence into account, we can see the outlines of an emerging picture of galaxy inflows, at least at low redshift. They arise in the massive reservoir of cold, metal enriched gas bound to a galaxy's potential well, and enter the disk in HVC-like clouds but also in smooth flows of ionized gas. There may be a metal-poor component that comes more directly from the IGM without spending much time in the CGM, or otherwise acquiring metals. All these aspects of the CGM---cold, bound, metal enriched, and accreting---align better with the phenomenon of ``recycled accretion'' better than the bimodal ``hot / cold'' accretion. Recycled accretion arises from the ejection of metal-enriched galactic winds that lack the energy to escape the halo entirely, or which encounter the CGM itself and lose energy to radiation from shocks and then eventually cool and re-enter the galaxy. It may be that ``recycling'', rather than ``accretion and feedback'' is the more accurate way of viewing how galaxies acquire their gas.


\section{The Paradox of Quenching}
\label{section_quenching}

Passive and/or quenched galaxies possess little if any cold gas in their ISM, and blaming the CGM merely relocates the problem: how and why do these massive galaxies that once possessed a cold ISM lose and not regain it? Presumably their dark matter halos continue to add mass, but the accompanying gas does not enter the ISM and form stars like it once did. How galaxies achieve this transition is a deep and abiding problem in astrophysics, and the array of possible mechanisms for consuming, removing, and/or heating cold gas are beyond the scope of our review. We address the phenomenon of quenching by considering the CGM as a factor in, and indicator of, the quenching process.

\begin{marginnote}[]
\entry{LRG}{Luminous Red Galaxies}
\end{marginnote}

\subsection{The Fate of Cold Accretion and The Problem with Recycling}

The accretion of gas into halos, its heating to around $T_{\rm vir}$, and eventual cooling and entry to the ISM was long the prevailing picture of galaxy fueling. In an important twist on this basic picture, \cite{Keres:2005} argued that star-forming galaxies are fed by ``cold accretion'' never reaches $T_{\rm vir}$ but entered a galaxy's disk via streams while remaining below $T \sim 10^5$ K. Above $\log \mstar/{\rm M}_{\odot} \sim 10.3-10.5$ (or $\mhalo \sim 10^{12}$\Msun), the dark matter halo has sufficient mass, and the CGM enough pressure, to support a virial shock and suppress the cold mode. The coincidence of this mass with the stellar mass that divides star-forming from passive galaxies (Figure~\ref{fig_problems}) drew great attention to this scenario \cite[e.g.][]{Dekel:2006}, leading to predictions that the halos of passive galaxies should possess little cold gas \citep{Stewart:2011a}. 

\begin{figure}[t]
\includegraphics[width=4.5in]{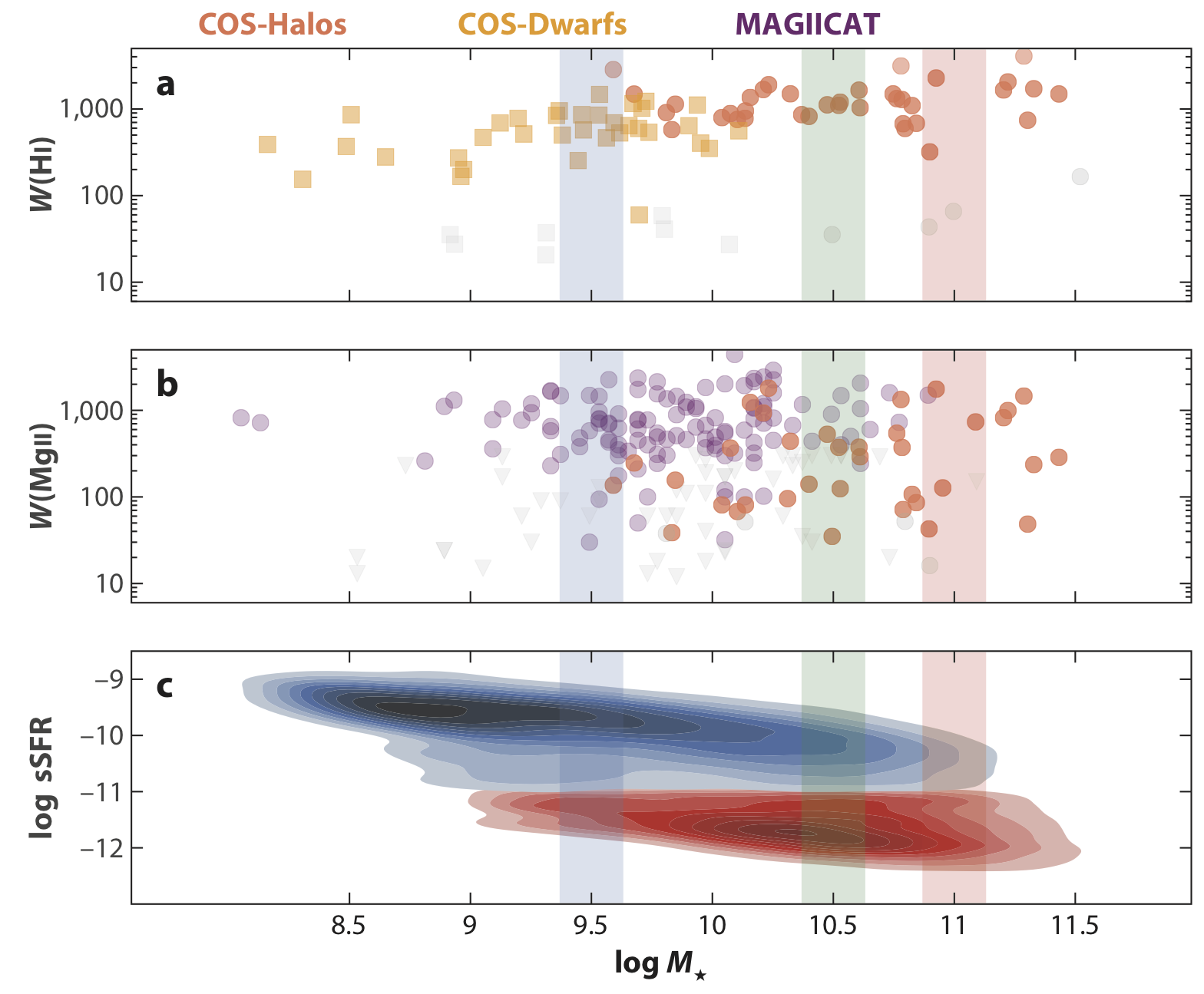}
\caption{Three views of the CGM and quenching. Top: a trend in \lya\ equivalent width over three decades in stellar mass from COS-Halos (\citealp{Tumlinson:2013}, purple) and COS-Dwarfs (\citealp{Bordoloi:2014b}, orange). As shown by \cite{Thom:2012}, the presence of \hi\ around red, passive galaxies indicates that their halos are not devoid of cold gas. Middle: \mgii\ from COS-Halos and MAGIICAT (\citealp{Nielsen:2016}, green). Bottom: the galaxy SFR bimodality from Figure 2. 
}
\label{fig_quenching}
\end{figure}

The observational picture belies the clean transition seen in simulations and the stark division of observed star formation rates. While COS-Halos did find a dramatic difference in highly ionized \ovi\ around star-forming and passive galaxies, the latter do not show as strong a deficit of CGM \hi. As shown in Figure \ref{fig_quenching}, the equivalent widths and covering fractions of \hi\ do not drop as stellar mass increases across the range $\log \mstar \simeq 10-11$ \citep{Thom:2012}. This is directly contrary to the expectation from, e.g. \citet{Stewart:2011a} that the covering fraction of strong \hi\ should drop to nearly zero as galaxies transition to the hot mode of accretion. The inner CGM ($< 50$\, kpc), however, is not well covered by these observations (Figure 4); it is possible that high pressure hot gas close to the galaxy prevents this cold material from accreting, as some models predict \citep{Schawinski:2014}. 

The presence of cool gas in the halos of massive red galaxies is now well-established by \mgii\ studies. \cite{Gauthier:2010} and \cite{Bowen:2011} found covering fraction of $f_c = 10-20$\% out to 100-200 kpc for $>1$\AA\ absorbers around LRGs. Using a sample of $\sim 4000$ foreground galaxies at $z = 0.5 - 0.9$ from the zCOSMOS survey, \cite{Bordoloi:2011} found that the \mgii\ equivalent width for blue galaxies is 8--10 times stronger at inner radii ($<50$ kpc) than for red galaxies, but even red galaxies possess evidence for cold gas. Using a new SDSS-based catalog of \mgii\ QSO absorbers and LRGs, \cite{Zhu:2014} mapped the mean profile out to $\gg 1$ Mpc scales, and argue that the mean profile at this mass scale is even stronger than found by \citeauthor{Bordoloi:2011},  extending at a detectable level out to 1 Mpc for LRGs. \cite{Johnson:2015b} have pointed out that strong \mgii\ absorbers are usually consistent with being bound to their host halos, meaning that the cold gas is contained with the dynamical influence of the galaxy. 

From a theoretical perspective, the quenching of galaxies is still a significant unsolved problem. Star formation must be curtailed, and later accretion and cooling of gas must be suppressed indefinitely to explain how galaxies remain passive for $>6$\,Gyr \citep{Gallazzi:2008}. Theories vary in how they accomplish this: some models artificially truncate star formation based on halo mass \citep{Somerville:2015b}, while others suppress the star-forming fuel by heating the CGM itself \citep[e.g.,][]{Gabor:2010, Gabor:2012}.
Thus the CGM itself can be the proximate cause of quenching, even if the source of CGM heating is not yet identified. Unfortunately models that manipulate the CGM directly cannot be tested against CGM observations, or at least, they must be modified somehow to recover the cold gas seen in passive galaxy halos.

By contrast, models that include self-consistent subgrid treatments of feedback, whether ``thermal'' \citep{Schaye:2015}, ``mechanical'' \citep{Choi:2015}, or a combination of thermal, mechanical, and radiative \citep{Vogelsberger:2014} can be compared to CGM observations as tests of their success. As an example, the mechanical feedback model implemented by \cite{Choi:2015} performed better than the ``standard'' \citep{Springel:2005, DiMatteo:2008} thermal feedback model in both suppressing galaxy formation and reducing the surface density of gas in the CGM by factors of 3--10 at 10--100\,kpc.

\cite{Suresh:2015a} addressed quenching using the Illustris simulations, which are tuned to the observed $\mstar/\mhalo$ and galaxy metallicities, but not the CGM. In Illustris, ``thermal'' AGN feedback is deposited locally, inside the galaxy, when the SMBH is in its energetic ``quasar'' mode. But in the $\sim 90$\% of the time when the SMBH is accreting quiescently, its ``radio mode'' feedback is deposited {\it non-locally} as thermal energy over 100 kpc scales. This amounts to direct heating of the CGM, shifting  cold gas to intermediate temperatures showing more \ovi, and otherwise warm gas to high temperatures showing \ion{O}{vii} and \ion{O}{viii}. The net effect is that the \cite{Tumlinson:2011} trend of strong \ovi\ around star forming galaxies and weak \ovi\ around passive galaxies is recovered. The ``cold'' CGM is reduced, but not completely destroyed. To be consistent, any visible effects of feedback would need to persist even when the AGN is not active, as the COS-Halos galaxies in question are not AGN at the time we observe them. The EAGLE simulations presented by \cite{Oppenheimer:2016} show a similar conclusion with models of thermal feedback and non-equilibrium cooling: at higher mass, with more feedback, \ovi\ is suppressed and the cold gas is depleted but not completely destroyed. These feedback effects force behaviors that generally resemble the data: they suppress star formation to create a red sequence, they force net gas loss from the inner CGM by heating gas that then bouyantly rises, and they shift the balance of gas ionization toward higher temperatures and higher ions. 

Despite these advances, the basic paradox of quenching remains: what happens to the halos of passive galaxies to quench their star formation, keep it quenched, and yet leave cold gas present in their halos? If passive galaxies possess cold gas and are not using it, can we be sure of the (naively obvious) conclusion that star-forming galaxies {\em are} using the diffuse gas they possess? Moreover, if the bulk of star formation at low-$z$ comes from recycled accretion, then to understand both how galaxies get their gas and how galaxies quench, we must understand how both the internal and external fuel supplies are shut off.

\subsection{The CGM of AGN and Quasars}

If feedback from AGN is effective at quenching their star formation and their cold CGM in simulations, it naturally suggests that this effect will be visible in the gaseous halos of galaxies with ongoing AGN activity. While hard radiation fields of AGN may leave distinctive ionization signatures in halo gas even long after the AGN fades \citep{Keel:2012, Oppenheimer:2013}, studies like COS-Halos with subsamples of passive galaxies have excluded active AGN for the most part, and even so have not seen any apparent signs of AGN effects on the CGM. No published study has systematically examined background QSO/foreground AGN pairs, though there is one such study underway with {\em Hubble}/COS\footnote{http://www.stsci.edu/cgi-bin/get-proposal-info?id=13774}. 

At $z>2$, the ``Quasars Probing Quasars'' (QPQ) program has seen clear evidence that galaxies hosting bright quasars show greatly enhanced gas budgets in \hi\ and low ions \citep{Prochaska:2014} though less excess in the high ions. This enhancement of neutral and low-ionization gas hints at a larger accretion rate for these robustly star-forming galaxies. AGN may even yield a net gain of cold gas in the CGM \cite{Faucher-Giguere:2016}. The \lya\ blobs observed at $z>2$ may  be gas accreting on to galaxies, with radiation powered by gravitational infall \citep{Goerdt:2010}, though these data may be more consistent with illumination from buried AGN \citep{Prescott:2015}. The higher gas masses only exacerbate the problem of feedback and quenching---there is more gas to be removed, and it is still not clear how that gas is removed or heated and accretion suppressed thereafter. Future work should focus on following such galaxies down through cosmic time as their QSOs fade, star formation is quenched, and the galaxies later evolve passively. Post-AGN and post-starburst galaxies should be examined for CGM gas as much as is practical. Understanding this process is critical to properly understanding the role of the CGM in creating or reflecting the birth of the red sequence.


\begin{issues}[Data in Need of More Theory]
\begin{enumerate}
\item Are there any clean observational tests or theoretical discriminants between the various heuristic models of feedback?
\item Are there self-consistent models of quenching that produce a red sequence of galaxies and yet leave a significant mass of cold CGM? How is the remaining cold gas kept from accreting? 
\item What do the detailed kinematic profiles of the multiphase suite of absorbing ions tell us about the physical and dynamic structure of the CGM?
\end{enumerate}
\end{issues}

\begin{issues}[Theory in Need of More Data]
\begin{enumerate}
\item What is the mass and composition of the CGM at high-redshift and in low-$z$ $\mstar < 10^{10}$ \Msun\ galaxies, and how do these constrain galaxy evolution models?  
\item What is the small-scale density and kinematic structure of the CGM, and what does it tell us about the physics? 
\item What does the CGM do as galaxies quench? Does cool, neutral gas extend into the inner CGM of passive galaxies?
\item Where are the metals that are still missing from the census? What are the elemental abundance ratios in CGM gas, and how do they depend on the galaxy's mass and star formation history?

\end{enumerate}
\end{issues}

\section{Open Problems, Future Prospects, and Final Thoughts} 

\subsection{Progress and Problems}

New instruments and new thinking reveal the CGM as a complex, dynamic gaseous environment that may close galactic baryon budgets and regulate gas accretion, star formation, and chemical enrichment. The observational studies that underlie the mass density profiles in Figure~\ref{fig_surface_density} and mass budgets in \S~4 and 5 have all been obtained since 2010. For years, questions about {\it how} and {\it when} gaseous halos influenced galaxy evolution consistently struggled with {\it what} was there. The bulk contents of the CGM are now better characterized than ever before. There remain missing pieces---the baryon and metals budget well below $L^*$ remain to be filled in (Figure~\ref{fig_baryons}), and many of the metals remain missing---but we can already see signs that the most urgent questions motivating new studies take {\it what} and {\it where} as known, and go on to ask {\it how} and {\it when}. These sort of questions strike more directly at physics than at phenomenology.

\subsubsection{The Scale Problem}

How a gaseous halo evolves is determined at any instant primarily by its density, temperature, metallicity, and radiation fields. But for an actual CGM (such as the simulated one in Figure 6) these physical quantities vary and evolve on many relevant scales, ranging from the sub-parsec sizes for single cold clouds to the $>$100 kpc size of the whole CGM and even $>$ Mpc scales in the IGM. If we are to answer the {\it hows} of accretion, feedback, recycling, and quenching, we must achieve a better understanding of the basic physical fields at higher spatial and kinematic resolution. This means finding ways to capture sub-parsec boundary layers and instabilities while also maintaining the $\gg$ kpc context. Yet this 5--6 order of magnitude range still cannot be captured simultaneously in numerical simulations. One approach would be to continue the development of physically rigorous analytic models \cite[e.g.,][]{Voit:2015b, Thompson:2016, fielding16, faerman17} that can isolate the key physical effects and then to incorporate these lessons into simulations at the subgrid level while their resolution improves with computing power. For instance, it might be possible to include subgrid models that account for unresolved interfaces between hot and cold gas, or to extract subgrid models for cosmological boxes from extremely high resolution idealized cloud simulations with carefully controlled physics. To complete the leap between phenomenology and physics, these intrinsically ``sub-grid'' processes must come under control while the proper cosmological and galactic context is maintained. 

The transport of metals and the information they provide would also benefit from addressing the scale problem. Metals trace feedback and drive cooling, so how they are distributed through CGM gas at small scales is a critical factor in a proper physical understanding of accretion and feedback. Dense CGM gas appears bimodal in metallicity, congruent with the idea of ``pristine accretion'' and ``recycled winds''. What does this tell us about the small-scale structure of the CGM, the relationship between accretion and feedback, and the mixing of diffuse gas? These are among the thorniest of open questions, because of the huge dynamic range in metallicity that must be captured. This problem will be addressed by larger absorber and galaxy surveys, but perhaps poses its stiffest challenges to numerical simulations, because many of the relevant physical mechanisms for mixing gas at boundaries and interfaces are still well below the ``sub-grid'' level of simulations. This is another case where coupling small-scale simulations of clouds to cosmological boxes could pay dividends. 

The ``scale'' problem exists also for data but might be better labeled a problem of resolution and confusion. In data, the rich multiphase and multiscale structures of CGM gas are seen through a complex rendering in absorption or emission lines from diagnostic ions. The line profiles of absorbers likely contain more information than we are currently able to extract and interpret. Systematic effects from line saturation, uncertain ionization and radiation fields, relative abundances, limited signal-to-noise, and finite spectral resolution all complicate the derivation of the true CGM density field, which in turn enters into mass estimates, energy balance, and timescales for the gas flows of interest. While we are learning to model and simulate the CGM at higher resolution with better physics, we should also aim to extract and use the full information available in the rich kinematic profiles of multiphase absorbers, which will likely require new analytic and statistical techniques. The importance and complexity of the CGM make it imperative to examine all of the information that Nature provides.

\subsubsection{Mass Flows and the Fate Problem}

The CGM matters to galaxies as long as it provides them with fuel and recycles their feedback. Ultimately this is what we care about most --- how does the CGM influence galaxy evolution?  The most fundamental questions with which we began are still not completely answered: How does cold gas accrete and form stars over billions of years, and why does this cycle stop in massive galaxies? Does the CGM empty out or get consumed when galaxies quench?  
How much star formation is fueled by recycling and how much by new accretion? Can we ever hope to identify particular absorbers as accretion, feedback, or recycling, or are we destined never to separate them? These questions will drive the field as it advances from phenomenology toward more sophisticated physical understanding. Properly explaining these phenomena in terms of the {\it hows} of accretion, feedback, recycling, and quenching requires that we follow mass flows, not merely mass budgets.

Now that we have a grip on the bulk contents of the CGM, it is time to develop and deploy the tools to probe these questions of how the gas flows operate. 
To follow flows, we will need to make at least three key advances. First, the mass budgets should be characterized more fully in all phases at stages of galaxy evolution, including those that are relatively short lived such as mergers and AGN. These analyses would additionally benefit from analyzing how outflows and inflows seen in down-the-barrel measurements relate to the kinematics viewed on transverse sightlines, an overdue synthesis deserving attention from both observations and theory. Second, we must attempt to directly constrain the timescales of CGM evolution using data alone---how do mass budgets and kinematics jointly constrain timescales? Third, we must look at simulations in a new way that focuses on the origins and evolution of the physical phases and how these appear in the data. A large measure of simulation work addressed to the CGM has focused on using column densities and kinematics to constrain uncertain mechanisms of feedback by matching real data to mocks from simulations. While these issues are being resolved, it is also valuable to look at simulations from a different phenomenological point of view. The study from \cite{Ford:2014} provides an example; that paper identified particles as ``pristine accretion'', ``recycled accretion'', ``young outflows'', and ``ancient outflows'' and followed their evolution over time. These insightful categories turn out to be correlated with observable signatures. We believe there is great potential in viewing models and data from this angle, trying to identify the more distinctive or even unique manifestations of key physical processes defined by their ``fate'' rather than their instantaneous properties or appearance. 

\subsection{Future Prospects for Data}

The next decade should bring a wide array of new instruments and numerical capabilities that will address these unsolved problems. 

While Hubble lasts (mid-2020s), UV absorber samples will grow, particularly those that focus on the $z > 0.5$ regime where a broader set of EUV ionization diagnostics is available (such as \ion{Ne}{viii}). This increase in coverage will in turn allow more careful treatments of ionization diagnostics component-by-component, hopefully with a better understanding of how CGM gas is spread across physical phases and across galaxy mass. COS remains the ideal instrument for this problem, and big advances are still possible in the metals budget, ionization and kinematic relationships of multiphase gas, and the relationships between CGM gas and special types of galaxies. Starting in 2018, the James Webb Space Telescope (JWST) will enable much deeper searches for faint galaxies near QSO sightlines, likely associating galaxies with samples of $z > 4$ absorbers that are already known \citep{Becker:2015, Matejek:2012}. 
Detections of \ion{H}{i} emission \citep[e.g.,][]{Martin:2015, ArrigoniBattaia:2015b, Cantalupo:2014} will provide useful tests of models for CGM mass and structure, but the problems of gas ionization state and metal transport will require much more challenging maps of emission from oxygen and carbon ions \citep[see][for a pioneering effort]{Hayes:2016}. Such maps might emerge from IFU spectrographs such as MUSE and KCWI, and their successors on 30m class telescopes; limits can be further improved by stacking of multiple galaxies. The optimal galaxies would be those where absorption line probes are also available, so that emission-line and pencil-beam measurements can be compared. Emission maps of metal-bearing CGM gas \citep[e.g.][]{Bertone:2010, Corlies:2016} are a key goal of the Large Ultraviolet/Optical/Near Infrared Surveyor (LUVOIR\footnote{http://asd.gsfc.nasa.gov/luvoir/}), which will push to 50x the UV point source sensitivity of Hubble/COS and 100-fold multiplexing in UV spectroscopy. Planned for launch in the 2030s, LUVOIR would be able to directly image the CGM in metal-line emission, map the most diffuse gas with weak absorbers, and resolve the multiphase kinematics of CGM gas with $R > 50\,000$ UV spectroscopy \citep{Dalcanton:2015}. The hot gas phase would be addressed by the ESA-planned X-ray flagship known as the Advanced Telescope for High ENergy Astrophysics ({\it ATHENA}\footnote{http://sci.esa.int/cosmic-vision/54517-athena/}) in 2028, with a significant focus on understanding the cosmic evolution of hot gas in the IGM and CGM. 

The size of our samples provide statistical power over the key galaxy variables: mass, redshift, shape, evolutionary state, and orientation to the sightline. Here, future UV absorber samples must be supplemented by optical absorber samples at $z \sim 3$, and by deeper galaxy surveys at all redshifts. This is a problem for the next generation of giant ground-based telescopes, which will advance high-$z$ CGM studies in rest-UV lines and support low-$z$ studies by obtaining redshifts of sub-$L^*$ galaxies near QSO sightlines at surveys at $z < 1$ to fill in the low-mass baryon and metals census, still a major missing piece. 

Massive fiber based surveys have proven effective at characterizing CGM gas and its flows with both intervening and down the-barrel measurements. This technique should only accelerate in the future, pushing to fainter sources, higher redshifts, and rarer foreground galaxies with future massively multiplexed spectrographs (e.g., eBOSS, PFS) on large telescopes. This technique excels at detecting weak signals in the CGM, and at examining more and more foreground galaxy properties with good statistics. With larger, deeper samples, we can look forward to addressing questions about the behavior of the cold/dense CGM in rarer galaxy types, such as quasars and AGN, mergers, and groups. 

\subsection{Final Thoughts}

Galaxies were understood as island universes long before astronomers discovered the interstellar gas that forms their stars. The intergalactic medium was added to the big picture with the discovery of QSO absorption lines and the development of the dark-matter cosmology. Because it is much fainter than stars, and much smaller than the IGM, the CGM is arguably the last major component of galaxies to be added but it has nevertheless become a vital frontier. As to {\it why}, it is clear that much has been learned by viewing galaxy evolution from the perspective of the CGM. The circumgalactic medium can even provoke fascination: might the heavy elements on Earth cycled back and forth through the Milky Way's CGM multiple times before the formation of the Solar System? It appears that the solution to major problems in galaxy formation that are still unsolved will run through this elusive region of the cosmos.

\section*{DISCLOSURE STATEMENT}
The authors are not aware of any affiliations, memberships, funding, or financial holdings that might be perceived as affecting the objectivity of this review. 

\section*{ACKNOWLEDGMENTS}
MSP and JT acknowledge support from NSF grant AST-1517908.  We are grateful to Ann Feild of STScI for her expert artistic contributions, to Joop Schaye and Ben Oppenheimer for use of the EAGLE simulation shown in Figures 2, 6, and 8, to Josh Suresh for data from the Illustris simulation shown in Figure 8, to Sasha Muratov for data from the FIRE simulation (Figure 8), and to Ben Oppenheimer for the data from the specially-analyzed EAGLE halos shown in Figure 9. We also thank Lauren Corlies, Matt McQuinn, Andrew Fox, Romeel Dav\'{e}, and John O'Meara for insightful comments on a draft of this article. We have made extensive use of NASA's Astrophysics Data System, astropy  \citep{Astropy},  matplotlib  \citep{Hunter:2007}, \texttt{yt} \citep{turk11}, and the python tools {\tt Colossus} from Benedikt Diemer and {\tt Seaborn} by Michael Waskom. 

\bibliographystyle{ar-style2}

\begin{thebibliography}{}
\expandafter\ifx\csname natexlab\endcsname\relax\def\natexlab#1{#1}\fi

\bibitem[{{Adelberger} et~al.(2003){Adelberger}, {Steidel}
  et~al.}]{Adelberger:2003}
{Adelberger} KL, {Steidel} CC, {Shapley} AE, {Pettini} M. 2003.
\textit{\apj} 584:45--75

\bibitem[{{Allen} et~al.(2008){Allen}, {Groves} et~al.}]{allen08}
{Allen} MG, {Groves} BA, {Dopita} MA, et~al. 2008.
\textit{\apjs} 178:20--55

\bibitem[{{Anderson} \& {Bregman}(2010)}]{Anderson:2010}
{Anderson} ME, {Bregman} JN. 2010.
\textit{\apj} 714:320--331

\bibitem[{{Anderson} \& {Bregman}(2011)}]{Anderson:2011}
{Anderson} ME, {Bregman} JN. 2011.
\textit{\apj} 737:22

\bibitem[{{Anderson}, {Bregman} \& {Dai}(2013)}]{Anderson:2013}
{Anderson} ME, {Bregman} JN, {Dai} X. 2013.
\textit{\apj} 762:106

\bibitem[{{Anderson}, {Churazov} \& {Bregman}(2016)}]{Anderson:2016}
{Anderson} ME, {Churazov} E, {Bregman} JN. 2016.
\textit{\mnras} 455:227--243

\bibitem[{{Armillotta} et~al.(2016){Armillotta}, {Werk} et~al.}]{armillotta16}
{Armillotta} L, {Werk} JK, {Prochaska} JX, et~al. 2016.
\textit{ArXiv e-prints}

\bibitem[{{Arrigoni Battaia} et~al.(2015){Arrigoni Battaia}, {Yang}
  et~al.}]{ArrigoniBattaia:2015b}
{Arrigoni Battaia} F, {Yang} Y, {Hennawi} JF, et~al. 2015.
\textit{\apj} 804:26

\bibitem[{{Bahcall} \& {Spitzer}(1969)}]{Bahcall:1969}
{Bahcall} JN, {Spitzer} Jr. L. 1969.
\textit{\apjl} 156:L63

\bibitem[{{Becker}, {Bolton} \& {Lidz}(2015)}]{Becker:2015}
{Becker} GD, {Bolton} JS, {Lidz} A. 2015.
\textit{\pasa} 32:e045

\bibitem[{{Begelman} \& {Fabian}(1990)}]{begelman90}
{Begelman} MC, {Fabian} AC. 1990.
\textit{\mnras} 244:26P--29P

\bibitem[{{Behroozi}, {Conroy} \& {Wechsler}(2010)}]{Behroozi:2010}
{Behroozi} PS, {Conroy} C, {Wechsler} RH. 2010.
\textit{\apj} 717:379--403

\bibitem[{{Benjamin}(1994)}]{benjamin94}
{Benjamin} RA. 1994.
\textit{{The Origin and Evolution of Galactic Halo Gas.}}
Ph.D. thesis, UT Austin

\bibitem[{{Bergeron}(1986)}]{Bergeron:1986a}
{Bergeron} J. 1986.
\textit{\aap} 155:L8--L11

\bibitem[{{Bergeron} \& {Boiss{\'e}}(1991)}]{Bergeron:1991}
{Bergeron} J, {Boiss{\'e}} P. 1991.
\textit{\aap} 243:344--366

\bibitem[{{Bergeron} \& {Stasi{\'n}ska}(1986)}]{bergeron:1986b}
{Bergeron} J, {Stasi{\'n}ska} G. 1986.
\textit{\aap} 169:1--13

\bibitem[{{Bertone} et~al.(2010){Bertone}, {Schaye} et~al.}]{Bertone:2010}
{Bertone} S, {Schaye} J, {Booth} CM, et~al. 2010.
\textit{\mnras} 408:1120--1138

\bibitem[{{Binney}, {Nipoti} \& {Fraternali}(2009)}]{Binney:2009}
{Binney} J, {Nipoti} C, {Fraternali} F. 2009.
\textit{\mnras} 397:1804--1815

\bibitem[{{Bland-Hawthorn} \& {Gerhard}(2016)}]{JBH:2016}
{Bland-Hawthorn} J, {Gerhard} O. 2016.
\textit{\araa} 54:529--596

\bibitem[{{Bland-Hawthorn} et~al.(2007){Bland-Hawthorn}, {Sutherland}
  et~al.}]{JBH:2007}
{Bland-Hawthorn} J, {Sutherland} R, {Agertz} O, {Moore} B. 2007.
\textit{\apjl} 670:L109--L112

\bibitem[{{Bogd{\'a}n} et~al.(2013){Bogd{\'a}n}, {Forman} et~al.}]{Bogdan:2013}
{Bogd{\'a}n} {\'A}, {Forman} WR, {Vogelsberger} M, et~al. 2013.
\textit{\apj} 772:97

\bibitem[{{Bordoloi}, {Heckman} \& {Norman}(2016)}]{bordoloi16}
{Bordoloi} R, {Heckman} TM, {Norman} CA. 2016.
\textit{ArXiv e-prints}

\bibitem[{{Bordoloi} et~al.(2014{\natexlab{a}}){Bordoloi}, {Lilly}
  et~al.}]{Bordoloi:2014b}
{Bordoloi} R, {Lilly} SJ, {Hardmeier} E, et~al. 2014{\natexlab{a}}.
\textit{\apj} 794:130

\bibitem[{{Bordoloi} et~al.(2011){Bordoloi}, {Lilly} et~al.}]{Bordoloi:2011}
{Bordoloi} R, {Lilly} SJ, {Knobel} C, et~al. 2011.
\textit{\apj} 743:10

\bibitem[{{Bordoloi} et~al.(2014{\natexlab{b}}){Bordoloi}, {Tumlinson}
  et~al.}]{Bordoloi:2014a}
{Bordoloi} R, {Tumlinson} J, {Werk} JK, et~al. 2014{\natexlab{b}}.
\textit{\apj} 796:136

\bibitem[{{Borthakur} et~al.(2015){Borthakur}, {Heckman}
  et~al.}]{Borthakur:2015}
{Borthakur} S, {Heckman} T, {Tumlinson} J, et~al. 2015.
\textit{\apj} 813:46

\bibitem[{{Bouch{\'e}} et~al.(2013){Bouch{\'e}}, {Murphy} et~al.}]{Bouche:2013}
{Bouch{\'e}} N, {Murphy} MT, {Kacprzak} GG, et~al. 2013.
\textit{Science} 341:50--53

\bibitem[{{Bowen} \& {Chelouche}(2011)}]{Bowen:2011}
{Bowen} DV, {Chelouche} D. 2011.
\textit{\apj} 727:47

\bibitem[{{Bowen} et~al.(2016){Bowen}, {Chelouche} et~al.}]{Bowen:2016}
{Bowen} DV, {Chelouche} D, {Jenkins} EB, et~al. 2016.
\textit{ArXiv e-prints}

\bibitem[{{Br{\"u}ns} et~al.(2000){Br{\"u}ns}, {Kerp} et~al.}]{bruns00}
{Br{\"u}ns} C, {Kerp} J, {Kalberla} PMW, {Mebold} U. 2000.
\textit{\aap} 357:120--128

\bibitem[{{Burchett} et~al.(2015){Burchett}, {Tripp} et~al.}]{Burchett:2015b}
{Burchett} JN, {Tripp} TM, {Bordoloi} R, et~al. 2015.
\textit{ArXiv e-prints}

\bibitem[{{Cantalupo} et~al.(2014){Cantalupo}, {Arrigoni-Battaia}
  et~al.}]{Cantalupo:2014}
{Cantalupo} S, {Arrigoni-Battaia} F, {Prochaska} JX, et~al. 2014.
\textit{\nat} 506:63--66

\bibitem[{{Chen} et~al.(2010){Chen}, {Helsby} et~al.}]{Chen:2010c}
{Chen} HW, {Helsby} JE, {Gauthier} JR, et~al. 2010.
\textit{\apj} 714:1521--1541

\bibitem[{{Chen} et~al.(1998){Chen}, {Lanzetta} et~al.}]{Chen:1998}
{Chen} HW, {Lanzetta} KM, {Webb} JK, {Barcons} X. 1998.
\textit{\apj} 498:77--94

\bibitem[{{Chen} \& {Mulchaey}(2009)}]{chen09}
{Chen} HW, {Mulchaey} JS. 2009.
\textit{\apj} 701:1219--1242

\bibitem[{{Choi} et~al.(2015){Choi}, {Ostriker} et~al.}]{Choi:2015}
{Choi} E, {Ostriker} JP, {Naab} T, et~al. 2015.
\textit{\mnras} 449:4105--4116

\bibitem[{{Christensen} et~al.(2016){Christensen}, {Dav{\'e}}
  et~al.}]{Christensen:2016}
{Christensen} CR, {Dav{\'e}} R, {Governato} F, et~al. 2016.
\textit{\apj} 824:57

\bibitem[{{Churchill} et~al.(2015){Churchill}, {Vander Vliet}
  et~al.}]{Churchill:2015}
{Churchill} CW, {Vander Vliet} JR, {Trujillo-Gomez} S, et~al. 2015.
\textit{\apj} 802:10

\bibitem[{{Cooper} et~al.(2015){Cooper}, {Simcoe} et~al.}]{Cooper:2015}
{Cooper} TJ, {Simcoe} RA, {Cooksey} KL, et~al. 2015.
\textit{\apj} 812:58

\bibitem[{{Corlies} \& {Schiminovich}(2016)}]{Corlies:2016}
{Corlies} L, {Schiminovich} D. 2016.
\textit{\apj} 827:148

\bibitem[{{Crighton}, {O'Meara} \& {Murphy}(2016)}]{Crighton:2016}
{Crighton} NHM, {O'Meara} JM, {Murphy} MT. 2016.
\textit{\mnras} 457:L44--L48

\bibitem[{{Dai} et~al.(2012){Dai}, {Anderson} et~al.}]{Dai:2012}
{Dai} X, {Anderson} ME, {Bregman} JN, {Miller} JM. 2012.
\textit{\apj} 755:107

\bibitem[{{Dalcanton} et~al.(2015){Dalcanton}, {Seager}
  et~al.}]{Dalcanton:2015}
{Dalcanton} J, {Seager} S, {Aigrain} S, et~al. 2015.
\textit{ArXiv e-prints}

\bibitem[{{Dav{\'e}}, {Finlator} \&
  {Oppenheimer}(2012{\natexlab{a}})}]{Dave:2012}
{Dav{\'e}} R, {Finlator} K, {Oppenheimer} BD. 2012{\natexlab{a}}.
\textit{\mnras} 421:98--107

\bibitem[{{Dav{\'e}}, {Finlator} \& {Oppenheimer}(2012{\natexlab{b}})}]{dave12}
{Dav{\'e}} R, {Finlator} K, {Oppenheimer} BD. 2012{\natexlab{b}}.
\textit{\mnras} 421:98--107

\bibitem[{{Dekel} \& {Birnboim}(2006)}]{Dekel:2006}
{Dekel} A, {Birnboim} Y. 2006.
\textit{\mnras} 368:2--20

\bibitem[{{Dekel} \& {Mandelker}(2014)}]{Dekel:2014}
{Dekel} A, {Mandelker} N. 2014.
\textit{\mnras} 444:2071--2084

\bibitem[{{Dekel} \& {Woo}(2003)}]{Dekel:2003}
{Dekel} A, {Woo} J. 2003.
\textit{\mnras} 344:1131--1144

\bibitem[{{Di Matteo} et~al.(2008){Di Matteo}, {Colberg}
  et~al.}]{DiMatteo:2008}
{Di Matteo} T, {Colberg} J, {Springel} V, et~al. 2008.
\textit{\apj} 676:33--53

\bibitem[{{Dopita} \& {Sutherland}(1996)}]{dopita96}
{Dopita} MA, {Sutherland} RS. 1996.
\textit{\apjs} 102:161

\bibitem[{{Edgar} \& {Chevalier}(1986)}]{edgar86}
{Edgar} RJ, {Chevalier} RA. 1986.
\textit{\apjl} 310:L27--L30

\bibitem[{{Faerman}, {Sternberg} \& {McKee}(2017)}]{faerman17}
{Faerman} Y, {Sternberg} A, {McKee} CF. 2017.
\textit{\apj} 835:52

\bibitem[{{Faucher-Giguere} et~al.(2016){Faucher-Giguere}, {Feldmann}
  et~al.}]{Faucher-Giguere:2016}
{Faucher-Giguere} CA, {Feldmann} R, {Quataert} E, et~al. 2016.
\textit{ArXiv e-prints}

\bibitem[{{Ferland} et~al.(2013){Ferland}, {Porter} et~al.}]{ferland13}
{Ferland} GJ, {Porter} RL, {van Hoof} PAM, et~al. 2013.
\textit{RMxAA} 49:137--163

\bibitem[{{Fielding} et~al.(2016){Fielding}, {Quataert} et~al.}]{fielding16}
{Fielding} D, {Quataert} E, {McCourt} M, {Thompson} TA. 2016.
\textit{ArXiv e-prints}

\bibitem[{{Finlator} \& {Dav{\'e}}(2008)}]{Finlator:2008}
{Finlator} K, {Dav{\'e}} R. 2008.
\textit{\mnras} 385:2181--2204

\bibitem[{{Ford} et~al.(2014){Ford}, {Dav{\'e}} et~al.}]{Ford:2014}
{Ford} AB, {Dav{\'e}} R, {Oppenheimer} BD, et~al. 2014.
\textit{\mnras} 444:1260--1281

\bibitem[{{Ford} et~al.(2013){Ford}, {Oppenheimer} et~al.}]{Ford:2013}
{Ford} AB, {Oppenheimer} BD, {Dav{\'e}} R, et~al. 2013.
\textit{\mnras} 432:89--112

\bibitem[{{Ford} et~al.(2016){Ford}, {Werk} et~al.}]{Ford:2016}
{Ford} AB, {Werk} JK, {Dav{\'e}} R, et~al. 2016.
\textit{\mnras} 459:1745--1763

\bibitem[{Fox \& Dav\'{e}(2017)}]{Fox:2017}
Fox A, Dav\'{e} R, eds. 2017.
\textit{Gas Accretion onto Galaxies}, vol. 430. Springer

\bibitem[{{Fox} et~al.(2015){Fox}, {Bordoloi} et~al.}]{Fox:2015}
{Fox} AJ, {Bordoloi} R, {Savage} BD, et~al. 2015.
\textit{\apjl} 799:L7

\bibitem[{{Fox} et~al.(2013){Fox}, {Lehner} et~al.}]{fox13}
{Fox} AJ, {Lehner} N, {Tumlinson} J, et~al. 2013.
\textit{\apj} 778:187

\bibitem[{{Fox} et~al.(2009){Fox}, {Prochaska} et~al.}]{fox09}
{Fox} AJ, {Prochaska} JX, {Ledoux} C, et~al. 2009.
\textit{\aap} 503:731--746

\bibitem[{{Fox}, {Savage} \& {Wakker}(2006)}]{fox06}
{Fox} AJ, {Savage} BD, {Wakker} BP. 2006.
\textit{\apjs} 165:229--255

\bibitem[{{Fox} et~al.(2014){Fox}, {Wakker} et~al.}]{Fox:2014a}
{Fox} AJ, {Wakker} BP, {Barger} KA, et~al. 2014.
\textit{\apj} 787:147

\bibitem[{{Fox} et~al.(2005){Fox}, {Wakker} et~al.}]{fox05}
{Fox} AJ, {Wakker} BP, {Savage} BD, et~al. 2005.
\textit{\apj} 630:332--354

\bibitem[{{Fraternali} \& {Binney}(2008)}]{fraternali08}
{Fraternali} F, {Binney} JJ. 2008.
\textit{\mnras} 386:935--944

\bibitem[{{Fraternali} et~al.(2015){Fraternali}, {Marasco}
  et~al.}]{Fraternali:2015}
{Fraternali} F, {Marasco} A, {Armillotta} L, {Marinacci} F. 2015.
\textit{\mnras} 447:L70--L74

\bibitem[{{Fumagalli}, {O'Meara} \& {Prochaska}(2011)}]{Fumagalli:2011a}
{Fumagalli} M, {O'Meara} JM, {Prochaska} JX. 2011.
\textit{Science} 334:1245

\bibitem[{{Fumagalli} et~al.(2011){Fumagalli}, {Prochaska}
  et~al.}]{Fumagalli:2011b}
{Fumagalli} M, {Prochaska} JX, {Kasen} D, et~al. 2011.
\textit{\mnras} 418:1796--1821

\bibitem[{{Gabor} \& {Dav{\'e}}(2012)}]{Gabor:2012}
{Gabor} JM, {Dav{\'e}} R. 2012.
\textit{\mnras} 427:1816--1829

\bibitem[{{Gabor} et~al.(2010){Gabor}, {Dav{\'e}} et~al.}]{Gabor:2010}
{Gabor} JM, {Dav{\'e}} R, {Finlator} K, {Oppenheimer} BD. 2010.
\textit{\mnras} 407:749--771

\bibitem[{{Gallazzi} et~al.(2008){Gallazzi}, {Brinchmann}
  et~al.}]{Gallazzi:2008}
{Gallazzi} A, {Brinchmann} J, {Charlot} S, {White} SDM. 2008.
\textit{\mnras} 383:1439--1458

\bibitem[{{Gauthier}, {Chen} \& {Tinker}(2010)}]{Gauthier:2010}
{Gauthier} JR, {Chen} HW, {Tinker} JL. 2010.
\textit{\apj} 716:1263--1268

\bibitem[{{Geha} et~al.(2012){Geha}, {Blanton} et~al.}]{Geha:2012}
{Geha} M, {Blanton} MR, {Yan} R, {Tinker} JL. 2012.
\textit{\apj} 757:85

\bibitem[{{Glidden} et~al.(2016){Glidden}, {Cooper} et~al.}]{Glidden:2016}
{Glidden} A, {Cooper} TJ, {Cooksey} KL, et~al. 2016.
\textit{\apj} 833:270

\bibitem[{{Gnat} \& {Sternberg}(2007{\natexlab{b}})}]{Gnat:2007}
{Gnat} O, {Sternberg} A. 2007{\natexlab{b}}.
\textit{\apjs} 168:213--230

\bibitem[{{Gnat} \& {Sternberg}(2009)}]{gnat09}
{Gnat} O, {Sternberg} A. 2009.
\textit{\apj} 693:1514--1542

\bibitem[{{Gnat}, {Sternberg} \& {McKee}(2010)}]{gnat10}
{Gnat} O, {Sternberg} A, {McKee} CF. 2010.
\textit{\apj} 718:1315--1331

\bibitem[{{Goerdt} et~al.(2010){Goerdt}, {Dekel} et~al.}]{Goerdt:2010}
{Goerdt} T, {Dekel} A, {Sternberg} A, et~al. 2010.
\textit{\mnras} 407:613--631

\bibitem[{{Greco} et~al.(2015){Greco}, {Hill} et~al.}]{Greco:2015}
{Greco} JP, {Hill} JC, {Spergel} DN, {Battaglia} N. 2015.
\textit{\apj} 808:151

\bibitem[{{Grimes} et~al.(2009){Grimes}, {Heckman} et~al.}]{grimes09}
{Grimes} JP, {Heckman} T, {Aloisi} A, et~al. 2009.
\textit{\apjs} 181:272--320

\bibitem[{{Gupta} et~al.(2012){Gupta}, {Mathur} et~al.}]{Gupta:2012a}
{Gupta} A, {Mathur} S, {Krongold} Y, et~al. 2012.
\textit{\apjl} 756:L8

\bibitem[{{Gutcke} et~al.(2017){Gutcke}, {Stinson} et~al.}]{Gutcke:2017}
{Gutcke} TA, {Stinson} GS, {Macci{\`o}} AV, et~al. 2017.
\textit{\mnras} 464:2796--2815

\bibitem[{{Haardt} \& {Madau}(2001)}]{hm01}
{Haardt} F, {Madau} P. 2001.
\textit{{Modelling the UV/X-ray cosmic background with CUBA}}. In
  \textit{Clusters of Galaxies and the High Redshift Universe Observed in
  X-rays}, eds. DM~{Neumann}, JTV {Tran}

\bibitem[{{Hafen} et~al.(2016){Hafen}, {Faucher-Giguere} et~al.}]{Hafen:2016}
{Hafen} Z, {Faucher-Giguere} CA, {Angles-Alcazar} D, et~al. 2016.
\textit{ArXiv e-prints}

\bibitem[{{Hayes} et~al.(2016){Hayes}, {Melinder} et~al.}]{Hayes:2016}
{Hayes} M, {Melinder} J, {{\"O}stlin} G, et~al. 2016.
\textit{ArXiv e-prints}

\bibitem[{{Heckman} et~al.(2015){Heckman}, {Alexandroff} et~al.}]{Heckman:2015}
{Heckman} TM, {Alexandroff} RM, {Borthakur} S, et~al. 2015.
\textit{\apj} 809:147

\bibitem[{{Heckman} \& {Borthakur}(2016)}]{Heckman:2016}
{Heckman} TM, {Borthakur} S. 2016.
\textit{\apj} 822:9

\bibitem[{{Heckman} et~al.(2002){Heckman}, {Norman} et~al.}]{Heckman:2002}
{Heckman} TM, {Norman} CA, {Strickland} DK, {Sembach} KR. 2002.
\textit{\apj} 577:691--700

\bibitem[{{Heitsch} \& {Putman}(2009)}]{heitsch09}
{Heitsch} F, {Putman} ME. 2009.
\textit{\apj} 698:1485--1496

\bibitem[{{Henry} et~al.(2015){Henry}, {Scarlata} et~al.}]{Henry:2015}
{Henry} A, {Scarlata} C, {Martin} CL, {Erb} D. 2015.
\textit{\apj} 809:19

\bibitem[{{Hill} et~al.(2016){Hill}, {Ferraro} et~al.}]{Hill:2016}
{Hill} JC, {Ferraro} S, {Battaglia} N, et~al. 2016.
\textit{Physical Review Letters} 117:051301

\bibitem[{{Hopkins} et~al.(2006){Hopkins}, {Hernquist} et~al.}]{Hopkins:2006}
{Hopkins} PF, {Hernquist} L, {Cox} TJ, et~al. 2006.
\textit{\apj} 639:700--709

\bibitem[{{Hopkins}, {Quataert} \& {Murray}(2012)}]{Hopkins:2012}
{Hopkins} PF, {Quataert} E, {Murray} N. 2012.
\textit{\mnras} 421:3522--3537

\bibitem[{{Hummels}, {Smith} \& {Silvia}(2016)}]{hummels16}
{Hummels} C, {Smith} B, {Silvia} D. 2016.
\textit{ArXiv e-prints}

\bibitem[{{Hummels} et~al.(2013){Hummels}, {Bryan} et~al.}]{Hummels:2013}
{Hummels} CB, {Bryan} GL, {Smith} BD, {Turk} MJ. 2013.
\textit{\mnras} 430:1548--1565

\bibitem[{{Humphrey} et~al.(2011){Humphrey}, {Buote} et~al.}]{Humphrey:2011}
{Humphrey} PJ, {Buote} DA, {Canizares} CR, et~al. 2011.
\textit{\apj} 729:53

\bibitem[{Hunter(2007)}]{Hunter:2007}
Hunter JD. 2007.
\textit{Computing In Science \& Engineering} 9:90--95

\bibitem[{{Johnson}, {Chen} \& {Mulchaey}(2015{\natexlab{a}})}]{Johnson:2015a}
{Johnson} SD, {Chen} HW, {Mulchaey} JS. 2015{\natexlab{a}}.
\textit{\mnras} 452:2553--2565

\bibitem[{{Johnson}, {Chen} \& {Mulchaey}(2015{\natexlab{b}})}]{Johnson:2015b}
{Johnson} SD, {Chen} HW, {Mulchaey} JS. 2015{\natexlab{b}}.
\textit{\mnras} 449:3263--3273

\bibitem[{{Johnson} et~al.(2014){Johnson}, {Chen} et~al.}]{Johnson:2014}
{Johnson} SD, {Chen} HW, {Mulchaey} JS, et~al. 2014.
\textit{\mnras} 438:3039--3048

\bibitem[{{Kacprzak} et~al.(2012){Kacprzak}, {Churchill}
  et~al.}]{Kacprzak:2012}
{Kacprzak} GG, {Churchill} CW, {Steidel} CC, et~al. 2012.
\textit{\mnras} 427:3029--3043

\bibitem[{{Keel} et~al.(2012){Keel}, {Lintott} et~al.}]{Keel:2012}
{Keel} WC, {Lintott} CJ, {Schawinski} K, et~al. 2012.
\textit{\aj} 144:66

\bibitem[{{Kere{\v s}} \& {Hernquist}(2009)}]{Keres:2009a}
{Kere{\v s}} D, {Hernquist} L. 2009.
\textit{\apjl} 700:L1--L5

\bibitem[{{Kere{\v s}} et~al.(2005){Kere{\v s}}, {Katz} et~al.}]{Keres:2005}
{Kere{\v s}} D, {Katz} N, {Weinberg} DH, {Dav{\'e}} R. 2005.
\textit{\mnras} 363:2--28

\bibitem[{{Kollmeier} et~al.(2014){Kollmeier}, {Weinberg} et~al.}]{kollmeier14}
{Kollmeier} JA, {Weinberg} DH, {Oppenheimer} BD, et~al. 2014.
\textit{\apjl} 789:L32

\bibitem[{{Kornei} et~al.(2012){Kornei}, {Shapley} et~al.}]{Kornei:2012a}
{Kornei} KA, {Shapley} AE, {Martin} CL, et~al. 2012.
\textit{\apj} 758:135

\bibitem[{{Kwak} \& {Shelton}(2010)}]{kwak10}
{Kwak} K, {Shelton} RL. 2010.
\textit{\apj} 719:523--539

\bibitem[{{Lanzetta} et~al.(1995){Lanzetta}, {Bowen} et~al.}]{Lanzetta:1995}
{Lanzetta} KM, {Bowen} DV, {Tytler} D, {Webb} JK. 1995.
\textit{\apj} 442:538--568

\bibitem[{{Lehner} \& {Howk}(2011)}]{Lehner:2011}
{Lehner} N, {Howk} JC. 2011.
\textit{Science} 334:955

\bibitem[{{Lehner} et~al.(2013){Lehner}, {Howk} et~al.}]{Lehner:2013}
{Lehner} N, {Howk} JC, {Tripp} TM, et~al. 2013.
\textit{\apj} 770:138

\bibitem[{{Lehner}, {Howk} \& {Wakker}(2015)}]{Lehner:2015}
{Lehner} N, {Howk} JC, {Wakker} BP. 2015.
\textit{\apj} 804:79

\bibitem[{{Lehner} et~al.(2014){Lehner}, {O'Meara} et~al.}]{Lehner:2014a}
{Lehner} N, {O'Meara} JM, {Fox} AJ, et~al. 2014.
\textit{\apj} 788:119

\bibitem[{{Lehner} et~al.(2016){Lehner}, {O'Meara} et~al.}]{Lehner:2016}
{Lehner} N, {O'Meara} JM, {Howk} JC, et~al. 2016.
\textit{ArXiv e-prints}

\bibitem[{{Lehner} et~al.(2009){Lehner}, {Prochaska} et~al.}]{lehner09}
{Lehner} N, {Prochaska} JX, {Kobulnicky} HA, et~al. 2009.
\textit{\apj} 694:734--750

\bibitem[{{Lehnert} et~al.(2013){Lehnert}, {Le Tiran} et~al.}]{Lehnert:2013}
{Lehnert} MD, {Le Tiran} L, {Nesvadba} NPH, et~al. 2013.
\textit{\aap} 555:A72

\bibitem[{{Liang} \& {Chen}(2014)}]{Liang:2014}
{Liang} CJ, {Chen} HW. 2014.
\textit{\mnras} 445:2061--2081

\bibitem[{{Lilly} et~al.(2013){Lilly}, {Carollo} et~al.}]{Lilly:2013}
{Lilly} SJ, {Carollo} CM, {Pipino} A, et~al. 2013.
\textit{\apj} 772:119

\bibitem[{{Lu}, {Blanc} \& {Benson}(2015)}]{Lu:2015a}
{Lu} Y, {Blanc} GA, {Benson} A. 2015.
\textit{\apj} 808:129

\bibitem[{{Maller} \& {Bullock}(2004)}]{Maller:2004}
{Maller} AH, {Bullock} JS. 2004.
\textit{\mnras} 355:694--712

\bibitem[{{Martin}(2005)}]{Martin:2005}
{Martin} CL. 2005.
\textit{\apj} 621:227--245

\bibitem[{{Martin} et~al.(2012){Martin}, {Shapley} et~al.}]{Martin:2012}
{Martin} CL, {Shapley} AE, {Coil} AL, et~al. 2012.
\textit{\apj} 760:127

\bibitem[{{Martin} et~al.(2015){Martin}, {Matuszewski} et~al.}]{Martin:2015}
{Martin} DC, {Matuszewski} M, {Morrissey} P, et~al. 2015.
\textit{\nat} 524:192--195

\bibitem[{{Matejek} \& {Simcoe}(2012)}]{Matejek:2012}
{Matejek} MS, {Simcoe} RA. 2012.
\textit{\apj} 761:112

\bibitem[{{Mathes} et~al.(2014){Mathes}, {Churchill} et~al.}]{Mathes:2014a}
{Mathes} NL, {Churchill} CW, {Kacprzak} GG, et~al. 2014.
\textit{\apj} 792:128

\bibitem[{{McCourt} et~al.(2012){McCourt}, {Sharma} et~al.}]{McCourt:2012}
{McCourt} M, {Sharma} P, {Quataert} E, {Parrish} IJ. 2012.
\textit{\mnras} 419:3319--3337

\bibitem[{{McGaugh} et~al.(2010){McGaugh}, {Schombert} et~al.}]{McGaugh:2010}
{McGaugh} SS, {Schombert} JM, {de Blok} WJG, {Zagursky} MJ. 2010.
\textit{\apjl} 708:L14--L17

\bibitem[{{Meiring} et~al.(2013){Meiring}, {Tripp} et~al.}]{Meiring:2013}
{Meiring} JD, {Tripp} TM, {Werk} JK, et~al. 2013.
\textit{\apj} 767:49

\bibitem[{{M{\'e}nard} et~al.(2010){M{\'e}nard}, {Scranton}
  et~al.}]{Menard:2010}
{M{\'e}nard} B, {Scranton} R, {Fukugita} M, {Richards} G. 2010.
\textit{\mnras} 405:1025--1039

\bibitem[{{Mo} \& {Miralda-Escude}(1996)}]{Mo:1996}
{Mo} HJ, {Miralda-Escude} J. 1996.
\textit{\apj} 469:589

\bibitem[{{Morton}(2003)}]{Morton:2003}
{Morton} DC. 2003.
\textit{\apjs} 149:205--238

\bibitem[{{M{\"u}nch} \& {Zirin}(1961)}]{Munch:1961}
{M{\"u}nch} G, {Zirin} H. 1961.
\textit{\apj} 133:11

\bibitem[{{Muratov} et~al.(2016){Muratov}, {Keres} et~al.}]{Muratov:2016}
{Muratov} AL, {Keres} D, {Faucher-Giguere} CA, et~al. 2016.
\textit{ArXiv e-prints}

\bibitem[{{Muratov} et~al.(2015){Muratov}, {Kere{\v s}} et~al.}]{Muratov:2015}
{Muratov} AL, {Kere{\v s}} D, {Faucher-Gigu{\`e}re} CA, et~al. 2015.
\textit{\mnras} 454:2691--2713

\bibitem[{{Murray}, {M{\'e}nard} \& {Thompson}(2011)}]{Murray:2011}
{Murray} N, {M{\'e}nard} B, {Thompson} TA. 2011.
\textit{\apj} 735:66

\bibitem[{{Murray}, {Quataert} \& {Thompson}(2005)}]{Murray:2005}
{Murray} N, {Quataert} E, {Thompson} TA. 2005.
\textit{\apj} 618:569--585

\bibitem[{{Muzahid} et~al.(2015){Muzahid}, {Kacprzak} et~al.}]{muzahid15}
{Muzahid} S, {Kacprzak} GG, {Churchill} CW, et~al. 2015.
\textit{\apj} 811:132

\bibitem[{{Muzahid} et~al.(2012){Muzahid}, {Srianand} et~al.}]{muzahid12}
{Muzahid} S, {Srianand} R, {Bergeron} J, {Petitjean} P. 2012.
\textit{\mnras} 421:446--467

\bibitem[{{Narayanan} et~al.(2011){Narayanan}, {Savage}
  et~al.}]{Narayanan:2011}
{Narayanan} A, {Savage} BD, {Wakker} BP, et~al. 2011.
\textit{\apj} 730:15

\bibitem[{{Narayanan} et~al.(2010){Narayanan}, {Wakker}
  et~al.}]{Narayanan:2010}
{Narayanan} A, {Wakker} BP, {Savage} BD, et~al. 2010.
\textit{\apj} 721:960--974

\bibitem[{{Nelson} et~al.(2013){Nelson}, {Vogelsberger} et~al.}]{Nelson:2013}
{Nelson} D, {Vogelsberger} M, {Genel} S, et~al. 2013.
\textit{\mnras} 429:3353--3370

\bibitem[{{Nicastro} et~al.(2005){Nicastro}, {Mathur} et~al.}]{nicastro05}
{Nicastro} F, {Mathur} S, {Elvis} M, et~al. 2005.
\textit{\nat} 433:495--498

\bibitem[{{Nielsen} et~al.(2013){Nielsen}, {Churchill} et~al.}]{Nielsen:2013a}
{Nielsen} NM, {Churchill} CW, {Kacprzak} GG, {Murphy} MT. 2013.
\textit{\apj} 776:114

\bibitem[{{Nielsen} et~al.(2015){Nielsen}, {Churchill} et~al.}]{Nielsen:2015a}
{Nielsen} NM, {Churchill} CW, {Kacprzak} GG, et~al. 2015.
\textit{\apj} 812:83

\bibitem[{{Nielsen} et~al.(2016){Nielsen}, {Churchill} et~al.}]{Nielsen:2016}
{Nielsen} NM, {Churchill} CW, {Kacprzak} GG, et~al. 2016.
\textit{\apj} 818:171

\bibitem[{{Oppenheimer} et~al.(2016{\natexlab{b}}){Oppenheimer}, {Crain}
  et~al.}]{Oppenheimer:2016}
{Oppenheimer} BD, {Crain} RA, {Schaye} J, et~al. 2016{\natexlab{b}}.
\textit{\mnras} 460:2157--2179

\bibitem[{{Oppenheimer} \& {Dav{\'e}}(2006)}]{Oppenheimer:2006}
{Oppenheimer} BD, {Dav{\'e}} R. 2006.
\textit{\mnras} 373:1265--1292

\bibitem[{{Oppenheimer} \& {Dav{\'e}}(2008)}]{Oppenheimer:2008}
{Oppenheimer} BD, {Dav{\'e}} R. 2008.
\textit{\mnras} 387:577--600

\bibitem[{{Oppenheimer} et~al.(2010){Oppenheimer}, {Dav{\'e}}
  et~al.}]{Oppenheimer:2010}
{Oppenheimer} BD, {Dav{\'e}} R, {Kere{\v s}} D, et~al. 2010.
\textit{\mnras} 406:2325--2338

\bibitem[{{Oppenheimer} \& {Schaye}(2013{\natexlab{a}})}]{Oppenheimer:2013}
{Oppenheimer} BD, {Schaye} J. 2013{\natexlab{a}}.
\textit{\mnras} 434:1063--1078

\bibitem[{{Oppenheimer} \& {Schaye}(2013{\natexlab{b}})}]{oppenheimer13}
{Oppenheimer} BD, {Schaye} J. 2013{\natexlab{b}}.
\textit{\mnras} 434:1043--1062

\bibitem[{{Peek}, {M{\'e}nard} \& {Corrales}(2015)}]{Peek:2015}
{Peek} JEG, {M{\'e}nard} B, {Corrales} L. 2015.
\textit{\apj} 813:7

\bibitem[{{Peeples} et~al.(2014){Peeples}, {Werk} et~al.}]{Peeples:2014}
{Peeples} MS, {Werk} JK, {Tumlinson} J, et~al. 2014.
\textit{\apj} 786:54

\bibitem[{{Planck Collaboration} et~al.(2013){Planck Collaboration}, {Ade}
  et~al.}]{Planck:2013}
{Planck Collaboration}, {Ade} PAR, {Aghanim} N, et~al. 2013.
\textit{\aap} 557:A52

\bibitem[{{Prescott}, {Martin} \& {Dey}(2015)}]{Prescott:2015}
{Prescott} MKM, {Martin} CL, {Dey} A. 2015.
\textit{\apj} 799:62

\bibitem[{{Prochaska} et~al.(2004){Prochaska}, {Bloom} et~al.}]{prochaska04}
{Prochaska} JX, {Bloom} JS, {Chen} HW, et~al. 2004.
\textit{\apj} 611:200--207

\bibitem[{{Prochaska}, {Lau} \& {Hennawi}(2014)}]{Prochaska:2014}
{Prochaska} JX, {Lau} MW, {Hennawi} JF. 2014.
\textit{\apj} 796:140

\bibitem[{{Prochaska} et~al.(2011{\natexlab{a}}){Prochaska}, {Weiner}
  et~al.}]{Prochaska:2011c}
{Prochaska} JX, {Weiner} B, {Chen} HW, et~al. 2011{\natexlab{a}}.
\textit{\apjs} 193:28

\bibitem[{{Prochaska} et~al.(2011{\natexlab{b}}){Prochaska}, {Weiner}
  et~al.}]{Prochaska:2011a}
{Prochaska} JX, {Weiner} B, {Chen} HW, et~al. 2011{\natexlab{b}}.
\textit{\apj} 740:91

\bibitem[{{Prochaska} et~al.(2017){Prochaska}, {Werk} et~al.}]{Prochaska:2017}
{Prochaska} JX, {Werk} JK, {Worseck} G, et~al. 2017.
\textit{\apj} 837:169

\bibitem[{{Putman}, {Peek} \& {Joung}(2012{\natexlab{b}})}]{Putman:2012}
{Putman} ME, {Peek} JEG, {Joung} MR. 2012{\natexlab{b}}.
\textit{\araa} 50:491--529

\bibitem[{{Rauch} \& {Haehnelt}(2011)}]{Rauch:2011}
{Rauch} M, {Haehnelt} MG. 2011.
\textit{\mnras} 412:L55--L57

\bibitem[{{Rauch}, {Sargent} \& {Barlow}(2001)}]{rauch01}
{Rauch} M, {Sargent} WLW, {Barlow} TA. 2001.
\textit{\apj} 554:823--840

\bibitem[{{Richter} et~al.(2004){Richter}, {Savage} et~al.}]{richter04}
{Richter} P, {Savage} BD, {Tripp} TM, {Sembach} KR. 2004.
\textit{\apjs} 153:165--204

\bibitem[{{Robitaille} et~al.(2013){Robitaille}, {Tollerud} et~al.}]{Astropy}
{Robitaille} TP, {Tollerud} EJ, {Greenfield} P, et~al. 2013.
\textit{\aap} 558:A33

\bibitem[{{Rubin} et~al.(2015){Rubin}, {Hennawi} et~al.}]{Rubin:2015}
{Rubin} KHR, {Hennawi} JF, {Prochaska} JX, et~al. 2015.
\textit{\apj} 808:38

\bibitem[{{Rubin} et~al.(2012){Rubin}, {Prochaska} et~al.}]{Rubin:2012}
{Rubin} KHR, {Prochaska} JX, {Koo} DC, {Phillips} AC. 2012.
\textit{\apjl} 747:L26

\bibitem[{{Rubin} et~al.(2014){Rubin}, {Prochaska} et~al.}]{Rubin:2014}
{Rubin} KHR, {Prochaska} JX, {Koo} DC, et~al. 2014.
\textit{\apj} 794:156

\bibitem[{{Rudie} et~al.(2012){Rudie}, {Steidel} et~al.}]{Rudie:2012}
{Rudie} GC, {Steidel} CC, {Trainor} RF, et~al. 2012.
\textit{\apj} 750:67

\bibitem[{{Salem}, {Bryan} \& {Corlies}(2016)}]{Salem:2016}
{Salem} M, {Bryan} GL, {Corlies} L. 2016.
\textit{\mnras} 456:582--601

\bibitem[{{Sargent} et~al.(1980){Sargent}, {Young} et~al.}]{Sargent:1980}
{Sargent} WLW, {Young} PJ, {Boksenberg} A, {Tytler} D. 1980.
\textit{\apjs} 42:41--81

\bibitem[{{Savage} et~al.(2014){Savage}, {Kim} et~al.}]{Savage:2014}
{Savage} BD, {Kim} TS, {Wakker} BP, et~al. 2014.
\textit{\apjs} 212:8

\bibitem[{{Savage}, {Lehner} \& {Narayanan}(2011)}]{Savage:2011a}
{Savage} BD, {Lehner} N, {Narayanan} A. 2011.
\textit{\apj} 743:180

\bibitem[{{Schawinski} et~al.(2014){Schawinski}, {Urry}
  et~al.}]{Schawinski:2014}
{Schawinski} K, {Urry} CM, {Simmons} BD, et~al. 2014.
\textit{\mnras} 440:889--907

\bibitem[{{Schaye}, {Carswell} \& {Kim}(2007)}]{schaye07}
{Schaye} J, {Carswell} RF, {Kim} TS. 2007.
\textit{\mnras} 379:1169--1194

\bibitem[{{Schaye} et~al.(2015){Schaye}, {Crain} et~al.}]{Schaye:2015}
{Schaye} J, {Crain} RA, {Bower} RG, et~al. 2015.
\textit{\mnras} 446:521--554

\bibitem[{{Schiminovich} et~al.(2010){Schiminovich}, {Catinella}
  et~al.}]{Schiminovich:2010}
{Schiminovich} D, {Catinella} B, {Kauffmann} G, et~al. 2010.
\textit{\mnras} 408:919--934

\bibitem[{{Sembach} et~al.(2004){Sembach}, {Tripp} et~al.}]{sembach04}
{Sembach} KR, {Tripp} TM, {Savage} BD, {Richter} P. 2004.
\textit{\apjs} 155:351--393

\bibitem[{{Sembach} et~al.(2003){Sembach}, {Wakker} et~al.}]{sembach03}
{Sembach} KR, {Wakker} BP, {Savage} BD, et~al. 2003.
\textit{\apjs} 146:165--208

\bibitem[{{Shapiro} \& {Field}(1976)}]{shapiro76}
{Shapiro} PR, {Field} GB. 1976.
\textit{\apj} 205:762--765

\bibitem[{{Shattow}, {Croton} \& {Bibiano}(2015)}]{Shattow:2015}
{Shattow} GM, {Croton} DJ, {Bibiano} A. 2015.
\textit{\mnras} 450:2306--2316

\bibitem[{{Shen} et~al.(2012){Shen}, {Madau} et~al.}]{Shen:2012a}
{Shen} S, {Madau} P, {Aguirre} A, et~al. 2012.
\textit{\apj} 760:50

\bibitem[{{Shen} et~al.(2013){Shen}, {Madau} et~al.}]{Shen:2013}
{Shen} S, {Madau} P, {Guedes} J, et~al. 2013.
\textit{\apj} 765:89

\bibitem[{{Silvia}(2013)}]{Silvia:2013}
{Silvia} DW. 2013.
\textit{{}}.
Ph.D. thesis, University of Colorado at Boulder

\bibitem[{{Slavin}, {Shull} \& {Begelman}(1993)}]{slavin93}
{Slavin} JD, {Shull} JM, {Begelman} MC. 1993.
\textit{\apj} 407:83--99

\bibitem[{{Somerville} \& {Dav{\'e}}(2015)}]{Somerville:2015b}
{Somerville} RS, {Dav{\'e}} R. 2015.
\textit{\araa} 53:51--113

\bibitem[{{Somerville}, {Popping} \& {Trager}(2015)}]{Somerville:2015a}
{Somerville} RS, {Popping} G, {Trager} SC. 2015.
\textit{\mnras} 453:4337--4367

\bibitem[{{Spitzer}(1956)}]{Spitzer:1956}
{Spitzer} Jr. L. 1956.
\textit{\apj} 124:20

\bibitem[{{Springel}, {Di Matteo} \& {Hernquist}(2005)}]{Springel:2005}
{Springel} V, {Di Matteo} T, {Hernquist} L. 2005.
\textit{\mnras} 361:776--794

\bibitem[{{Steidel} et~al.(2010){Steidel}, {Erb} et~al.}]{Steidel:2010}
{Steidel} CC, {Erb} DK, {Shapley} AE, et~al. 2010.
\textit{\apj} 717:289--322

\bibitem[{{Stern} et~al.(2016){Stern}, {Hennawi} et~al.}]{Stern:2016}
{Stern} J, {Hennawi} JF, {Prochaska} JX, {Werk} JK. 2016.
\textit{ArXiv e-prints}

\bibitem[{{Stewart} et~al.(2011){Stewart}, {Kaufmann} et~al.}]{Stewart:2011a}
{Stewart} KR, {Kaufmann} T, {Bullock} JS, et~al. 2011.
\textit{\apj} 738:39

\bibitem[{{Stinson} et~al.(2012){Stinson}, {Brook} et~al.}]{Stinson:2012}
{Stinson} GS, {Brook} C, {Prochaska} JX, et~al. 2012.
\textit{\mnras} 425:1270--1277

\bibitem[{{Stocke} et~al.(2013){Stocke}, {Keeney} et~al.}]{Stocke:2013}
{Stocke} JT, {Keeney} BA, {Danforth} CW, et~al. 2013.
\textit{\apj} 763:148

\bibitem[{{Stocke} et~al.(2006){Stocke}, {Penton} et~al.}]{Stocke:2006}
{Stocke} JT, {Penton} SV, {Danforth} CW, et~al. 2006.
\textit{\apj} 641:217--228

\bibitem[{{Strickland} \& {Heckman}(2009)}]{strickland09}
{Strickland} DK, {Heckman} TM. 2009.
\textit{\apj} 697:2030--2056

\bibitem[{{Suresh} et~al.(2015){Suresh}, {Rubin} et~al.}]{Suresh:2015a}
{Suresh} J, {Rubin} KHR, {Kannan} R, et~al. 2015.
\textit{ArXiv e-prints}

\bibitem[{{Suresh} et~al.(2017){Suresh}, {Rubin} et~al.}]{suresh17}
{Suresh} J, {Rubin} KHR, {Kannan} R, et~al. 2017.
\textit{\mnras} 465:2966--2982

\bibitem[{{Tejos} et~al.(2016){Tejos}, {Prochaska} et~al.}]{Tejos:2016}
{Tejos} N, {Prochaska} JX, {Crighton} NHM, et~al. 2016.
\textit{\mnras} 455:2662--2697

\bibitem[{{Tepper-Garc{\'{\i}}a}, {Bland-Hawthorn} \&
  {Sutherland}(2015)}]{Tepper-Garcia:2015}
{Tepper-Garc{\'{\i}}a} T, {Bland-Hawthorn} J, {Sutherland} RS. 2015.
\textit{\apj} 813:94

\bibitem[{{Thom} et~al.(2012){Thom}, {Tumlinson} et~al.}]{Thom:2012}
{Thom} C, {Tumlinson} J, {Werk} JK, et~al. 2012.
\textit{\apjl} 758:L41

\bibitem[{{Thompson} et~al.(2016){Thompson}, {Quataert} et~al.}]{Thompson:2016}
{Thompson} TA, {Quataert} E, {Zhang} D, {Weinberg} DH. 2016.
\textit{\mnras} 455:1830--1844

\bibitem[{{Tinsley}(1980)}]{tinsley80a}
{Tinsley} BM. 1980.
\textit{Fund. Cosm. Phys.} 5:287--388

\bibitem[{{Tremonti} et~al.(2004){Tremonti}, {Heckman} et~al.}]{Tremonti:2004}
{Tremonti} CA, {Heckman} TM, {Kauffmann} G, et~al. 2004.
\textit{\apj} 613:898--913

\bibitem[{{Tripp} et~al.(2001){Tripp}, {Giroux} et~al.}]{tripp01}
{Tripp} TM, {Giroux} ML, {Stocke} JT, et~al. 2001.
\textit{\apj} 563:724--735

\bibitem[{{Tripp} et~al.(2011){Tripp}, {Meiring} et~al.}]{Tripp:2011}
{Tripp} TM, {Meiring} JD, {Prochaska} JX, et~al. 2011.
\textit{Science} 334:952

\bibitem[{{Tripp} et~al.(2008){Tripp}, {Sembach} et~al.}]{Tripp:2008}
{Tripp} TM, {Sembach} KR, {Bowen} DV, et~al. 2008.
\textit{\apjs} 177:39--102

\bibitem[{{Tumlinson} et~al.(2005){Tumlinson}, {Shull} et~al.}]{tumlinson05}
{Tumlinson} J, {Shull} JM, {Giroux} ML, {Stocke} JT. 2005.
\textit{\apj} 620:95--112

\bibitem[{{Tumlinson} et~al.(2011){Tumlinson}, {Thom} et~al.}]{Tumlinson:2011}
{Tumlinson} J, {Thom} C, {Werk} JK, et~al. 2011.
\textit{Science} 334:948

\bibitem[{{Tumlinson} et~al.(2013){Tumlinson}, {Thom} et~al.}]{Tumlinson:2013}
{Tumlinson} J, {Thom} C, {Werk} JK, et~al. 2013.
\textit{\apj} 777:59

\bibitem[{{Turk} et~al.(2011){Turk}, {Smith} et~al.}]{turk11}
{Turk} MJ, {Smith} BD, {Oishi} JS, et~al. 2011.
\textit{\apjs} 192:9

\bibitem[{{Turner} et~al.(2016){Turner}, {Schaye} et~al.}]{Turner:2016}
{Turner} ML, {Schaye} J, {Crain} RA, et~al. 2016.
\textit{\mnras} 462:2440--2464

\bibitem[{{Turner} et~al.(2014){Turner}, {Schaye} et~al.}]{Turner:2014}
{Turner} ML, {Schaye} J, {Steidel} CC, et~al. 2014.
\textit{\mnras} 445:794--822

\bibitem[{{Turner} et~al.(2015){Turner}, {Schaye} et~al.}]{Turner:2015}
{Turner} ML, {Schaye} J, {Steidel} CC, et~al. 2015.
\textit{\mnras} 450:2067--2082

\bibitem[{{van de Voort} et~al.(2012){van de Voort}, {Schaye}
  et~al.}]{vandeVoort:2012b}
{van de Voort} F, {Schaye} J, {Altay} G, {Theuns} T. 2012.
\textit{\mnras} 421:2809--2819

\bibitem[{{Veilleux}, {Cecil} \& {Bland-Hawthorn}(2005)}]{Veilleux:2005}
{Veilleux} S, {Cecil} G, {Bland-Hawthorn} J. 2005.
\textit{\araa} 43:769--826

\bibitem[{{Vikhlinin} et~al.(2006){Vikhlinin}, {Kravtsov}
  et~al.}]{Vikhlinin:2005}
{Vikhlinin} A, {Kravtsov} A, {Forman} W, et~al. 2006.
\textit{\apj} 640:691--709

\bibitem[{{Vogelsberger} et~al.(2014){Vogelsberger}, {Genel}
  et~al.}]{Vogelsberger:2014}
{Vogelsberger} M, {Genel} S, {Springel} V, et~al. 2014.
\textit{\mnras} 444:1518--1547

\bibitem[{{Voit} et~al.(2015{\natexlab{a}}){Voit}, {Bryan} et~al.}]{Voit:2015d}
{Voit} GM, {Bryan} GL, {O'Shea} BW, {Donahue} M. 2015{\natexlab{a}}.
\textit{\apjl} 808:L30

\bibitem[{{Voit} et~al.(2015{\natexlab{b}}){Voit}, {Donahue}
  et~al.}]{Voit:2015b}
{Voit} GM, {Donahue} M, {Bryan} GL, {McDonald} M. 2015{\natexlab{b}}.
\textit{\nat} 519:203--206

\bibitem[{{Wakker} et~al.(2015){Wakker}, {Hernandez} et~al.}]{Wakker:2015}
{Wakker} BP, {Hernandez} AK, {French} DM, et~al. 2015.
\textit{\apj} 814:40

\bibitem[{{Wakker} et~al.(2012){Wakker}, {Savage} et~al.}]{wakker12}
{Wakker} BP, {Savage} BD, {Fox} AJ, et~al. 2012.
\textit{\apj} 749:157

\bibitem[{{Walker}, {Bagchi} \& {Fabian}(2015)}]{Walker:2015}
{Walker} SA, {Bagchi} J, {Fabian} AC. 2015.
\textit{\mnras} 449:3527--3534

\bibitem[{{Wang}(1995)}]{wang95}
{Wang} B. 1995.
\textit{\apj} 444:590--609

\bibitem[{{Wang} \& {Yao}(2012)}]{Wang:2012}
{Wang} QD, {Yao} Y. 2012.
\textit{ArXiv e-prints}

\bibitem[{{Werk} et~al.(2016){Werk}, {Prochaska} et~al.}]{werk16}
{Werk} JK, {Prochaska} JX, {Cantalupo} S, et~al. 2016.
\textit{ArXiv e-prints}

\bibitem[{{Werk} et~al.(2012){Werk}, {Prochaska} et~al.}]{Werk:2012}
{Werk} JK, {Prochaska} JX, {Thom} C, et~al. 2012.
\textit{\apjs} 198:3

\bibitem[{{Werk} et~al.(2013){Werk}, {Prochaska} et~al.}]{Werk:2013}
{Werk} JK, {Prochaska} JX, {Thom} C, et~al. 2013.
\textit{\apjs} 204:17

\bibitem[{{Werk} et~al.(2014){Werk}, {Prochaska} et~al.}]{Werk:2014}
{Werk} JK, {Prochaska} JX, {Tumlinson} J, et~al. 2014.
\textit{\apj} 792:8

\bibitem[{{Whitaker} et~al.(2012){Whitaker}, {van Dokkum} et~al.}]{whitaker12}
{Whitaker} KE, {van Dokkum} PG, {Brammer} G, {Franx} M. 2012.
\textit{\apjl} 754:L29

\bibitem[{{Williams} et~al.(2005){Williams}, {Mathur} et~al.}]{williams05}
{Williams} RJ, {Mathur} S, {Nicastro} F, et~al. 2005.
\textit{\apj} 631:856--867

\bibitem[{{Wotta} et~al.(2016){Wotta}, {Lehner} et~al.}]{Wotta:2016}
{Wotta} CB, {Lehner} N, {Howk} JC, et~al. 2016.
\textit{ArXiv e-prints}

\bibitem[{{Wu}, {Fabian} \& {Nulsen}(2001)}]{Wu:2001}
{Wu} KKS, {Fabian} AC, {Nulsen} PEJ. 2001.
\textit{\mnras} 324:95--107

\bibitem[{{Yao} et~al.(2012){Yao}, {Shull} et~al.}]{Yao:2012}
{Yao} Y, {Shull} JM, {Wang} QD, {Cash} W. 2012.
\textit{\apj} 746:166

\bibitem[{{Yao} et~al.(2010){Yao}, {Wang} et~al.}]{Yao:2010a}
{Yao} Y, {Wang} QD, {Penton} SV, et~al. 2010.
\textit{\apj} 716:1514--1521

\bibitem[{{York} et~al.(2006){York}, {Khare} et~al.}]{York:2006}
{York} DG, {Khare} P, {Vanden Berk} D, et~al. 2006.
\textit{\mnras} 367:945--978

\bibitem[{{Zahid} et~al.(2012){Zahid}, {Bresolin} et~al.}]{Zahid:2012b}
{Zahid} HJ, {Bresolin} F, {Kewley} LJ, et~al. 2012.
\textit{\apj} 750:120

\bibitem[{{Zhang} et~al.(2016){Zhang}, {Zaritsky} et~al.}]{Zhang:2016}
{Zhang} H, {Zaritsky} D, {Zhu} G, et~al. 2016.
\textit{\apj} 833:276

\bibitem[{{Zheng} et~al.(2017){Zheng}, {Peek} et~al.}]{zheng17}
{Zheng} Y, {Peek} JEG, {Werk} JK, {Putman} ME. 2017.
\textit{\apj} 834:179

\bibitem[{{Zhu} \& {M{\'e}nard}(2013{\natexlab{a}})}]{Zhu:2013b}
{Zhu} G, {M{\'e}nard} B. 2013{\natexlab{a}}.
\textit{\apj} 773:16

\bibitem[{{Zhu} \& {M{\'e}nard}(2013{\natexlab{b}})}]{Zhu:2013c}
{Zhu} G, {M{\'e}nard} B. 2013{\natexlab{b}}.
\textit{\apj} 770:130

\bibitem[{{Zhu} et~al.(2014){Zhu}, {M{\'e}nard} et~al.}]{Zhu:2014}
{Zhu} G, {M{\'e}nard} B, {Bizyaev} D, et~al. 2014.
\textit{\mnras} 439:3139--3155

\bibitem[{{Zhu} et~al.(2013){Zhu}, {Feng} et~al.}]{Zhu:2013a}
{Zhu} W, {Feng} Ll, {Xia} Y, et~al. 2013.
\textit{\apj} 777:48

\bibitem[{{Zu} et~al.(2011){Zu}, {Weinberg} et~al.}]{Zu:2011}
{Zu} Y, {Weinberg} DH, {Dav{\'e}} R, et~al. 2011.
\textit{\mnras} 412:1059--1069

\end{thebibliography}

\end{document}